\PassOptionsToPackage{unicode}{hyperref}
\PassOptionsToPackage{hyphens}{url}
\PassOptionsToPackage{dvipsnames,svgnames,x11names}{xcolor}
\documentclass[
  letterpaper,
]{article}
\usepackage{xcolor}
\usepackage[margin=1in]{geometry}
\usepackage{amsmath,amssymb}
\usepackage{iftex}
\ifPDFTeX
  \usepackage[T1]{fontenc}
  \usepackage{textcomp} % provide euro and other symbols
\else % if luatex or xetex
  \usepackage{unicode-math} % this also loads fontspec
  \defaultfontfeatures{Scale=MatchLowercase}
  \defaultfontfeatures[\rmfamily]{Ligatures=TeX,Scale=1}
\fi
\usepackage{lmodern}
\ifPDFTeX\else
\fi
\IfFileExists{upquote.sty}{\usepackage{upquote}}{}
\IfFileExists{microtype.sty}{% use microtype if available
  \usepackage[]{microtype}
  \UseMicrotypeSet[protrusion]{basicmath} % disable protrusion for tt fonts
}{}
\makeatletter
\@ifundefined{KOMAClassName}{% if non-KOMA class
  \IfFileExists{parskip.sty}{%
    \usepackage{parskip}
  }{% else
    \setlength{\parindent}{0pt}
    \setlength{\parskip}{6pt plus 2pt minus 1pt}}
}{% if KOMA class
  \KOMAoptions{parskip=half}}
\makeatother
\makeatletter
\ifx\paragraph\undefined\else
  \let\oldparagraph\paragraph
  \renewcommand{\paragraph}{
    \@ifstar
      \xxxParagraphStar
      \xxxParagraphNoStar
  }
  \newcommand{\xxxParagraphStar}[1]{\oldparagraph*{#1}\mbox{}}
  \newcommand{\xxxParagraphNoStar}[1]{\oldparagraph{#1}\mbox{}}
\fi
\ifx\subparagraph\undefined\else
  \let\oldsubparagraph\subparagraph
  \renewcommand{\subparagraph}{
    \@ifstar
      \xxxSubParagraphStar
      \xxxSubParagraphNoStar
  }
  \newcommand{\xxxSubParagraphStar}[1]{\oldsubparagraph*{#1}\mbox{}}
  \newcommand{\xxxSubParagraphNoStar}[1]{\oldsubparagraph{#1}\mbox{}}
\fi
\makeatother

\usepackage{longtable,booktabs,array}
\usepackage{multirow}
\usepackage{calc} % for calculating minipage widths
\usepackage{etoolbox}
\makeatletter
\patchcmd\longtable{\par}{\if@noskipsec\mbox{}\fi\par}{}{}
\makeatother
\IfFileExists{footnotehyper.sty}{\usepackage{footnotehyper}}{\usepackage{footnote}}
\makesavenoteenv{longtable}
\usepackage{graphicx}
\makeatletter
\newsavebox\pandoc@box
\newcommand*\pandocbounded[1]{% scales image to fit in text height/width
  \sbox\pandoc@box{#1}%
  \Gscale@div\@tempa{\textheight}{\dimexpr\ht\pandoc@box+\dp\pandoc@box\relax}%
  \Gscale@div\@tempb{\linewidth}{\wd\pandoc@box}%
  \ifdim\@tempb\p@<\@tempa\p@\let\@tempa\@tempb\fi% select the smaller of both
  \ifdim\@tempa\p@<\p@\scalebox{\@tempa}{\usebox\pandoc@box}%
  \else\usebox{\pandoc@box}%
  \fi%
}
\def\fps@figure{htbp}
\makeatother

\providecommand{\tightlist}{%
  \setlength{\itemsep}{0pt}\setlength{\parskip}{0pt}}

\usepackage[]{natbib}
\DeclareUnicodeCharacter{2212}{\ensuremath{-}}
\DeclareUnicodeCharacter{2013}{--}
\DeclareUnicodeCharacter{2014}{---}
\DeclareUnicodeCharacter{2009}{\,}
\DeclareUnicodeCharacter{00A0}{~}
\DeclareUnicodeCharacter{00D7}{\ensuremath{\times}}
\DeclareUnicodeCharacter{2264}{\ensuremath{\leq}}
\DeclareUnicodeCharacter{2265}{\ensuremath{\geq}}
\DeclareUnicodeCharacter{2260}{\ensuremath{\neq}}
\DeclareUnicodeCharacter{2248}{\ensuremath{\approx}}
\DeclareUnicodeCharacter{03B1}{\ensuremath{\alpha}}
\DeclareUnicodeCharacter{03B2}{\ensuremath{\beta}}
\DeclareUnicodeCharacter{03C3}{\ensuremath{\sigma}}
\DeclareUnicodeCharacter{03C7}{\ensuremath{\chi}}
\DeclareUnicodeCharacter{2032}{\ensuremath{'}}

\usepackage{caption}
\setcitestyle{round}

\AtBeginEnvironment{apptbl}{\footnotesize}
\makeatletter
\newcommand\table@defaultsize{\footnotesize}
\let\table@size\table@defaultsize
\newcommand\nexttablesize[1]{\gdef\table@size{#1}}
\AtBeginEnvironment{longtable}{\table@size\global\let\table@size\table@defaultsize}
\makeatother

\makeatletter
\let\floatnote@text\@empty
\newcommand\floatnote[1]{\gdef\floatnote@text{#1}}
\newcommand\fs@plainnote{\fs@plain
  \def\@fs@post{\ifx\floatnote@text\@empty\else
    \vspace\abovecaptionskip\noindent\floatnote@text\par
    \global\let\floatnote@text\@empty\fi}}
\makeatother

\AtBeginDocument{%
  \floatstyle{plainnote}\restylefloat{appfig}%
  \floatplacement{appfig}{H}\floatplacement{apptbl}{H}}

\AtBeginDocument{}
\makeatletter
\renewcommand\@maketitle{%
  \newpage\null\vskip 0.5em
  \begin{center}
    {\Large\bfseries \@title\par}
    \vskip 1.5em
    {\normalsize \@author\par}
  \end{center}
  \par\vskip 1em}
\newcommand\aclauthor[3]{% name, affiliation lines, email
  \begin{tabular}[t]{@{}c@{}}
    \textbf{#1}\\[1pt]
    #2\\[1pt]
    {\small\texttt{#3}}
  \end{tabular}}
\makeatother

\AtBeginDocument{%
  \author{%
    \aclauthor{Davood Wadi}
      {Desautels Faculty of Management\\ McGill University\\ Montreal, Quebec, Canada}
      {davood.wadi@hec.ca}%
    \hspace{3em}%
    \aclauthor{Yu Ma}
      {Desautels Faculty of Management\\ McGill University\\ Montreal, Quebec, Canada}
      {yu.ma@mcgill.ca}%
  }%
}
\makeatletter
\@ifpackageloaded{float}{}{\usepackage{float}}
\floatstyle{plain}
\@ifundefined{c@chapter}{\newfloat{apptbl}{h}{loapptbl}}{\newfloat{apptbl}{h}{loapptbl}[chapter]}
\floatname{apptbl}{Table W}
\floatstyle{plaintop}
\restylefloat{apptbl}
\newcommand*\quartoapptblref[1]{Table \hyperref[#1]{W\ref{#1}}}
\@ifpackageloaded{caption}{}{\usepackage{caption}}
\DeclareCaptionLabelFormat{quartoapptblreflabelformat}{#1#2}
\makeatother
\makeatletter
\@ifpackageloaded{float}{}{\usepackage{float}}
\floatstyle{plain}
\@ifundefined{c@chapter}{\newfloat{appfig}{h}{loappfig}}{\newfloat{appfig}{h}{loappfig}[chapter]}
\floatname{appfig}{Figure W}
\newcommand*\quartoappfigref[1]{Figure \hyperref[#1]{W\ref{#1}}}
\@ifpackageloaded{caption}{}{\usepackage{caption}}
\DeclareCaptionLabelFormat{quartoappfigreflabelformat}{#1#2}
\makeatother
\makeatletter
\@ifpackageloaded{float}{}{\usepackage{float}}
\floatstyle{plain}
\@ifundefined{c@chapter}{\newfloat{appenv}{h}{loapp}}{\newfloat{appenv}{h}{loapp}[chapter]}
\floatname{appenv}{Web Appendix}
\floatstyle{plaintop}
\restylefloat{appenv}
\newcommand*\quartoappref[1]{Web \hyperref[#1]{Appendix\ref{#1}}}
\@ifpackageloaded{caption}{}{\usepackage{caption}}
\DeclareCaptionLabelFormat{quartoappreflabelformat}{#1#2}
\makeatother
\makeatletter
\@ifpackageloaded{caption}{}{\usepackage{caption}}
\AtBeginDocument{%
\ifdefined\contentsname
  \renewcommand*\contentsname{Table of contents}
\else
  \newcommand\contentsname{Table of contents}
\fi
\ifdefined\listfigurename
  \renewcommand*\listfigurename{List of Figures}
\else
  \newcommand\listfigurename{List of Figures}
\fi
\ifdefined\listtablename
  \renewcommand*\listtablename{List of Tables}
\else
  \newcommand\listtablename{List of Tables}
\fi
\ifdefined\figurename
  \renewcommand*\figurename{Figure}
\else
  \newcommand\figurename{Figure}
\fi
\ifdefined\tablename
  \renewcommand*\tablename{Table}
\else
  \newcommand\tablename{Table}
\fi
}
\@ifpackageloaded{float}{}{\usepackage{float}}
\floatstyle{ruled}
\@ifundefined{c@chapter}{\newfloat{codelisting}{h}{lop}}{\newfloat{codelisting}{h}{lop}[chapter]}
\floatname{codelisting}{Listing}

\makeatother
\makeatletter
\@ifpackageloaded{caption}{}{\usepackage{caption}}
\@ifpackageloaded{subcaption}{}{\usepackage{subcaption}}
\makeatother
\usepackage{bookmark}
\IfFileExists{xurl.sty}{\usepackage{xurl}}{} % add URL line breaks if available
\makeatletter
\@ifundefined{xmpquote}{}{}
\makeatother
\hypersetup{
  pdftitle={Shopping by algorithm: How agentic AI deploys human heuristics as a surrogate consumer},
  pdfauthor={Davood Wadi; Yu Ma},
  colorlinks=true,
  linkcolor={blue},
  filecolor={Maroon},
  citecolor={Blue},
  urlcolor={Blue},
  pdfcreator={LaTeX via pandoc}}

\title{Shopping by algorithm: How agentic AI deploys human heuristics as
a surrogate consumer}
\author{Davood Wadi \and Yu Ma}
\date{}
\begin{document}
\maketitle

\textbf{Abstract}\\
Consumers increasingly delegate purchasing decisions to Large Language
Models (LLMs) acting as surrogate consumers. We develop ``Tool-Lab,'' an
adaptation of information-board process tracing that places product
attributes behind costly tool calls, and use it to examine how marketing
pricing cues (i.e., just-below pricing and promotional framing)
influence AI shopping agents. Across eight commercially deployed LLMs
from three providers, we trace pre-choice information acquisition. Under
zero cost, pricing cues rarely mislead. Imposing acquisition costs under
a vague goal prompt leads LLMs to omit diagnostic attributes required to
compute unit price and make suboptimal choices resembling human
heuristics. Relative to a specific goal prompt that largely preserves
diagnostic search and choice optimality, a vague goal prompt under
constraints creates a search-mediated vulnerability for AI agents. This
research demonstrates that the efficacy of marketing cues in delegated
AI shopping is governed by storefront information architecture combined
with consumers' instructions, not necessarily immutable LLM
characteristics.


\section{Introduction}\label{introduction}

The evaluation of market alternatives has historically relied on human
judgment. Increasingly, however, consumers are delegating end-to-end
purchasing tasks to autonomous Large Language Models (LLMs) deployed as
surrogate consumers
\citep{adobe2025genai, aw2025future, hasselwander2026toward, puntoni2024being},
that is, an external entity to whom a principal relinquishes decision
authority to navigate the marketplace \citep{solomon1986missing}. Major
platforms have begun deploying agents that search, compare, and complete
purchases on consumers' behalf
\citep{verdon2026retailers, sataua2026meta, kahn2026shopping}. While
surrogate consumers have traditionally been humans, delegating this role
to AI agents represents a fundamental shift in who evaluates the market
\citep{dholakia2025artificial}.

Since for decades, retail pricing strategies, such as just-below pricing
\citep{lacetera2012heuristic, stiving1997empirical, thomas2005penny} and
promotional discount framing
\citep{inman1990promotion, thaler1985mental}, have been engineered to
exploit human bounded rationality, practitioners already advise
retailers to structure pricing and product attributes so that AI agents
can evaluate them \citep{cooper2026visa}. Still, when the buyer is an
autonomous AI agent, it remains an open question whether pricing cues
built for human cognitive limitations continue to influence purchasing,
and if so, how.

Research evaluating LLMs as economic actors has found them to produce
both rational
\citep{brand2023using, cillo2025generative, horton2023large} and
heuristic-like behavior
\citep{binz2023using, filandrianos2025bias, hermann2025generative}.
Reconciling this duality has been proven difficult because the existing
literature predominantly treats heuristic-like behavior as an inherent,
static defect hardwired into LLMs during training. As a result, research
has largely documented whether LLMs exhibit heuristic-like behavior, but
not what produces it.

We propose that heuristic-like purchasing in delegated AI shopping can
be a property of the information state that the AI agent holds at the
moment it commits to a choice. This pre-choice information state can be
jointly shaped by retailers and consumers. Retailers determine the cost
of retrieving product attributes through platform design, and consumers
determine, through the specificity of their delegating prompt, which
attributes the AI agent treats as diagnostic. The combination of costly
retrieval and vague goal prompts leads AI agents to truncate their
information search and choose suboptimally.

To test this claim, we introduce Tool-Lab, an experimental paradigm that
adapts classical information-board process tracing
\citep{johnson1989monitoring, payne1993adaptive, willemsen2011visiting}
to multi-turn LLM interactions. Rather than presenting product
attributes simultaneously within a single prompt, Tool-Lab embeds an LLM
within an interactive environment where product attributes are hidden
behind costly tool calls. To inspect an attribute, the LLM must execute
a tool call (\emph{inspect\_cell}), which incurs an explicit acquisition
cost. Tool-Lab records the observable sequence of information
acquisitions prior to choice and allows researchers to measure search
depth and composition directly. Therefore, Tool-Lab allows us to
manipulate retrieval cost and prompt specificity while observing the
search they produce. This makes the pathway from environmental
constraint to final choice through information acquisition measurable,
rather than inferred from choice outcomes alone.

Across two primary studies, several robustness tests, and two validation
studies (\(N = 33{,}704\) total trials), we evaluate eight commercially
deployed LLMs from three major providers (Google, OpenAI, and
Anthropic). Each study establishes a full information baseline, then
contrasts LLM choices with a constrained Tool-Lab environment where
information acquisition incurs financial costs (\(c = \$0\)
vs.~\(\$10.00\) or \(\$50.00\)), and the consumer prompt's goal is
specific or vague (``find the lowest price per ounce'' vs.~``find the
best deal'').

The empirical results document two consistent findings. First, under
full information and specific goal prompt, the evaluated LLMs make
decisions that largely align with normatively optimal behavior. Second,
when acquisition costs are imposed under a vague goal prompt, most LLMs
decrease search depth, and the impact on choice optimality depends on
whether the omitted attributes are diagnostic (i.e., they are necessary
to calculate the unit price). Because vague prompts lack a clear
normative objective, LLMs abandon some diagnostic attributes (i.e.,
cents or product weight). Without the attributes that are required to
calculate unit prices, the AI agents make suboptimal decisions that are
consistent with human heuristic decision making. Validation experiments
(\(N = 5{,}200\)) show that choice suboptimality stems from missing
diagnostic attributes rather than arithmetic calculation failures for
most LLMs.

This research makes three contributions to marketing theory of agentic
AI shopping. First, it challenges the assumption that heuristic-like
behavior in LLMs as shopping assistants is solely inherited from
training data, by showing that susceptibility to marketing cues depends
on the information state in which AI agents choose. Second, it
identifies the principal's delegation instruction as a determinant of
whether marketing cues influence the AI agent's choice. Third, it
introduces Tool-Lab, a process-tracing method that makes the information
acquisition of AI agents observable and allows researchers to study how
AI agents make decisions.

These findings have implications for four stakeholders. Retailers should
make diagnostic attributes (e.g., ``price'', ``weight'') readily
available to AI shopping agents. Consumers should clearly state the
criterion that defines a good choice (e.g., ``lowest price per ounce'')
rather than issuing vague requests such as ``find the best deal.'' AI
agent developers should design agents to clarify vague objectives before
searching and should audit the specific LLM they deploy for acquisition
truncation. Finally, regulators should consider extending disclosure
rules, such as unit-price requirements, to the product information made
available to AI agents.

\section{Literature review}\label{literature-review}

\subsection{AI agents as surrogate
consumers}\label{ai-agents-as-surrogate-consumers}

Marketing scholarship has examined Generative Artificial Intelligence
(GenAI) primarily through firm-side strategy
\citep{prasanna2025generative, riandhi2025ai, grewal2025generative, kumar2026transformative},
yet the critical frontier in marketing theory centers on the autonomous
AI agent itself
\citep{aw2025future, dholakia2025artificial, puntoni2024being}.
Behavioral research in this domain predominantly treats Large Language
Models (LLMs) as passive ``silicon samples'' or synthetic respondent
pools \citep{joerling2025integrating, sarstedt2024using} to emulate
human survey responses and aid in marketing research
\citep{arora2025ai, li2025consumer, Liu2026WhoDT, wang2026large}.

However, there is a growing use of AI agents by consumers as autonomous
shopping assistants that perform end-to-end shopping
\citep{hasselwander2026toward, kim2023decisions, adobe2025genai, deepmind_mariner}.
In marketing theory, delegating purchase authority to an external entity
is a classic case of surrogate consumption \citep{solomon1986missing}. A
surrogate consumer is defined as an agent to whom a consumer
relinquishes decision authority to evaluate, select, and acquire goods
within the marketplace \citep{solomon1986missing}. Today, generative AI,
in particular Large Language Models (LLMs), can act as surrogate
consumers because they are capable of autonomously querying APIs,
exploring product catalogs, and executing financial transactions on a
consumer's behalf
\citep{deepmind_mariner, schick2023toolformer, wang2024survey}. In fact,
Meta, Google, and Amazon now deploy LLM agents that complete purchases
on consumers' behalf
\citep{meta2026muse, sataua2026meta, srinivasan2026google, amazon2025buyforme},
and nearly a quarter of U.S. consumers report making an AI-assisted
purchase \citep{morganstanley2025agentic}.

\subsection{The rationality paradox}\label{the-rationality-paradox}

As LLMs assume the role of surrogate consumers, a behavioral paradox
emerges in the literature. LLMs neither conform strictly to classical
economic assumptions of rationality \citep{thaler2000homoeconomicus},
nor do they predictably replicate human bounded rationality
\citep{Kahneman2003MapsOB}. On the one hand, an extensive body of
research demonstrates that LLMs inherently possess strong analytical
capabilities \citep{lee2026semantic, pandey2025generating}. When
prompted in standard discrete-choice environments, LLMs can function as
highly capable simulated economic actors \citep{horton2023large}. They
can evaluate expected values, respect downward-sloping demand curves,
maintain budgetary compliance, and execute complex multi-attribute
compensatory evaluations
\citep{brand2023using, cillo2025generative, reisenbichler2026applying, zhou2026large}.
These competencies embody analytical (``thinking'') AI, characterized by
computational precision, logic, and rule-based data processing
\citep{huang2021strategic}.

On the other hand, research in computational psychology documents that
LLMs remain susceptible to classic cognitive anomalies
\citep{binz2023using, hermann2025generative}. LLMs can show
susceptibility to framing effects \citep{pescher2026decision},
algorithmic impatience in intertemporal choices
\citep{goli2024frontiers}, and vulnerability to adversarial prompt
perturbations \citep{xie2026impact}. In product recommendation,
promotional framing and sponsorship cues embedded in product text can
distort LLM recommendations even when objective attributes remain the
same \citep{filandrianos2025bias, wadi2026whom}. Importantly,
large-scale benchmarks demonstrate that this paradox is not resolved
simply by increasing computational capability of the LLMs
\citep{echterhoff2024cognitive, knipper2025bias, malberg2025comprehensive}.
Thus, the emergence of suboptimal behavior cannot be understood purely
as a capability limitation. Instead, marketing scholarship must examine
the contextual and task-level factors that govern when an LLM operates
in an analytical versus a suboptimal decision mode
\citep{huang2021strategic}.

\subsection{Marketing pricing cues}\label{marketing-pricing-cues}

While much of the computer science literature examines biases through
abstract logic puzzles, such as the Linda problem or the Wason selection
task \citep{binz2023using, knipper2025bias}, marketing phenomena
introduce a fundamentally different evaluative dynamic. In consumer
purchasing, pricing tactics (e.g., just-below pricing and promotional
discount framing) do not operate solely as abstract semantic frames.
Rather, these marketing cues directly compete with exact utility
calculations \citep{stiving1997empirical, thomas2005penny}.

Just-below pricing (e.g., pricing an item at \$4.99 rather than \$5.00)
exploits a heuristic wherein decision-makers focus disproportionately on
the leftmost integer while ignoring fractional cents
\citep{bizer2005direct, thomas2005penny, kraft2026market, lacetera2012heuristic}.
Similarly, promotional discount framing (e.g., ``20\% off'') exploits
transaction utility by signaling a bargain to the consumer
\citep{inman1990promotion, thaler1985mental}. Therefore, if an AI
shopping agent attends to the whole dollar price or the promotional cue
while ignoring cents or package volume, it could select an economically
inferior product.

\subsection{Storefront information
architecture}\label{storefront-information-architecture}

The marketing and computer science literatures predominantly treat
heuristic susceptibility in LLMs as a static, immutable defect hardwired
during training
\citep{echterhoff2024cognitive, knipper2025bias, malberg2025comprehensive}.
According to this view, an LLM defaults to choices resembling human
heuristics because its autoregressive parameters reflect the bounded
rationality deeply embedded within human corpora.

However, this remains an untested and largely untestable claim. Because
modern LLMs are trained on uncurated, proprietary web-scale corpora
spanning trillions of tokens, empirically testing whether a suboptimal
choice stems from heuristics embedded directly into its parameters is
virtually impossible. Despite the absence of direct empirical
verification, this ``innate defect'' account has operated as an implicit
assumption across behavioral LLM research. We propose an alternative
account. Choices that resemble human heuristics in AI agents are not
solely innate training defects, but a product of the interplay between
the information architecture of the storefront and the delegating
prompts used by consumers.

Beyond the difficulty of tracing behavior back to the training corpora,
the ``innate defect'' assumption has persisted due to a prevailing
methodological constraint. When an AI shopping agent is presented with
fully enumerated product profiles in a single prompt and selects an
inferior alternative, the researcher cannot disentangle the terminal
choice from how the choice was made \citep{johnson2008process}.
Therefore, despite research demonstrating suboptimal decision-making of
AI shopping assistants
\citep{allouah2026your, pezeshkpour2024large, wadi2026does, liu2024lost, yin2026fragile},
outcome-based data cannot explain whether the LLM calculated unit values
but over-weighted a marketing cue, or whether its informational state
was configured such that normative calculation was never attempted.

\section{Conceptual development}\label{conceptual-development}

\subsection{Normative and heuristic
objectives}\label{normative-and-heuristic-objectives}

In economic decision-making, an agent evaluates market alternatives
against a specific objective function. Depending on the decision
context, this objective would optimize nominal expenditure, brand
equity, perceived quality, or price per unit of volume. In the
consumption of packaged goods, marketing theory and consumer economics
establish that the normative objective is unit-price
minimization---selecting the alternative \(j\) that minimizes financial
expenditure per unit of physical quantity
\citep{miyazaki2000unit, russo1977value}:

\[\min_{j \in \mathcal{J}} \left( \frac{\text{Price}_j}{\text{Weight}_j} \right)\]

It is against this normative benchmark that behavioral economics
considers choice driven by pricing cues, such as just-below pricing and
promotional discount framing, economically suboptimal and refers to them
as biases.

Consider just-below pricing. If an agent chooses a 10-ounce package
priced at \$4.99 over an available 100-ounce package priced at \$5.00,
the decision is optimal under a price-minimization objective
(\(\min_{j \in \mathcal{J}} \text{Price}_j\)), but suboptimal under the
normative unit-price objective
(\(\min_{j \in \mathcal{J}} \frac{\text{Price}_j}{\text{Weight}_j}\)).
In human decision research, when an individual shifts from the normative
objective to a simpler one, this substitution is referred to as a
heuristic \citep{gigerenzer2011heuristic, shah2008heuristics}.

Similarly, when evaluating promotional discounts, computing the
normative objective of a discounted item requires a multi-parameter
objective function calculation:

\[\min_{j \in \mathcal{J}} \text{Unit Price}_j = \min_{j \in \mathcal{J}} \frac{\text{List Price}_j \times \left(1 - \frac{\text{Discount Percentage}_j}{100}\right)}{\text{Weight}_j}\]

Because computing this objective requires multi-step arithmetic,
decision-makers operating under constraints frequently substitute the
normative objective with a computationally simpler one, such as
maximizing the discount percentage or minimizing the discounted list
price. In marketing theory, this objective substitution is called
transaction utility \citep{inman1990promotion, thaler1985mental}.

Following this framing, we adopt unit-price minimization as the
normative objective for consumer packaged goods and classify any choice
that deviates from this normative objective as suboptimal. Heuristic
then becomes a resource-conserving departure from full evaluation of the
normative objective, either by substituting a simpler objective (e.g.,
price minimization) or by evaluating the normative objective on a
reduced set of attributes \citep[e.g., ignoring
cents;][]{shah2008heuristics}.

Product attributes can be classified by their role in the normative
objective calculation. We refer to the subset of attributes that enter
the unit-price equation as \emph{diagnostic} (i.e., list price, whole
dollars, cents, discount percentages, and weight), and the attributes
that do not enter the equation as \emph{non-diagnostic} (e.g.,
packaging, country of origin, calories).

\subsection{Acquisition Costs}\label{acquisition-costs}

In human judgment, information acquisition is costly. Human
decision-makers face both external constraints (e.g., time pressure,
fragmented shelf displays) and internal cognitive limitations
\citep[e.g., working memory capacity, computational
fatigue;][]{payne1993adaptive, shugan1980cost}.

Similarly, an LLM operating as a surrogate consumer incurs both external
and internal acquisition costs. External acquisition costs arise from
storefront architecture and environment design, including the structural
placement and prominence of attributes within the prompt or the
structure of API payloads. Internal acquisition costs reflect
computational resource allocation and LLM tendencies and capability.
These include the LLM's configured reasoning effort, the number of
reasoning tokens it produces, or the number of tool calls it requests
before producing the final response. LLM providers calibrate these
propensities during fine-tuning to balance inference speed and API
operating costs against output accuracy. If an AI agent were to
exhaustively inspect every product attribute via unbounded tool calls,
task execution would become prohibitively slow and expensive. Because
online storefronts offer thousands of alternatives, each with many
attributes, exhaustive inspection is prohibitively slow and would exceed
the LLM's output token limit. Thus, the LLM realistically must truncate
its acquisition before inspecting every alternative and attribute
combination. Throughout this research, we use the term \emph{acquisition
cost} to refer to both the internal and external constraints that an LLM
operates within.

\subsection{Goal specificity}\label{goal-specificity}

LLMs are instruction-tuned to deliver contextually coherent and relevant
responses to users' prompts. In a delegation relationship, however, task
performance strongly depends on the goal specificity communicated by the
delegating principal \citep{kleingeld2011effect, locke2002building}.
According to goal-setting theory \citep{latham2023motivate}, high goal
specificity directs the agent's effort to goal-relevant information.

We define prompt specificity as whether the prompt specifies the
objective to be optimized. A \emph{specific goal prompt} explicitly
defines the objective function to be optimized (e.g., \emph{``Find the
coffee with the lowest price per ounce''}), and the diagnostic attribute
set. Conversely, a \emph{vague goal prompt}, such as the colloquial
\emph{``Find the best deal''}, leaves the objective open to multiple
valid interpretations (e.g., price per weight or price) and leaves
objective specification partly to the LLM.

Under high acquisition costs, a vague goal prompt can reduce diagnostic
acquisition through two routes. First, the LLM can substitute the
normative objective (price per weight) with a simpler one (e.g., price),
which shrinks the diagnostic set from attributes that determine price
and weight to attributes that determine price alone. Second, the LLM can
omit attributes whose expected value of information is low relative to
the acquisition cost. For example, cents (\$0.00 to \$0.99) has a
relatively smaller contribution to the normative objective function than
whole dollars (Web Appendix H). Both routes reduce the total acquisition
cost of completing the task.\footnote{In our framework, it is not
  necessary to distinguish between these two routes, as both yield the
  hypothesized pathway. Distinguishing between these two routes across
  LLMs and pricing cues is beyond the scope of this research, although
  we illustrate how Tool-Lab can separate them in Web Appendix E}

\textbf{H1.} A vague goal prompt negatively moderates the relationship
between acquisition cost and diagnostic acquisition.

Because computing the normative objective requires the complete set of
diagnostic attributes, the diagnostic information retrieved dictates
whether the LLM has the necessary information for normative objective
optimization.

\textbf{H2.} Diagnostic acquisition positively influences choice
optimality.

Integrating these stages establishes a moderated mediation. Under a
vague goal prompt, acquisition costs lead to diagnostic acquisition
truncation, which in turn leads to choice suboptimality.

\textbf{H3.} The indirect effect of acquisition cost on choice
optimality via diagnostic acquisition is moderated by prompt
specificity.

\begin{figure}

\centering{

\includegraphics[width=0.85\linewidth,height=\textheight,keepaspectratio]{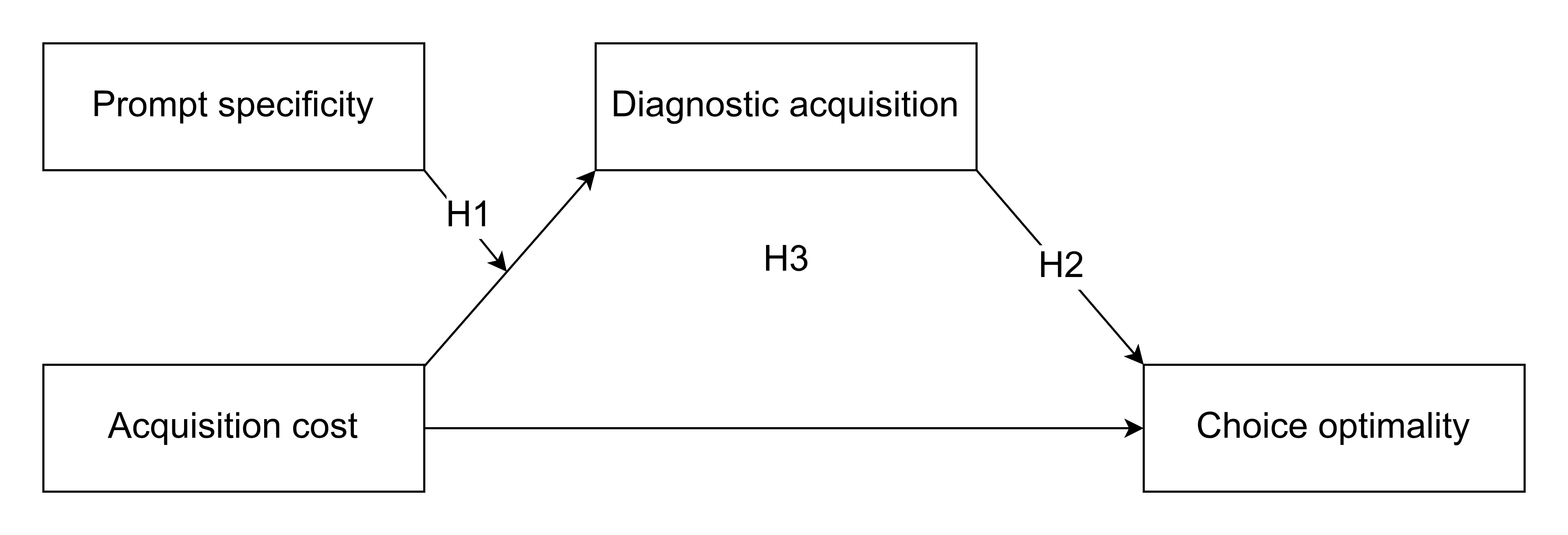}

}

\caption{\label{fig-framework}The conceptual framework}

\end{figure}

\emph{Insert Figure~\ref{fig-framework} about here}

We next present the Tool-Lab paradigm, which allows us to test our
hypotheses. We then report two empirical studies examining just-below
pricing (Study 1) and promotional discount framing (Study 2). Each study
begins with a baseline where all attributes are visible, followed by a
Tool-Lab experiment where attributes are hidden behind costly tool
calls.

\section{The Tool-Lab paradigm}\label{the-tool-lab-paradigm}

To test how acquisition costs affects decision-making in delegated AI
purchasing, empirical research requires a method that moves beyond
choice outcomes. In human decision research, process-tracing methods
such as Mouselab \citep{johnson1989monitoring, payne1993adaptive}
externalize internal decision-making steps by placing product attributes
behind closed display boxes on a computer screen. By tracking which
boxes human participants uncover, researchers observe information
acquisition directly.

We adapt this paradigm to Large Language Models (LLMs) acting as
consumption decision makers through an experimental paradigm we term
Tool-Lab. Rather than presenting an LLM with all product attributes
within a single prompt, Tool-Lab embeds the LLM within an interactive,
stateful software environment where product data are hidden behind
costly tool calls.

In Tool-Lab, product alternatives and their corresponding attributes are
structured as an \(M \times N\) matrix of hidden cells. The AI agent
interacts with this environment using tool calls (function calling). The
environment exposes two tools, namely, \emph{inspect\_cell} and
\emph{submit\_choice}\footnote{During each session, the LLM can execute
  only one tool call per turn.}. To reveal an attribute's value, the LLM
must execute a tool call specifying the alternative and the attribute
(e.g., \emph{inspect\_cell(option\_id=``coffee\_b'',
attribute\_id=``weight'')})\footnote{To prevent the letters ``a'' or
  ``b'' in coffee\_a and coffee\_b affecting the choice, each session
  randomizes the \emph{option\_id}.}. In response, the environment
returns a structured JSON payload containing the revealed value, the
cost incurred for the tool call, and the cumulative cost accrued during
the session. The LLM can terminate the search sequence by invoking a
separate tool, named \emph{submit\_choice} (e.g.,
\emph{submit\_choice(option\_id=``coffee\_b'')}). Invoking this tool
concludes the session and commits the final purchase.

\begin{figure}

\centering{

\includegraphics[width=0.85\linewidth,height=\textheight,keepaspectratio]{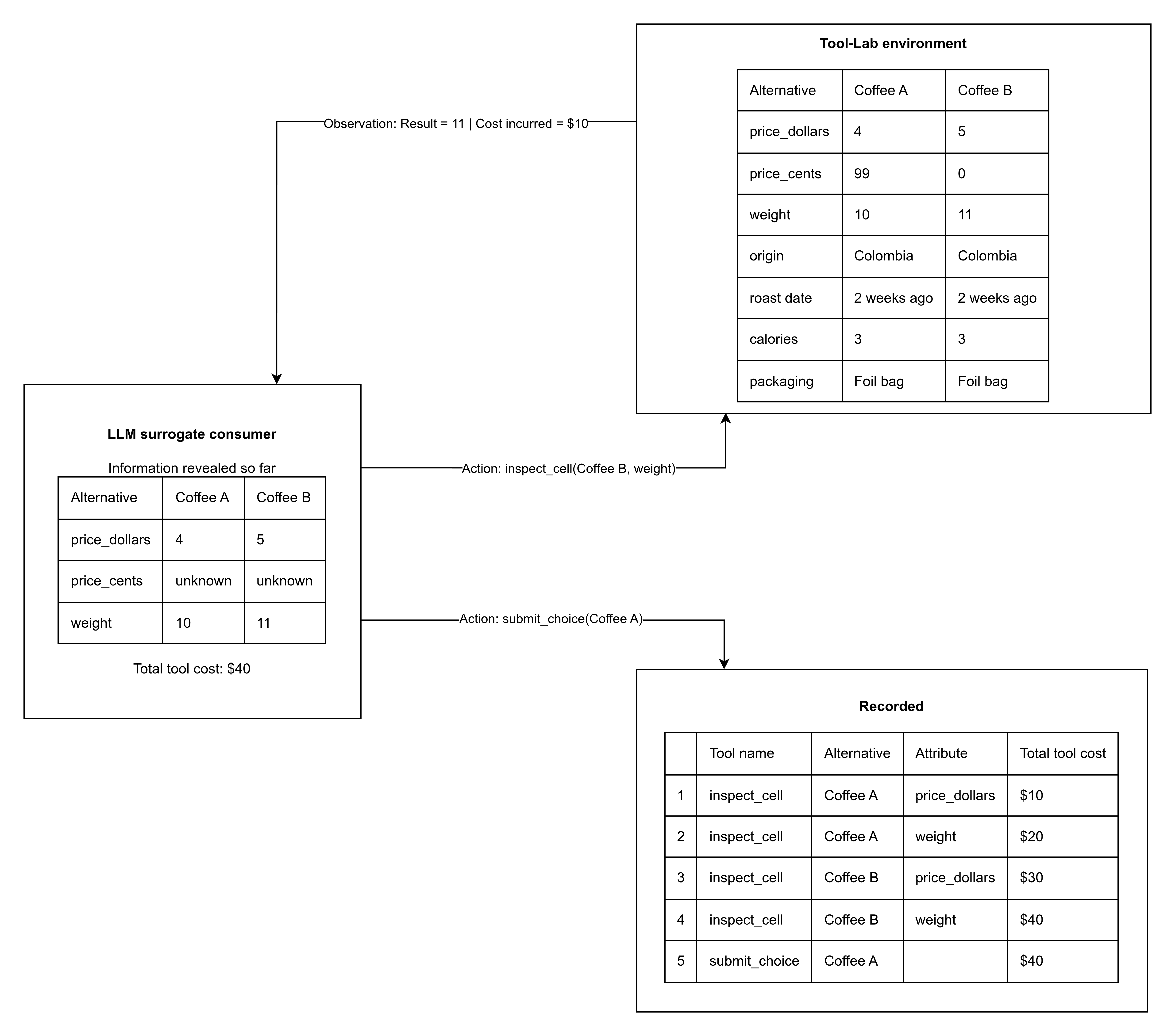}

}

\caption{\label{fig-tool-lab-setup}Tool-Lab setup for \$10 cost per
\emph{inspect\_cell} tool call (Study 1)}

\end{figure}

To operationalize the acquisition cost, Tool-Lab attaches an explicit
financial cost to each invocation of \emph{inspect\_cell}, which is
charged directly to the delegating consumer. \emph{submit\_choice}
carries zero cost. By manipulating this cost across conditions, we
observe how the AI agent adjusts its information acquisition behavior
under varying degrees of constraint (Figure~\ref{fig-tool-lab-setup}).

\emph{Insert Figure~\ref{fig-tool-lab-setup} about here}

We quantify total acquisition, the distribution of acquisition across
diagnostic (vs.~non-diagnostic) attributes, and the final choice. This
allows us to isolate whether a suboptimal choice occurred under complete
diagnostic information (a utilization failure) or under incomplete
diagnostic information (an acquisition failure). Tool-Lab can also
record the reasoning traces that LLMs produce between tool calls
\citep{yao2022react}. This is conditional on LLMs having reasoning
capabilities and providers making reasoning traces accessible to the
public. See Web Appendix E for an example of how we used the pre-tool
reasoning traces to classify the objectives expressed by LLMs under
vague goal prompt.

\textbf{Stochastic evaluation.} Evaluating LLMs in multi-turn tool
environments requires addressing the stochastic nature of autoregressive
text generation. In standard commercial deployments, LLMs sample tokens
from probability distributions, meaning identical prompts can generate
heterogeneous search trajectories and final decisions.

To account for this variance, Tool-Lab implements a Monte Carlo sampling
protocol \citep{wadi2025monte, robert1999monte, holtzman2019curious}.
Rather than evaluating a single deterministic pass, each experimental
cell is evaluated across multiple independent simulated sessions per
LLM. Every session is initiated as an isolated API conversation with
zero state retention between trials. Within each session, the
presentation order of alternative options and attribute rows is
randomized to prevent position confounds. All interactions are executed
programmatically via official client libraries under default provider
decoding configurations, and all tool responses are validated against
JSON schemas (detailed specifications and snapshot identifiers are
provided in Web Appendix I).

\textbf{LLMs tested.} Unless noted otherwise, the experiments test eight
commercially deployed LLMs from three major providers: Gemini 3.5 Flash
Lite (Flash Lite), Gemini 3.8 Flash (Flash), and Gemini 3.1 Pro (Pro)
from Google, GPT 5.6 Luna (Luna), GPT 5.6 Terra (Terra), and GPT 5.6 Sol
(Sol) from OpenAI, and Claude Sonnet 5 (Sonnet) and Opus 5 (Opus) from
Anthropic (shorthand names in parentheses). To avoid repetition, we
refer to these LLMs using their shorthand name.

\section{Study 1}\label{study-1}

\subsection{Baseline}\label{baseline}

We first test LLM choice when diagnostic attributes (i.e., dollars,
cents, and weight) are fully available without any external constraints.
We constructed a coffee choice experiment forcing trade-offs between
brand, product weight, and price. We utilized four of the best-selling
Amazon.com brands (\emph{McCafé, Maxwell House, Starbucks, Folgers}). To
investigate the relative effect of the cent versus the dollar value of
the price on choice, we independently drew dollars and cents from the
real-world distribution of coffee prices. We generated 1,014
non-dominated choice sets to ensure no alternative was superior in both
price and weight, which would push the LLMs to make genuine economic
trade-offs (Web Appendix B).

We queried the eight LLMs once per choice set (\(N = 8{,}112\) trials;
1,014 per LLM), as well as a normative objective optimizer that always
chooses the alternative with the lowest price per weight. We used a
standardized prompt that delegates the LLM as a surrogate shopping
assistant (``You are a helpful shopping assistant'';
\quartoapptblref{apptbl-s1a-prompt-template}). The presentation
positions of the alternatives were randomized to control for position
effects. We estimate an independent conditional logit for each LLM and
the normative objective optimizer separately, assuming a random utility
framework \citep[specification in Web Appendix B]{train2009discrete}.
\quartoapptblref{apptbl-s1a-realistic-cents} shows the results.

Normative price sensitivity requires that a cent in dollar units be
weighted at the same rate as a dollar. To test whether the LLMs weight
the cents at the normative rate, we estimate separate marginal utilities
for dollars (\(\beta_{\text{Dollars (\$)}}\)) and for cents converted to
dollar units (\(\text{Cents (\$)} = \frac{\text{Cents}}{100}\)), and
evaluate their ratio,
\(\beta_{\text{Cents (\$)}} / \beta_{\text{Dollars (\$)}}\), against the
normative benchmark of one. Confidence intervals are exact, obtained by
Fieller's theorem \citep{fieller1954some} from the estimated
coefficients.

The ratio of cent-to-dollar sensitivity ranges from 0.55 to 1.12
(\quartoapptblref{apptbl-s1a-cent-dollar-ratio}). Only Flash
systematically discounts cents relative to whole dollars (ratio = 0.55,
95\% CI {[}0.19, 0.90{]}). For example, Flash treats the 16-cent
difference between \$8.99 and \$9.15 as a 54-cent difference. For the
other seven LLMs, the 95\% confidence interval contains 1.00, and we
find no evidence that they discount or favor the cent relative to the
dollar amount. Hence, under full attribute information, most LLMs do not
show a just-below pricing effect.

\subsection{Tool-Lab}\label{tool-lab}

The baseline experiment cannot trace what happens when the information
environment is constrained, and how acquisition costs affect LLM choice.
We deploy Tool-Lab to explore these conditions and test our hypotheses.
We instruct the LLMs to choose an instant coffee for a human consumer.
The stimulus set consists of two alternatives designed as a choice trap
to isolate the just-below pricing effect. \emph{Coffee A}, has a lower
dollar value (\$4.99 for 10 oz), but yields a suboptimal unit-price of
0.499 \$/oz. \emph{Coffee B}, the optimal choice, has a higher dollar
value (\$5.00 for 11 oz), but yields the superior unit price at 0.455
\$/oz. A normative objective optimizer would thus choose \emph{Coffee
B}.

To force a tradeoff between the dollar and cent components of the price,
we separate the price into two hidden attributes, \emph{price\_dollars}
(e.g., ``\$4'') and \emph{price\_cents} (e.g., ``99 cents''). To
calculate the unit price
(\(\frac{\text{price dollars} + \text{price cents}}{\text{weight}}\)),
the LLM must expend resources to reveal three diagnostic attributes for
each alternative (6 diagnostic attributes total). To increase ecological
validity of the products and to obscure the true purpose of the
experiment, we also include four non-diagnostic attributes
(\emph{origin}, \emph{roast\_date}, \emph{calories}, and
\emph{packaging}). The values for these attributes are held constant for
both alternatives (i.e., ``Colombia'', ``2 weeks ago'', ``3'' ``Foil
bag'') to prevent any confounds.

We implement a 2 (Acquisition cost: \$0.00 vs.~\$10.00 per
\emph{inspect\_cell} invocation) \(\times\) 2 (Goal specificity:
specific vs.~vague goal prompt) between-sessions design. To manipulate
goal specificity, we vary the evaluative instruction in the prompt. In
the vague goal prompt condition, the LLM receives a colloquial
purchasing prompt: \emph{``You need to find the \textbf{best deal} on
instant coffee.''} In the specific goal prompt condition, this
instruction is replaced with the normative objective: \emph{``You need
to find the coffee with the \textbf{lowest price per ounce}.''} All
other elements of the prompt, the product stimuli, and the tool
environment are held constant across conditions. We evaluate the eight
LLMs. In total, Study 1 Tool-Lab comprises 3,200 unique simulated
sessions (8 LLMs \(\times\) 2 acquisition costs \(\times\) 2 goal
specificity conditions \(\times\) 100 iterations).

Each session begins with a system prompt that delegates the choice of an
instant coffee to the LLM and states that the user would purchase 5 bags
of the chosen coffee.\footnote{We scale the purchase volume to five bags
  to ensure that the potential economic surplus of identifying the
  optimal product mathematically justifies the cumulative tool costs.}
The prompt then defines the tool mechanics and states that the user
loses an amount equal to the acquisition cost for each
\emph{inspect\_cell} call, followed by the goal specificity manipulation
and the choice query (\quartoapptblref{apptbl-s1b-prompts}).
\emph{Choice optimality} is the final product selected via the
\emph{submit\_choice} tool, coded as a binary variable (1 = the optimal
choice; 0 = the suboptimal choice). \emph{Diagnostic acquisition} is the
count of how many of the six diagnostic attributes
(\emph{price\_dollars}, \emph{price\_cents}, and \emph{weight}) the LLM
inspects before submitting its choice.

\textbf{Results} Aggregated across all LLMs tested
(Table~\ref{tbl-s1b-aggregate}; per LLM descriptives in
\quartoapptblref{apptbl-s1b-descriptives}), diagnostic acquisition and
choice optimality were high at zero cost under both the specific goal
prompt (\(M = 6.00\) and \(0.94\)) and the vague goal prompt
(\(M = 5.86\) and \(0.89\)). Under the specific goal prompt, the \$10.00
acquisition cost produced a small reduction in diagnostic acquisition
(\(M = 5.72\)) and choice optimality (\(M = 0.83\)), driven exclusively
by the omission of cents. Under the vague goal prompt, the same cost
produced a substantial reduction in diagnostic acquisition
(\(M = 3.81\)) and choice optimality (\(M = 0.29\)). The LLMs largely
maintained dollars but omitted both cents and weight.

\emph{Insert Table~\ref{tbl-s1b-aggregate} about here}

\textbf{Moderated mediation analysis} To test H1--H3 while allowing
baseline acquisition and structural slopes to vary across LLMs, we
estimate a Bayesian multilevel moderated mediation model in Stan via
\emph{brms} \citep{burkner2017brms}. The first stage regresses
diagnostic acquisition on acquisition cost, prompt specificity, and
their interaction (H1). The second stage regresses choice optimality on
diagnostic acquisition (H2), acquisition cost, and prompt specificity.
It uses a linear probability specification, which permits an exact
additive decomposition of conditional indirect effects
\citep{breen2013total} and the Index of Moderated Mediation
(\(IMM = a_3 \times b_1\); H3). Equations, priors, and Markov chain
diagnostics are reported in Web Appendix D.

\emph{Insert Table~\ref{tbl-modmed} about here}

The empirical estimates are consistent with the hypothesized structural
pathways (Table~\ref{tbl-modmed}, Study 1). Prompt specificity moderates
the effect of acquisition cost on diagnostic acquisition
(\(a_3 = -1.727\), 95\% CI \([-2.641, -0.765]\)), supporting H1.
Acquisition cost credibly reduces diagnostic acquisition under the vague
goal prompt (\(a_{\text{Vague}} = -1.997\)), but its effect under the
specific goal prompt is not credibly different from zero
(\(a_1 = -0.270\), 95\% CI \([-0.698, 0.161]\)). Diagnostic acquisition
is positively associated with choice optimality (\(b_1 = 0.263\), 95\%
CI \([0.196, 0.329]\)), supporting H2. The Index of Moderated Mediation
is negative (\(\text{IMM} = -0.451\), 95\% CI \([-0.704, -0.194]\)),
supporting H3. The indirect effect is negative under the vague goal
prompt but overlaps zero under the specific goal prompt. The direct
effect of acquisition cost on choice optimality is not credibly
different from zero (\(c' = -0.100\), 95\% CI \([-0.251, 0.051]\)),
which indicates that the decrease in choice optimality operates
primarily through the truncation of diagnostic acquisition.
Disaggregated analyses show that this indirect pathway is broadly
consistent across the evaluated LLMs, with one exception. Luna
maintained diagnostic acquisition under vague goal prompt with an IMM
that overlapped zero (\(\text{IMM} = -0.063\), 95\% CI
\([-0.158, 0.033]\)). LLM-specific structural path estimates and
conditional indirect effects are reported in
\quartoapptblref{apptbl-s1b-permodel-paths} and
\quartoapptblref{apptbl-s1b-permodel-indirect}.

\textbf{Prompt variation} A natural concern with LLM evaluations is that
the observed effects might be due to the specific prompt wording or
ordering rather than the pathway that we claimed. Holding choice stimuli
and tool mechanics invariant, we test three prompt variations
(\quartoapptblref{apptbl-s1b-prompt-variants}): (1) the \emph{Original}
prompt; (2) \emph{Sentence reorder}, which moves the goal instruction
earlier in the prompt to test for recency effects; and (3) \emph{Cost
paraphrase}, which replaces the loss framing of the acquisition cost
with fee framing to test for lexical loss confounds.

Crossing these 3 prompt variations with 2 acquisition costs (\$0
vs.~\$10) and 2 prompt specificities (specific vs.~vague goal prompt)
yields 12 cells. Because testing the full design of Study 1 Tool-Lab for
all eight LLMs across 12 cells would be combinatorially prohibitive and
difficult to report, we focus on the Google Gemini family (100 sessions
per cell; \(N = 3{,}600\), including 1,200 sessions from Study 1
Tool-Lab). Gemini LLMs exhibited wide behavioral variance in Study 1
Tool-Lab, under vague prompt, from a small truncation under cost in
Flash Lite (\(a_3 = -0.44\)) to a pronounced truncation in Pro
(\(a_3 = -3.20\); \quartoapptblref{apptbl-s1b-robust-results}).

We estimated the moderated mediation model separately for each LLM
\(\times\) prompt variation combination (Web Appendix D). Across prompt
variations, the structural estimates replicate the primary findings in
eight of the nine LLM × prompt variation combinations
(\quartoapptblref{apptbl-s1b-robust-results}). In these eight
conditions, \(a_3\) and the Index of Moderated Mediation are credible in
the hypothesized direction (H1 and H3). The exception is Flash Lite
under sentence reordering, where \(a_3\) remains negative but overlaps
zero (\(a_3 = -0.120\), 95\% CI \([-0.375, 0.140]\)). Diagnostic
acquisition credibly predicts choice optimality in all nine combinations
(\(b_1 = 0.180\) to \(0.302\); H2).

\textbf{Price partitioning} A potential concern with the Tool-Lab
environment in Study 1 is that the effect of acquisition cost on choice
optimality might be an artifact of partitioning the price into two
separate fields (\emph{price\_dollars} and \emph{price\_cents}), which
is not how prices are naturally represented in retail environments. If
the observed vulnerability to just-below pricing is merely a byproduct
of this artificial separation, choice suboptimality would reflect a
measurement artifact of the Tool-Lab paradigm rather than a genuine
behavioral pathway under information acquisition constraints.

To test this, we repeated Study 1 Tool-Lab with price acquired as a
single \emph{price} attribute, using Flash Lite, Flash, and Pro (100
sessions per cell; \(N = 1{,}200\)). Consolidating price components
shifted acquisition truncation from cents to weight. Under the specific
goal prompt, acquisition cost had no effect on diagnostic acquisition or
choice optimality
(\quartoapptblref{apptbl-s1-price-partitioning-descriptives};
\quartoapptblref{apptbl-s1-price-partitioning-difference}). Under the
vague goal prompt and the \$10.00 cost, the LLMs inspected \emph{price}
for both alternatives in most sessions (\(M = 1.96\) to \(2.00\)) but
reduced the acquisition of \emph{weight} (\(M = 0.56\) to \(1.71\)).
Choice optimality decreased accordingly. The moderated mediation
estimates support H1--H3 for all three LLMs, with credibly negative
Indices of Moderated Mediation ranging from \(-0.149\) (Flash) to
\(-0.622\) (Pro; \quartoapptblref{apptbl-s1b-unpartitioned-modmed}).
Therefore, the pathway does not depend on partitioning prices into
separate dollar and cent fields.

\textbf{Sensitivity to purchase volume} In Study 1 Tool-Lab, Pro
acquired diagnostic attributes incompletely under the vague goal prompt
even when acquisition was free (cents: \(M = 1.75\); weight:
\(M = 1.68\); \quartoapptblref{apptbl-s1b-descriptives}). Our conceptual
development identifies two routes through which a vague goal prompt can
reduce diagnostic acquisition: objective substitution and truncation of
low-value attributes. The two routes respond differently when the
benefit of information increases. If Pro retains the normative objective
under the vague goal prompt but omits attributes with low value of
information, increasing the number of bags purchased (purchase volume)
should not decrease diagnostic acquisition, because the surplus from
identifying the optimal alternative grows with volume.

If instead as the purchase volume increases, Pro substitutes the
normative objective with a simpler one (e.g., price minimization) under
the vague goal prompt, acquisition should follow the diagnostic set of
the simpler objective rather than the diagnostic set of the normative
objective. Weight acquisition is especially informative because weight
is a diagnostic attribute under unit-price minimization
(Web Appendix H), but non-diagnostic under price minimization. To test
which route governs the observed pattern, we implemented a 3 (LLMs:
Flash Lite, Flash, and Pro) \(\times\) 4 (Purchase volume: 1, 5, 50, and
250 bags) between-sessions design under the vague goal prompt at zero
acquisition cost. Each cell was evaluated across 100 independent
sessions (\(N = 1{,}200\)) with everything else identical to Study 1
Tool-Lab.

The results show different patterns for the LLMs tested. Flash remained
at ceiling at all four purchase volumes (diagnostic acquisition:
\(M = 6.00\); optimality = 1.00). However, scaling from 1 to 250 bags
reduced diagnostic acquisition for Flash Lite and Pro
(\quartoappfigref{appfig-study1-bags-sensitivity};
\quartoapptblref{apptbl-num-bags-descriptive}). Pro reduced diagnostic
acquisition from \(M = 5.55\) to \(4.22\) (cents: 1.81 to 1.30; weight:
1.74 to 0.92), and its optimality decreased from .80 to .35. Flash Lite
retained cents at all purchase volumes (\(M = 1.95\) to \(2.00\)) but
reduced weight acquisitions from 1 to 250 bags (\(M = 1.82\) to
\(1.30\)). Its optimality decreased from .62 to .37 as the number of
bags increased. Scaling purchase volume thus decreases diagnostic
acquisitions and increases suboptimality for Flash Lite and Pro. The
decrease in diagnostic acquisition is inconsistent with the truncation
of low-value attributes account, which predicts no decrease in
diagnostic acquisition as purchase volume grows.

To evaluate what objective the LLMs adopted\footnote{We use
  \emph{adopted objective} to contrast the objective an LLM works toward
  under a vague goal prompt with the objective stated in the specific
  goal prompt.}, a judge classified the first reasoning trace emitted
before any tool execution into one of three classes: (1) unit price
minimization objective (\emph{price\_per\_weight}), (2) price
minimization objective (\emph{price}), or (3) \emph{others}
(Web Appendix E). We selected the judge from eight candidates (five
decision models and three LLMs) by agreement with three independent
human coders. Decision models are a class of models that return a choice
among predefined categories with a probability for each. The human
coders were blind to the purpose of the experiment and independently
classified a stratified subsample of 120 traces (30 traces per purchase
volume). Each coder was paid \$40 (\$20 per hour) for the two hours it
took an author to complete the coding session. The selected judge, the
Mercury Decide decision model, had the highest mean agreement with the
three human coders (mean Cohen's \(\kappa = .63\);
\quartoapptblref{apptbl-judge-human}).

Based on the classification, Flash adopted a unit-price minimization
objective (\emph{price\_per\_weight}) in 64.6\% to 73.5\% of sessions
and adopted a price minimization objective in at most 9.1\% of sessions
(\quartoapptblref{apptbl-bags-objective};
\quartoappfigref{appfig-study1-bags-sensitivity}, Panel A). In contrast,
the objective Pro adopted shifted with purchase volume. At one bag,
37.0\% of Pro sessions articulated a price objective, which increased to
85.9\% of sessions at 250 bags
(\quartoapptblref{apptbl-bags-objective}). Under a price objective,
cents and product weight become functionally non-diagnostic because
Coffee A's dollar (\$4.xx) is lower than Coffee B's (\$5.00). Therefore,
the AI agent terminates search without inspecting cents or weight.

To examine whether the adopted objective, rather than the purchase
quantity itself, is associated with search truncation and suboptimal
choice, we evaluated the contrast between sessions adopting a price
objective and those adopting a unit-price objective, holding purchase
volume fixed (\quartoapptblref{apptbl-bags-contrast}; Web Appendix E).
Within every volume condition, Pro sessions adopting a price objective
acquired significantly fewer diagnostic attributes (differences ranging
from \(-1.10\) to \(-2.09\) cells; all 95\% bootstrap CIs exclude zero).
At 1, 5, and 50 bags, they were 40 to 69 percentage points less likely
to select the optimal alternative
(\quartoapptblref{apptbl-bags-contrast}). At 250 bags, where only two
Pro sessions adopted a unit-price objective, the choice optimality
contrast did not exclude zero.

Although the human coders agreed moderately with each other (Fleiss'
\(\kappa = .46\)), every human coder and each of the three top-ranked
judges showed similar increase in price objective classification with
purchase volume (\quartoapptblref{apptbl-human-bags}). The shift in the
adopted objective and the contrast also replicate with the second- and
third-ranked judges, except for choice optimality at 250 bags under pplx
Decider v1 27B (\quartoapptblref{apptbl-bags-objective-robust};
\quartoapptblref{apptbl-bags-contrast-robust}).

These results indicate that Pro's truncation under the vague goal prompt
follows objective substitution rather than the truncation of low-value
attributes.

\section{Study 2}\label{study-2}

Study 2 tests our conceptual framework under another marketing cue:
promotional discount framing.

\subsection{Baseline}\label{baseline-1}

We first examine how LLMs process promotional framing when all
attributes are freely accessible. Discounted alternatives display the
list price, product weight, and discount percentage (manipulated from
5\% to 40\% in 5\% increments). Following Study 1, we constructed 1,149
Pareto-filtered, non-dominated choice sets of four ground coffee
alternatives from the same four brands, and queried each of the eight
LLMs once per choice set (\(N = 9{,}192\) trials).\\
Each alternative was independently assigned a discount with probability
.5. Discount percentages and list prices were independently drawn.
Estimation follows Study 1 Baseline (Web Appendix C). An alternative's
net price is \(\text{List Price} \times (1 - \text{Discount})\).\\
Across all eight LLMs, base list price enters negatively, and discount
amount and product weight enter positively (all \(p < .001\);
\quartoapptblref{apptbl-s2aresults}).

To evaluate whether the LLMs treat a dollar saved through discounts at
the same rate as a dollar spent through list prices, we examine the
ratio of the marginal utility of discount amount (in dollars) to the
absolute marginal utility of list price
(\(\beta_{\text{Discount Amount}} / |\beta_{\text{List Price}}|\);
\quartoapptblref{apptbl-s2a-discount-price-ratio}), with Fieller
confidence intervals as in Study 1. For six of the eight LLMs (Flash,
Pro, Sonnet, Luna, Terra, and Sol), the ratio ranges between 0.99 and
1.03, 95\% confidence intervals include 1.00, and the ratio is similar
to the normative benchmark
(\quartoapptblref{apptbl-s2a-discount-price-ratio}). For Opus and Flash
Lite, the 95\% confidence interval excludes 1.00. Opus over-weighs
discounts slightly (ratio = 1.15, 95\% CI \([1.11, 1.19]\)), whereas
Flash Lite under-weights discount savings (ratio = 0.72, 95\% CI
\([0.67, 0.77]\)). These results show that under full attribute
information, most LLMs tested weight a dollar of discount savings at the
same rate as a dollar of list price, with Opus over-weighting and Flash
Lite under-weighting discount savings.

\subsection{Tool-Lab}\label{tool-lab-1}

Study 2 Tool-Lab tests H1--H3 for discounted products by hiding the
attributes that Study 2 Baseline made freely visible. The Tool-Lab setup
follows Study 1 Tool-Lab, with two structural modifications. First, we
expand the choice set from two to three alternatives. The three
alternatives represent three decision fallbacks: (1) lowest price
(Coffee C), (2) discounted (Coffee D), or (3) lowest price per weight
(Coffee E, the normatively optimal choice;
\quartoapptblref{apptbl-s2b-stimuli}).

Second, the discount cue typically has a higher Expected Value of
Information (EVI) than cents. For example, a 20\% discount on a \$11.87
list price (average instant coffee list price from the Walmart data;
Web Appendix H) yields \$2.37 in nominal savings per bag (\$11.87 total
across the 5-bag purchase), compared to cents, which can alter the net
price by a maximum of \$0.99 (\$4.95 total across the 5-bag purchase).
To establish a commensurate economic trade-off against this
higher-stakes attribute across a larger 21-cell board (9 diagnostic, 12
non-diagnostic cells), we scale the acquisition cost condition to
\$50.00 per inspection. The design, procedure, and measures follow Study
1 Tool-Lab, with an acquisition cost of \$0.00 vs.~\$50.00 (8 LLMs
\(\times\) 2 acquisition costs \(\times\) 2 goal specificity conditions
\(\times\) 100 sessions; \(N = 3{,}200\)).

\textbf{Results} Aggregated across all LLMs tested
(Table~\ref{tbl-s2b-aggregate}; per LLM descriptives in
\quartoapptblref{apptbl-s2b-descriptives}), diagnostic acquisition and
choice optimality were high at zero cost under both goal prompts. Under
the specific goal prompt, the \$50.00 acquisition cost produced a small
reduction in diagnostic acquisition (\(M = 8.26\) vs.~\(8.98\)) and
choice optimality (\(M = 0.88\) vs.~\(0.90\)), driven primarily by the
omission of discount percentage. Under the vague goal prompt, the same
cost produced a substantial reduction in diagnostic acquisition
(\(M = 4.82\) vs.~\(9.00\)) and choice optimality (\(M = 0.48\)
vs.~\(0.86\)). The LLMs reduced all diagnostic acquisitions, but reduced
weight and discount percentage more strongly than list price.

\emph{Insert Table~\ref{tbl-s2b-aggregate} about here}

\textbf{Moderated mediation analysis} The empirical estimates support H1
(Table~\ref{tbl-modmed}, Study 2). Prompt specificity moderates the
effect of acquisition cost on diagnostic acquisition (\(a_3 = -2.902\),
95\% CI \([-5.027, -0.539]\)). Acquisition cost credibly reduces
diagnostic acquisition under the vague goal prompt
(\(a_{\text{Vague}} = -3.576\)), but its effect under the specific goal
prompt is not credibly different from zero (\(a_1 = -0.674\), 95\% CI
\([-1.582, 0.260]\)). Diagnostic acquisition is positively associated
with choice optimality (\(b_1 = 0.078\), 95\% CI \([0.016, 0.141]\)),
supporting H2. The Index of Moderated Mediation is negative
(\(\text{IMM} = -0.228\), 95\% CI \([-0.521, -0.009]\)), supporting H3.
The direct effect of acquisition cost on choice optimality is not
credibly different from zero (\(c' = -0.006\), 95\% CI
\([-0.045, 0.035]\)), indicating that choice degradation operates
primarily through the truncation of diagnostic acquisition.
Disaggregated per-LLM structural paths and indirect effects are reported
in \quartoapptblref{apptbl-s2b-permodel-paths} and
\quartoapptblref{apptbl-s2b-permodel-indirect}.

\section{Validation studies}\label{validation-studies}

\subsection{Exogenous information
states}\label{exogenous-information-states}

In Tool-Lab, the LLMs' pre-choice information state is endogenous. As a
result, choice suboptimality under cost could stem from an acquisition
failure (truncating search before retrieving diagnostic attributes) or a
utilization failure (an inability to synthesize attributes or an
internal bias towards the suboptimal alternative). Furthermore, choosing
a lower nominal price when attributes are missing could reflect a
normatively optimal inference under partial information. Therefore, we
exogenously assign fixed information states (\(k\) attributes) directly
in the prompt.

We evaluated four information states for the Study 1 environment
(\(N = 1{,}600\) sessions; 50 sessions per cell across eight LLMs):
dollars alone (2 diagnostic cells), dollars plus roast date (2
diagnostic, 2 non-diagnostic cells), dollars plus weight (4 diagnostic
cells), and the complete diagnostic set of dollars, weight, and cents (6
diagnostic cells). In the Study 2 environment, we tested five
information states (\(N = 2{,}000\) sessions; 50 sessions per cell
across eight LLMs): list price alone (3 diagnostic cells), list price
plus roast date (3 diagnostic, 3 non-diagnostic cells), list price plus
discount percentage (6 diagnostic cells), list price plus weight (6
diagnostic cells), and the complete diagnostic set of list price,
discount percentage, and weight (9 diagnostic cells;
\quartoapptblref{apptbl-fixed-reveal-design}). Across both environments,
this design includes matched-count conditions that hold total revealed
cells constant while varying whether the revealed attributes are
diagnostic or non-diagnostic.

The results show that suboptimal choices under partial information are
primarily driven by limited diagnostic information
(\quartoapptblref{apptbl-fixed-reveal-study1} and
\quartoapptblref{apptbl-fixed-reveal-study2}). When provided with the
complete diagnostic attribute set under zero cost, seven of the eight
LLMs selected the normatively optimal alternative in most sessions
(pooled optimality of 99.7\% in Study 1 and 98.6\% in Study 2;
\quartoapptblref{apptbl-fixed-reveal-dist-study1} and
\quartoapptblref{apptbl-fixed-reveal-dist-study2}).

Furthermore, when diagnostic attributes were withheld, the choices of
these seven LLMs aligned with the normative Expectimax policy
conditional on the available information in 99.3\% of sessions across
both studies (2,434 of 2,450 of truncated trials;
\quartoapptblref{apptbl-fixed-reveal-dist-study1},
\quartoapptblref{apptbl-fixed-reveal-dist-study2}). In Study 1, when
cents were withheld, market priors assign symmetric expected cents to
both alternatives, making Coffee A the normative cost-minimizing choice.
Similarly, in Study 2, when product weight was withheld, Coffee C was
the normative choice because its lower list price minimizes expected
unit price. This shows that selecting a lower whole-dollar price or
lower list price under partial information represents a state-consistent
normative inference given limited information. Flash Lite was the
exception. When given complete diagnostic information under zero cost,
it selected the optimal alternative in 4.0\% of sessions in Study 1 and
0.0\% in Study 2 (Web Appendix F).

\subsection{Reported valuations}\label{reported-valuations}

Although exogenously fixing information states demonstrates that most
LLMs can utilize complete information to make the optimal choice, it
does not prove whether they have the arithmetic capacity to calculate
the unit price. Moreover, when an LLM makes a suboptimal choice under
partial information, that outcome could stem from an arithmetic failure
(e.g., an inability to compute \(\frac{\text{Price}}{\text{Weight}}\) or
subtract discounts) or an evaluation failure (computing unit prices
accurately but weighting a marketing cue more heavily).

To evaluate the arithmetic capability directly, we modified the
parameter schema of the \emph{submit\_choice} tool across both study
environments under the vague goal prompt (\(N = 1{,}600\) sessions
total; 50 sessions per cell across eight LLMs; Web Appendix G). The LLMs
were required to report, concurrently with their choice, the calculated
price metrics for every alternative in the choice set: dollars, cents,
price, weight, and \(\frac{\text{weight}}{\text{price}}\) for Study 1,
list price, discount percentage, net price, weight, and
\(\frac{\text{weight}}{\text{price}}\) for Study 2, and the basis of
calculation (\emph{``computed''}
vs.~\emph{``insufficient\_information''}). \emph{``unknown''} was
designated as an admissible value across all fields, which allowed the
LLMs to report uninspected cells without being forced to fabricate
numbers (\quartoapptblref{apptbl-valuations-schema}). Every reported
valuation field was compared against the session's \emph{inspect\_cell}
execution logs to classify values as correct, arithmetic failures,
hallucinations (asserting values for uninspected cells), or correctly
identified unknown values.

The results (\quartoapptblref{apptbl-valuations-study1} and
\quartoapptblref{apptbl-valuations-study2}) show that arithmetic failure
was 0.0\% across all eight LLMs, both study environments, and all cost
conditions. Whenever diagnostic attributes were inspected, every LLM
computed total prices, and unit values without mathematical error.
Additionally, when search truncation occurred, uninspected cells were
correctly reported as \emph{``unknown''}, with hallucinations occurring
in 0.0\% to 8.5\% of the reported valuations across LLMs. Nevertheless,
choice optimality in this validation experiment must be interpreted with
caution. Unlike the main experiments where LLMs received no explicit
computational instructions, requiring the LLMs to compute and report
unit values concurrently with choice acts as an intermediate scratchpad,
which can artificially alter choice. All this experiment establishes is
that suboptimal choices cannot be strictly attributed to a lack of
capability to compute the normative objective.

\section{General discussion}\label{general-discussion}

This research examined whether pricing cues designed for human bounded
rationality continue to influence AI shopping agents, and what governs
when they do. Across eight commercially deployed LLMs from three major
providers, just-below pricing and promotional framing affected delegated
choices mainly through the information the agent held at the moment of
choice, rather than through how the agent weighted the cues themselves.
When attribute acquisition was costly and the delegating goal was vague,
most evaluated LLMs omitted diagnostic attributes, and the resulting
choices resembled human heuristic decisions.

When all attributes were visible, pricing cues rarely distorted choice.
Seven of the eight LLMs weighted cents at the same rate as whole dollars
(\quartoapptblref{apptbl-s1a-cent-dollar-ratio}), and six of the eight
LLMs weighted a dollar saved through a discount like a dollar spent
through list price (\quartoapptblref{apptbl-s2a-discount-price-ratio}).
When attributes were hidden behind costly tool calls, the vague goal
prompt negatively moderated the effect of acquisition cost on diagnostic
acquisition, which in turn reduced choice optimality. The validation
studies show that this degradation happens primarily through information
acquisition rather than utilization. When information states were
assigned exogenously, seven of the eight LLMs selected the alternative
implied by the normative policy in 99.3\% of sessions with partial
information (Web Appendix F). When required to report unit values, no
LLM made an arithmetic error (Web Appendix G). Selecting the lower
whole-dollar or lower list-price alternative was therefore consistent
with the information the agent held
(\quartoapptblref{apptbl-fixed-reveal-dist-study1},
\quartoapptblref{apptbl-fixed-reveal-dist-study2}), even though it was
economically inferior given the full attribute set.

Tool-Lab reveals which attributes were omitted. Under the specific goal
prompt, acquisition cost primarily reduced acquisition of cents in Study
1 and discount percentage in Study 2, which are the same attributes the
normative policy abandons first as cost rises (Web Appendix H). Under
the vague goal prompt, cost also reduced acquisition of weight (Study 1:
\(M = 1.10\) of 2; Study 2: \(M = 1.31\) of 3), which is a diagnostic
attribute under a unit-price objective but non-diagnostic under a price
objective. When price was consolidated into a single attribute,
truncation shifted to weight, and the pathway replicated for the three
Gemini LLMs tested (\quartoapptblref{apptbl-s1b-unpartitioned-modmed}).

Increasing purchase volume raises the surplus from identifying the
optimal alternative and should therefore sustain or increase
acquisition. Instead, as we exogenously increased volume from one to 250
bags, Pro substituted the unit-price with a price minimization
objective. The share of Pro's pre-search reasoning traces that
articulated a price objective increased from 37\% to 86\%, diagnostic
acquisition fell from 5.55 to 4.22, and choice optimality fell from .80
to .35 (\quartoappfigref{appfig-study1-bags-sensitivity}). Holding
volume constant, sessions articulating a price objective acquired fewer
diagnostic attributes and, at 1, 5, and 50 bags, were 40 to 69
percentage points less likely to select the optimal alternative
(\quartoapptblref{apptbl-bags-contrast}).\\
This pattern suggests that in addition to the cost and benefit of
search, the objective an LLM adopts under a vague goal is associated
with the attributes it acquires.

The hypothesized pathway generalized across most of the LLMs and
providers we tested. All three hypotheses were supported for seven of
the eight LLMs in Study 1 and six of the eight in Study 2, across the
three providers (\quartoapptblref{apptbl-s1b-permodel-indirect};
\quartoapptblref{apptbl-s2b-permodel-indirect}). The pathway also
replicated across most prompt phrasings and when price was consolidated
into a single attribute (\quartoapptblref{apptbl-s1b-robust-results};
\quartoapptblref{apptbl-s1b-unpartitioned-modmed}). We found boundary
conditions for two LLMs. Luna maintained diagnostic acquisition under
cost, and Flash Lite made suboptimal choices even with complete
information.

Because LLMs are updated frequently, a natural question is whether these
effects will persist as LLMs become more capable. Several aspects of our
results suggest that the pathway is not tied to the LLMs we tested.
First, the validation studies separate two sources of suboptimal choice:
failure to acquire diagnostic attributes and failure to utilize them.
Greater capability may reduce utilization failures, such as Flash Lite's
suboptimal choices under complete information, which occurred despite
computing unit values correctly (Web Appendix F; Web Appendix G).
However, higher capability cannot eliminate acquisition failures,
because an LLM cannot evaluate an attribute that it did not acquire,
however capable it is. Even the normative Expectimax policy omits cents
when acquisition is costly (Web Appendix H). Second, acquisition cost
and goal specificity are set outside the LLM, by the retailer and
consumer respectively. A vague goal prompt leaves the objective open,
and no improvement in the LLM can supply an objective the consumer did
not state. Third, the pathway held across eight LLMs from three
providers, consistent with benchmarks finding that biases in LLMs are
not resolved by increasing capability
\citep{echterhoff2024cognitive, knipper2025bias, malberg2025comprehensive}.
Finally, our Tool-Lab experiments can be re-executed with the same
design on newer LLMs as they are released.

\section{Theoretical implications}\label{theoretical-implications}

Research on biases in LLMs commonly attributes heuristic-like behavior
to patterns inherited from human-generated training data and treats such
behavior as a stable property of the LLM
\citep{binz2023using, hermann2025generative, echterhoff2024cognitive, knipper2025bias, malberg2025comprehensive, filandrianos2025bias}.
Our findings challenge this assumption by showing that the
susceptibility of AI agents to marketing cues depends also on the
information state in which they choose. This shifts the unit of analysis
for research on AI surrogate consumption from the LLM in isolation to
the agent operating within an information architecture. Training could
still shape LLM behavior, but it does not by itself determine whether
marketing cues influence delegated choices.

Research on pricing cues explains their effectiveness through the
cognitive limitations of the consumer who encounters them
\citep{inman1990promotion, thomas2005penny}. In surrogate consumption,
however, the consumer transfers decision authority to an agent
\citep{solomon1986missing} and never encounters the cue. Our findings
show that when acquisition is costly, the specificity of the principal's
instruction determines whether marketing cues influence the AI agent's
choice. In AI-delegated purchasing, cue effectiveness is therefore
jointly determined by the retailer, through the information
architecture, and by the consumer, through the delegation instruction.
This extends pricing cue research on AI shopping assistants by showing
that cue effectiveness can arise from the delegation instruction and the
information architecture, not just from the limitations of the agent
that encounters the cue.

Behavioral research on LLMs mainly evaluates their final outputs
\citep{joerling2025integrating, wadi2026every, sarstedt2024using}, which
cannot attribute suboptimal choices to failure in information
acquisition or utilization \citep{johnson2008process}. By extending
information-board process tracing
\citep{johnson1989monitoring, payne1993adaptive} to AI agents, Tool-Lab
showed that most suboptimal choices in our studies arose from incomplete
acquisition rather than from flawed use of information. This separation
allows researchers to evaluate an AI agent's choice against two
standards, that is, whether the choice is optimal given the information
the agent held, and whether it is optimal given the full attribute set.
Heuristic-like choices can therefore be assessed without being equated
with utilization failures, and AI agent search can be benchmarked
against a normative policy. Tool-Lab observes the information
acquisition and choice, but not the internal mechanisms of the neural
network. Nevertheless, within this scope, it offers a reusable
methodology for studying AI agents when information must be acquired
before choice.

\section{Managerial implications}\label{managerial-implications}

\textbf{Retailers.} Pricing cues designed for human shoppers showed
little influence on AI agents when they acquired the complete product
information, so retailers should not rely on such cues when the buyer is
an AI agent. Instead, retailers should make the diagnostic attributes
readily available to AI agents, because obscuring these attributes could
steer agents toward suboptimal choices.

\textbf{Consumers.} Lack of specificity in the delegation prompt can
lead to suboptimal choices, particularly, if consumers assume that AI
agents have the common-sense to understand vague prompts (e.g., ``find
the best deal''). Vague prompts allow AI agents to adopt simpler
objectives under constraints and can lead to suboptimal purchases.
Consumers should therefore explicitly state the criterion that defines a
good choice for them, such as the ``lowest price per unit''.

\textbf{AI agent developers.} Because vague instructions made agents
vulnerable to marketing cues, developers should design agentic systems
that ask consumers to clarify vague objectives before recommending an
alternative. Developers should also ensure that agents acquire the
attributes required by the consumer's objective, possibly through an
independent verification mechanism, before committing to a purchase.
Because vulnerability varied across LLMs, developers should audit the
specific LLM they deploy rather than assume that results from one LLM
transfer to another.

\textbf{Regulators.} Our findings show that AI agents tested make
suboptimal choices when the attributes needed to compare value (e.g.,
weight and unit price) are costly to acquire. Current disclosure rules,
such as unit-price requirements, govern what is displayed to human
shoppers, not what is made available to AI agents. A retailer could
therefore comply with these rules while making value-relevant attributes
difficult for AI agents to acquire. Regulators should consider requiring
that diagnostic attributes be easily accessible to AI agents.

\section{Limitations and future
research}\label{limitations-and-future-research}

Although the studies include robustness tests and validation
experiments, several limitations qualify our findings and suggest
directions for future research.

First, Tool-Lab records the observable sequence of tool calls and, when
providers expose them, the reasoning traces that precede those calls.
Like concurrent think-aloud protocols in human decision research
\citep{ericsson1998study, payne1993adaptive}, reasoning traces are
verbal reports of how an LLM describes its task, not direct measures of
the neural network activations that produce its behavior. Therefore,
Tool-Lab shows what information an agent acquired and which objective it
articulated, but not the internal mechanism that generated acquisitions
and choice. Also, our results establish when search truncation produces
heuristic-like choices, but they do not rule out rival accounts of why
truncation occurs, such as statistical regularities in training data or
limits on planning. In addition, reasoning traces were available only
for LLMs whose providers expose them, and we classified traces only in
the purchase volume study. Our evidence distinguishing objective
substitution from value-based omission is therefore suggestive rather
than conclusive. Future research could combine Tool-Lab with mechanistic
interpretability methods, such as activation patching on open-weight
LLMs, to examine how internal representations relate to observed
acquisition. Future studies could also design manipulations that
separate the two routes directly, for example by varying the expected
value of specific attributes while holding the stated objective
constant.

Our hypotheses are tested with a sample of LLMs we tested. One result
ran counter to the framework. Flash Lite selected the suboptimal
alternative even with complete diagnostic information. We do not attempt
to explain differences between individual LLMs, because the
characteristics that distinguish commercial LLMs (e.g., training data
and tool-use tuning) are undisclosed and vary together. Future research
could identify LLM-level moderators of the acquisition and choice
pathways.

Tool-Lab operationalizes acquisition cost as an explicit monetary cost
per inspection, charged to the delegating consumer. This makes cost
exogenous and measurable, but it differs from the internal costs AI
agents encounter in deployment. Future research could test whether costs
in live e-commerce environments produces similar diagnostic truncation.
It could also examine whether manipulating internal costs (e.g., through
reasoning-effort settings) has effects comparable to explicit costs.

Our primary outcome, choice optimality, evaluates whether the AI agent
selected the alternative with the lowest unit price. It does not account
for the search costs incurred. Under high costs, several LLMs acquired
most diagnostic attributes and chose optimally while incurring costs
that exceeded the purchase cost. Choice optimality therefore captures
decision quality but not consumer welfare. Future research could
evaluate AI agent performance on welfare-inclusive measures that combine
product value, search costs, and time. Such measures would show when the
choices of AI agents are economically defensible and when they are
normatively suboptimal.

We studied a single product category in which the normative objective of
unit-price minimization is unambiguous. This allowed us to define choice
optimality precisely and to benchmark search against a normative policy
(Web Appendix H). Many categories, however, involve compensatory
trade-offs among price, quality, and brand, or depend on hedonic and
experiential evaluations for which no objective benchmark exists. The
Tool-Lab studies also used one stimulus configuration per study, with
two alternatives in Study 1 and three in Study 2. Larger choice sets,
different price levels, and different magnitudes of unit-price advantage
may change which attributes AI agents omit. Similarly, our specific goal
prompt named the normative objective exactly. Its protective effect is
therefore likely an upper bound on the benefit of goal specificity,
because consumers rarely state such precise criteria. Future research
could test more naturalistic delegation instructions, including
multi-attribute preferences and affective requests such as asking an
agent to find a product the consumer will ``love''. Such requests would
extend the study of AI agents from analytical to more intuitive and
emotional modes of decision making \citep{huang2021strategic}.
Researchers could also examine whether agents that ask clarifying
questions before searching reduce the vulnerability we document.

Finally, our studies placed a single AI shopping agent in a static
environment. In agentic commerce, seller-side agents may adapt product
information in response to how buyer agents search, creating strategic
interactions that our design does not capture
\citep{calvano2020artificial}. Future research could examine these
interactions, including whether seller agents learn to exploit the
omission patterns we document.

\section{Statements and declarations}\label{statements-and-declarations}

\textbf{Compliance with ethical standards.} The research does not
involve human subjects.\\
\textbf{Competing interests.} The authors have no relevant financial or
non-financial interests to disclose.\\
\textbf{Authorship principles.} All authors contributed to the
manuscript. All authors read and approved the final manuscript.\\
\textbf{Data availability.} The raw session data for all experiments and
the code to reproduce every analysis, table, and figure in the article
and web appendix are available in an anonymized repository for peer
review:
https://osf.io/qs5uz/?view\_only=4231ad6a60d2402d8bc554080700c080.

\section{References}\label{references}

\renewcommand{\bibsection}{}
\bibliography{refs/ai-marketing.bib,refs/acl-refs.bib,refs/marketing.bib,refs/others.bib,refs/myrefs.bib}

\section{Tables and figures}\label{tables-and-figures}

\emph{Note.} H3: The indirect effect of acquisition cost on choice
optimality via diagnostic acquisition is moderated by prompt
specificity.

\begin{longtable}[]{@{}
  >{\raggedright\arraybackslash}p{(\linewidth - 12\tabcolsep) * \real{0.1546}}
  >{\raggedright\arraybackslash}p{(\linewidth - 12\tabcolsep) * \real{0.0825}}
  >{\raggedright\arraybackslash}p{(\linewidth - 12\tabcolsep) * \real{0.1340}}
  >{\raggedright\arraybackslash}p{(\linewidth - 12\tabcolsep) * \real{0.1340}}
  >{\raggedright\arraybackslash}p{(\linewidth - 12\tabcolsep) * \real{0.1340}}
  >{\raggedright\arraybackslash}p{(\linewidth - 12\tabcolsep) * \real{0.1443}}
  >{\raggedright\arraybackslash}p{(\linewidth - 12\tabcolsep) * \real{0.2165}}@{}}
\caption{Aggregate diagnostic information acquisition and choice
optimality across LLMs (Study 1
Tool-Lab)}\label{tbl-s1b-aggregate}\tabularnewline
\toprule\noalign{}
\begin{minipage}[b]{\linewidth}\raggedright
Goal Prompt
\end{minipage} & \begin{minipage}[b]{\linewidth}\raggedright
Cost
\end{minipage} & \begin{minipage}[b]{\linewidth}\raggedright
Cents
\end{minipage} & \begin{minipage}[b]{\linewidth}\raggedright
Weight
\end{minipage} & \begin{minipage}[b]{\linewidth}\raggedright
Dollars
\end{minipage} & \begin{minipage}[b]{\linewidth}\raggedright
Diagnostic
\end{minipage} & \begin{minipage}[b]{\linewidth}\raggedright
Choice Optimality
\end{minipage} \\
\midrule\noalign{}
\endfirsthead
\toprule\noalign{}
\begin{minipage}[b]{\linewidth}\raggedright
Goal Prompt
\end{minipage} & \begin{minipage}[b]{\linewidth}\raggedright
Cost
\end{minipage} & \begin{minipage}[b]{\linewidth}\raggedright
Cents
\end{minipage} & \begin{minipage}[b]{\linewidth}\raggedright
Weight
\end{minipage} & \begin{minipage}[b]{\linewidth}\raggedright
Dollars
\end{minipage} & \begin{minipage}[b]{\linewidth}\raggedright
Diagnostic
\end{minipage} & \begin{minipage}[b]{\linewidth}\raggedright
Choice Optimality
\end{minipage} \\
\midrule\noalign{}
\endhead
\bottomrule\noalign{}
\endlastfoot
Specific & \$0 & 2.00 (0.00) & 2.00 (0.00) & 2.00 (0.00) & 6.00 (0.00) &
0.94 (0.16) \\
Specific & \$10 & 1.72 (0.59) & 2.00 (0.00) & 2.00 (0.00) & 5.72 (0.59)
& 0.83 (0.30) \\
Vague & \$0 & 1.97 (0.09) & 1.90 (0.18) & 2.00 (0.00) & 5.86 (0.24) &
0.89 (0.19) \\
Vague & \$10 & 0.77 (0.78) & 1.10 (0.64) & 1.94 (0.09) & 3.81 (1.30) &
0.29 (0.28) \\
\end{longtable}

\emph{Note.} \(N = 3{,}200\) sessions across eight LLMs (\(n = 100\)
independent Monte Carlo sessions per cell). Entries are cross-LLM means
with cross-LLM standard deviations in parentheses across the eight
evaluated LLMs. Diagnostic acquisitions are the combined count of
\emph{price\_dollars}, \emph{price\_cents}, and \emph{weight} across all
alternatives (maximum = 6). Choice Optimality is the proportion of
sessions selecting the normatively optimal alternative.

\begin{longtable}[]{@{}
  >{\raggedright\arraybackslash}p{(\linewidth - 12\tabcolsep) * \real{0.3060}}
  >{\raggedleft\arraybackslash}p{(\linewidth - 12\tabcolsep) * \real{0.0970}}
  >{\raggedleft\arraybackslash}p{(\linewidth - 12\tabcolsep) * \real{0.0746}}
  >{\raggedright\arraybackslash}p{(\linewidth - 12\tabcolsep) * \real{0.1567}}
  >{\raggedleft\arraybackslash}p{(\linewidth - 12\tabcolsep) * \real{0.0970}}
  >{\raggedleft\arraybackslash}p{(\linewidth - 12\tabcolsep) * \real{0.0746}}
  >{\raggedright\arraybackslash}p{(\linewidth - 12\tabcolsep) * \real{0.1567}}@{}}
\caption{Multilevel moderated mediation of acquisition cost and prompt
specificity on choice optimality (Studies 1 and 2
Tool-Lab)}\label{tbl-modmed}\tabularnewline
\toprule\noalign{}
\multirow{2}{=}{\begin{minipage}[b]{\linewidth}\raggedright
\end{minipage}} &
\multicolumn{3}{>{\raggedleft\arraybackslash}p{(\linewidth - 12\tabcolsep) * \real{0.3284} + 4\tabcolsep}}{%
\begin{minipage}[b]{\linewidth}\raggedleft
Study 1
\end{minipage}} &
\multicolumn{3}{>{\raggedleft\arraybackslash}p{(\linewidth - 12\tabcolsep) * \real{0.3284} + 4\tabcolsep}@{}}{%
\begin{minipage}[b]{\linewidth}\raggedleft
Study 2
\end{minipage}} \\
& \begin{minipage}[b]{\linewidth}\raggedleft
Estimate
\end{minipage} & \begin{minipage}[b]{\linewidth}\raggedleft
\emph{SD}
\end{minipage} & \begin{minipage}[b]{\linewidth}\raggedright
95\% CI
\end{minipage} & \begin{minipage}[b]{\linewidth}\raggedleft
Estimate
\end{minipage} & \begin{minipage}[b]{\linewidth}\raggedleft
\emph{SD}
\end{minipage} & \begin{minipage}[b]{\linewidth}\raggedright
95\% CI
\end{minipage} \\
\midrule\noalign{}
\endfirsthead
\toprule\noalign{}
\multirow{2}{=}{\begin{minipage}[b]{\linewidth}\raggedright
\end{minipage}} &
\multicolumn{3}{>{\raggedleft\arraybackslash}p{(\linewidth - 12\tabcolsep) * \real{0.3284} + 4\tabcolsep}}{%
\begin{minipage}[b]{\linewidth}\raggedleft
Study 1
\end{minipage}} &
\multicolumn{3}{>{\raggedleft\arraybackslash}p{(\linewidth - 12\tabcolsep) * \real{0.3284} + 4\tabcolsep}@{}}{%
\begin{minipage}[b]{\linewidth}\raggedleft
Study 2
\end{minipage}} \\
& \begin{minipage}[b]{\linewidth}\raggedleft
Estimate
\end{minipage} & \begin{minipage}[b]{\linewidth}\raggedleft
\emph{SD}
\end{minipage} & \begin{minipage}[b]{\linewidth}\raggedright
95\% CI
\end{minipage} & \begin{minipage}[b]{\linewidth}\raggedleft
Estimate
\end{minipage} & \begin{minipage}[b]{\linewidth}\raggedleft
\emph{SD}
\end{minipage} & \begin{minipage}[b]{\linewidth}\raggedright
95\% CI
\end{minipage} \\
\midrule\noalign{}
\endhead
\bottomrule\noalign{}
\endlastfoot
\multicolumn{7}{@{}>{\raggedright\arraybackslash}p{(\linewidth - 12\tabcolsep) * \real{0.9627} + 12\tabcolsep}@{}}{%
\textbf{Stage 1: Diagnostic acquisition}} \\
Acquisition cost (\(a_1\)) & -0.270 & 0.217 & {[}-0.698, 0.161{]} &
-0.674 & 0.462 & {[}-1.582, 0.260{]} \\
Vague goal prompt (\(a_2\)) & -0.135 & 0.117 & {[}-0.371, 0.095{]} &
0.015 & 0.072 & {[}-0.125, 0.156{]} \\
Acquisition cost \(\times\) Vague goal prompt (\(a_3\)) & -1.727* &
0.467 & {[}-2.641, -0.765{]} & -2.902* & 1.121 & {[}-5.027, -0.539{]} \\
Acquisition cost under vague goal prompt (\(a_{\text{Vague}}\)) &
-1.997* & 0.487 & {[}-2.952, -1.003{]} & -3.576* & 1.218 & {[}-5.925,
-1.062{]} \\
\multicolumn{7}{@{}>{\raggedright\arraybackslash}p{(\linewidth - 12\tabcolsep) * \real{0.9627} + 12\tabcolsep}@{}}{%
\textbf{Stage 2: Choice optimality}} \\
Diagnostic acquisition (\(b_1\)) & 0.263* & 0.034 & {[}0.196, 0.329{]} &
0.078* & 0.031 & {[}0.016, 0.141{]} \\
Acquisition cost (\(c'\)) & -0.100 & 0.076 & {[}-0.251, 0.051{]} &
-0.006 & 0.020 & {[}-0.045, 0.035{]} \\
Vague goal prompt (\(c_2\)) & -0.056* & 0.019 & {[}-0.094, -0.018{]} &
-0.083* & 0.039 & {[}-0.162, -0.003{]} \\
\multicolumn{7}{@{}>{\raggedright\arraybackslash}p{(\linewidth - 12\tabcolsep) * \real{0.9627} + 12\tabcolsep}@{}}{%
\textbf{Conditional indirect effects}} \\
Specific goal prompt (\(a_1 b_1\)) & -0.070 & 0.057 & {[}-0.184,
0.044{]} & -0.053 & 0.044 & {[}-0.156, 0.021{]} \\
Vague goal prompt (\(a_{\text{Vague}} b_1\)) & -0.522* & 0.132 &
{[}-0.783, -0.256{]} & -0.280* & 0.154 & {[}-0.625, -0.026{]} \\
Index of Moderated Mediation (\(a_3 \times b_1\)) & -0.451* & 0.128 &
{[}-0.704, -0.194{]} & -0.228* & 0.132 & {[}-0.521, -0.009{]} \\
\multicolumn{7}{@{}>{\raggedright\arraybackslash}p{(\linewidth - 12\tabcolsep) * \real{0.9627} + 12\tabcolsep}@{}}{%
\textbf{Total effects}} \\
Specific goal prompt & -0.170 & 0.106 & {[}-0.377, 0.048{]} & -0.058 &
0.047 & {[}-0.164, 0.025{]} \\
Vague goal prompt & -0.621* & 0.155 & {[}-0.933, -0.306{]} & -0.286* &
0.154 & {[}-0.632, -0.027{]} \\
\multicolumn{7}{@{}>{\raggedright\arraybackslash}p{(\linewidth - 12\tabcolsep) * \real{0.9627} + 12\tabcolsep}@{}}{%
\textbf{Between-LLM heterogeneity (\(\tau\))}} \\
\(\tau_{a_1}\) (Acquisition cost) & 0.594* & 0.182 & {[}0.358, 1.048{]}
& 1.255* & 0.382 & {[}0.732, 2.186{]} \\
\(\tau_{a_2}\) (Vague goal prompt) & 0.300* & 0.108 & {[}0.157, 0.571{]}
& 0.052* & 0.046 & {[}0.002, 0.167{]} \\
\(\tau_{a_3}\) (Acquisition cost \(\times\) Vague goal prompt) & 1.292*
& 0.352 & {[}0.800, 2.165{]} & 3.350* & 0.792 & {[}2.165, 5.236{]} \\
\(\tau_{b_1}\) (Diagnostic acquisition) & 0.089* & 0.030 & {[}0.050,
0.164{]} & 0.083* & 0.028 & {[}0.048, 0.153{]} \\
\(\tau_{c'}\) (Acquisition cost) & 0.203* & 0.067 & {[}0.119, 0.374{]} &
0.041* & 0.025 & {[}0.004, 0.099{]} \\
\(\tau_{c_2}\) (Vague goal prompt) & 0.042* & 0.020 & {[}0.012, 0.091{]}
& 0.103* & 0.039 & {[}0.053, 0.200{]} \\
\end{longtable}

\emph{Note.} \(N = 3{,}200\) sessions per study across eight LLMs from
three providers (\(n = 100\) per cell). Entries are posterior means,
posterior standard deviations, and 95\% equal-tailed credible intervals
from Hamiltonian Monte Carlo estimation (12,000 post-warmup draws across
four independent chains). First-stage paths are in cell counts (maximum
6 in Study 1 and 9 in Study 2). Second-stage paths, indirect effects,
direct effects, and total effects are in probability units. Group-level
standard deviations (\(\tau\)) estimate cross-LLM heterogeneity. \(a_1\)
is under the specific goal prompt. \(a_{\text{Vague}}\) is computed as
\(a_1 + a_3\).\\
\emph{*} 95\% credible interval excludes zero.

\begin{longtable}[]{@{}
  >{\raggedright\arraybackslash}p{(\linewidth - 12\tabcolsep) * \real{0.1531}}
  >{\raggedright\arraybackslash}p{(\linewidth - 12\tabcolsep) * \real{0.0816}}
  >{\raggedright\arraybackslash}p{(\linewidth - 12\tabcolsep) * \real{0.1429}}
  >{\raggedright\arraybackslash}p{(\linewidth - 12\tabcolsep) * \real{0.1327}}
  >{\raggedright\arraybackslash}p{(\linewidth - 12\tabcolsep) * \real{0.1327}}
  >{\raggedright\arraybackslash}p{(\linewidth - 12\tabcolsep) * \real{0.1429}}
  >{\raggedright\arraybackslash}p{(\linewidth - 12\tabcolsep) * \real{0.2143}}@{}}
\caption{Aggregate diagnostic information acquisition and choice
optimality across LLMs (Study 2
Tool-Lab)}\label{tbl-s2b-aggregate}\tabularnewline
\toprule\noalign{}
\begin{minipage}[b]{\linewidth}\raggedright
Goal Prompt
\end{minipage} & \begin{minipage}[b]{\linewidth}\raggedright
Cost
\end{minipage} & \begin{minipage}[b]{\linewidth}\raggedright
List Price
\end{minipage} & \begin{minipage}[b]{\linewidth}\raggedright
Discount
\end{minipage} & \begin{minipage}[b]{\linewidth}\raggedright
Weight
\end{minipage} & \begin{minipage}[b]{\linewidth}\raggedright
Diagnostic
\end{minipage} & \begin{minipage}[b]{\linewidth}\raggedright
Choice Optimality
\end{minipage} \\
\midrule\noalign{}
\endfirsthead
\toprule\noalign{}
\begin{minipage}[b]{\linewidth}\raggedright
Goal Prompt
\end{minipage} & \begin{minipage}[b]{\linewidth}\raggedright
Cost
\end{minipage} & \begin{minipage}[b]{\linewidth}\raggedright
List Price
\end{minipage} & \begin{minipage}[b]{\linewidth}\raggedright
Discount
\end{minipage} & \begin{minipage}[b]{\linewidth}\raggedright
Weight
\end{minipage} & \begin{minipage}[b]{\linewidth}\raggedright
Diagnostic
\end{minipage} & \begin{minipage}[b]{\linewidth}\raggedright
Choice Optimality
\end{minipage} \\
\midrule\noalign{}
\endhead
\bottomrule\noalign{}
\endlastfoot
Specific & \$0 & 3.00 (0.00) & 2.98 (0.03) & 3.00 (0.00) & 8.98 (0.03) &
0.90 (0.25) \\
Specific & \$50 & 2.90 (0.25) & 2.50 (0.59) & 2.86 (0.35) & 8.26 (1.03)
& 0.88 (0.23) \\
Vague & \$0 & 3.00 (0.00) & 3.00 (0.00) & 3.00 (0.00) & 9.00 (0.00) &
0.86 (0.31) \\
Vague & \$50 & 1.94 (1.24) & 1.58 (1.23) & 1.31 (1.14) & 4.82 (3.49) &
0.48 (0.29) \\
\end{longtable}

\emph{Note.} \(N = 3{,}200\) sessions across eight LLMs from three
providers (\(n = 100\) independent Monte Carlo sessions per cell).
Entries are cross-model means with cross-model standard deviations in
parentheses across the eight evaluated LLMs. Diagnostic acquisitions are
the combined count of \emph{list\_price}, \emph{discount\_percentage},
and \emph{weight} across all three alternatives (maximum = 9). Choice
Optimality is the proportion of sessions selecting the normatively
optimal alternative (\emph{Coffee E}: \$10.00 list price, 0\% discount,
10.9 oz; \$0.917/oz).

\newpage{}

\section*{Web appendix A: Supplementary figures and
tables}\label{sec-appendix-tables}
\addcontentsline{toc}{section}{Web appendix A: Supplementary figures and
tables}

\subsection{Study 1 - Baseline}\label{study-1---baseline}

{\def\LTcaptype{apptbl}
\begin{longtable}[]{@{}
  >{\raggedright\arraybackslash}p{(\linewidth - 18\tabcolsep) * \real{0.1264}}
  >{\raggedright\arraybackslash}p{(\linewidth - 18\tabcolsep) * \real{0.1264}}
  >{\raggedright\arraybackslash}p{(\linewidth - 18\tabcolsep) * \real{0.0934}}
  >{\raggedright\arraybackslash}p{(\linewidth - 18\tabcolsep) * \real{0.0934}}
  >{\raggedright\arraybackslash}p{(\linewidth - 18\tabcolsep) * \real{0.0934}}
  >{\raggedright\arraybackslash}p{(\linewidth - 18\tabcolsep) * \real{0.0934}}
  >{\raggedright\arraybackslash}p{(\linewidth - 18\tabcolsep) * \real{0.0934}}
  >{\raggedright\arraybackslash}p{(\linewidth - 18\tabcolsep) * \real{0.0934}}
  >{\raggedright\arraybackslash}p{(\linewidth - 18\tabcolsep) * \real{0.0934}}
  >{\raggedright\arraybackslash}p{(\linewidth - 18\tabcolsep) * \real{0.0934}}@{}}
\caption{{\label{apptbl-s1a-realistic-cents}}Conditional logit
estimates when all attributes are visible (Study 1 Baseline)}\tabularnewline
\toprule\noalign{}
\begin{minipage}[b]{\linewidth}\raggedright
\end{minipage} & \begin{minipage}[b]{\linewidth}\raggedright
Normative Benchmark
\end{minipage} & \begin{minipage}[b]{\linewidth}\raggedright
Flash Lite
\end{minipage} & \begin{minipage}[b]{\linewidth}\raggedright
Flash
\end{minipage} & \begin{minipage}[b]{\linewidth}\raggedright
Pro
\end{minipage} & \begin{minipage}[b]{\linewidth}\raggedright
Sonnet
\end{minipage} & \begin{minipage}[b]{\linewidth}\raggedright
Opus
\end{minipage} & \begin{minipage}[b]{\linewidth}\raggedright
Luna
\end{minipage} & \begin{minipage}[b]{\linewidth}\raggedright
Terra
\end{minipage} & \begin{minipage}[b]{\linewidth}\raggedright
Sol
\end{minipage} \\
\midrule\noalign{}
\endfirsthead
\toprule\noalign{}
\begin{minipage}[b]{\linewidth}\raggedright
\end{minipage} & \begin{minipage}[b]{\linewidth}\raggedright
Normative Benchmark
\end{minipage} & \begin{minipage}[b]{\linewidth}\raggedright
Flash Lite
\end{minipage} & \begin{minipage}[b]{\linewidth}\raggedright
Flash
\end{minipage} & \begin{minipage}[b]{\linewidth}\raggedright
Pro
\end{minipage} & \begin{minipage}[b]{\linewidth}\raggedright
Sonnet
\end{minipage} & \begin{minipage}[b]{\linewidth}\raggedright
Opus
\end{minipage} & \begin{minipage}[b]{\linewidth}\raggedright
Luna
\end{minipage} & \begin{minipage}[b]{\linewidth}\raggedright
Terra
\end{minipage} & \begin{minipage}[b]{\linewidth}\raggedright
Sol
\end{minipage} \\
\midrule\noalign{}
\endhead
\bottomrule\noalign{}
\endlastfoot
Weight (oz) & 2.78*** (0.21) & 1.51*** (0.09) & 1.15*** (0.07) & 2.04***
(0.13) & 2.61*** (0.19) & 1.06*** (0.06) & 2.60*** (0.19) & 2.78***
(0.21) & 2.78*** (0.21) \\
Dollars (\$) & -4.28*** (0.31) & -2.14*** (0.13) & -1.69*** (0.10) &
-3.11*** (0.20) & -3.99*** (0.28) & -1.54*** (0.09) & -4.00*** (0.29) &
-4.29*** (0.31) & -4.29*** (0.31) \\
Cents (\$) & -4.74*** (0.51) & -1.81*** (0.30) & -0.92** (0.31) &
-3.18*** (0.37) & -4.45*** (0.48) & -1.25*** (0.29) & -4.45*** (0.48) &
-4.75*** (0.51) & -4.75*** (0.51) \\
Brand (Maxwell House) & -0.24 (0.23) & -0.18 (0.18) & -0.56** (0.21) &
-0.30 (0.20) & -0.09 (0.22) & -0.46* (0.20) & -0.21 (0.22) & -0.24
(0.23) & -0.24 (0.23) \\
Brand (McCafé) & -0.35 (0.22) & 0.17 (0.18) & 0.95*** (0.19) & -0.03
(0.20) & -0.19 (0.21) & 2.10*** (0.19) & -0.26 (0.21) & -0.37 (0.22) &
-0.37 (0.22) \\
Brand (Starbucks) & -0.12 (0.22) & 0.91*** (0.18) & 4.44*** (0.28) &
1.03*** (0.20) & 0.26 (0.21) & 3.50*** (0.23) & -0.08 (0.21) & -0.10
(0.22) & -0.10 (0.22) \\
Position (2) & 0.17 (0.22) & -1.12*** (0.17) & -0.51** (0.18) & -0.33
(0.19) & 0.16 (0.21) & -0.02 (0.17) & 0.12 (0.21) & 0.16 (0.22) & 0.16
(0.22) \\
Position (3) & 0.09 (0.22) & -1.68*** (0.19) & -0.67*** (0.17) & -0.19
(0.19) & 0.20 (0.21) & 0.29 (0.17) & 0.09 (0.21) & 0.07 (0.22) & 0.07
(0.22) \\
Position (4) & 0.09 (0.21) & -2.38*** (0.21) & -0.90*** (0.18) & -0.23
(0.19) & 0.15 (0.21) & 0.37* (0.17) & 0.10 (0.21) & 0.11 (0.21) & 0.11
(0.21) \\
Choice Sets (N) & 1014 & 1014 & 1014 & 1014 & 1014 & 1014 & 1014 & 1014
& 1014 \\
Log-Likelihood & -212.06 & -344.99 & -330.50 & -274.90 & -224.43 &
-379.88 & -225.12 & -211.77 & -211.77 \\
\end{longtable}
}

\emph{Note.} Standard errors in parentheses.\\
Cents are drawn independently from the dollars. Cents (\$) represents
the cent attribute normalized to a dollar unit
(\(\text{Cents (\$)} = \frac{\text{Cents}}{100}\)) For Terra and Sol the
choices are identical.\\
* p \textless{} .05, ** p \textless{} .01, *** p \textless{} .001.

{\def\LTcaptype{apptbl}
\begin{longtable}[]{@{}
  >{\raggedright\arraybackslash}p{(\linewidth - 8\tabcolsep) * \real{0.1795}}
  >{\raggedleft\arraybackslash}p{(\linewidth - 8\tabcolsep) * \real{0.1795}}
  >{\raggedleft\arraybackslash}p{(\linewidth - 8\tabcolsep) * \real{0.2051}}
  >{\raggedleft\arraybackslash}p{(\linewidth - 8\tabcolsep) * \real{0.3162}}
  >{\raggedright\arraybackslash}p{(\linewidth - 8\tabcolsep) * \real{0.1197}}@{}}
\caption{{\label{apptbl-s1a-cent-dollar-ratio}}Ratio of cent to dollar
price sensitivity (Study 1 Baseline)}\tabularnewline
\toprule\noalign{}
\begin{minipage}[b]{\linewidth}\raggedright
LLM
\end{minipage} & \begin{minipage}[b]{\linewidth}\raggedleft
\(\beta_{Dollars (\$)}\)
\end{minipage} & \begin{minipage}[b]{\linewidth}\raggedleft
\(\beta_{Cents (\$)}\)
\end{minipage} & \begin{minipage}[b]{\linewidth}\raggedleft
\(\beta_{Cents (\$)} / \beta_{Dollars (\$)}\)
\end{minipage} & \begin{minipage}[b]{\linewidth}\raggedright
95\% CI
\end{minipage} \\
\midrule\noalign{}
\endfirsthead
\toprule\noalign{}
\begin{minipage}[b]{\linewidth}\raggedright
LLM
\end{minipage} & \begin{minipage}[b]{\linewidth}\raggedleft
\(\beta_{Dollars (\$)}\)
\end{minipage} & \begin{minipage}[b]{\linewidth}\raggedleft
\(\beta_{Cents (\$)}\)
\end{minipage} & \begin{minipage}[b]{\linewidth}\raggedleft
\(\beta_{Cents (\$)} / \beta_{Dollars (\$)}\)
\end{minipage} & \begin{minipage}[b]{\linewidth}\raggedright
95\% CI
\end{minipage} \\
\midrule\noalign{}
\endhead
\bottomrule\noalign{}
\endlastfoot
Normative Benchmark & -4.28 & -4.74 & 1.11 & {[}0.94, 1.27{]} \\
Flash Lite & -2.14 & -1.81 & 0.84 & {[}0.58, 1.11{]} \\
Flash & -1.69 & -0.92 & 0.55 & {[}0.19, 0.90{]} \\
Pro & -3.11 & -3.18 & 1.02 & {[}0.83, 1.22{]} \\
Sonnet & -3.99 & -4.45 & 1.12 & {[}0.94, 1.28{]} \\
Opus & -1.54 & -1.25 & 0.81 & {[}0.45, 1.17{]} \\
Luna & -4 & -4.45 & 1.11 & {[}0.94, 1.28{]} \\
Terra & -4.29 & -4.75 & 1.11 & {[}0.94, 1.27{]} \\
Sol & -4.29 & -4.75 & 1.11 & {[}0.94, 1.27{]} \\
\end{longtable}
}

\emph{Note.} \(N = 1{,}014\) choice sets (\(4{,}056\) observations) per
LLM. Fractional cents are scaled to dollars
(\(\text{Cents (\$)} = \text{Cents}/100\)), hence both coefficients
express the marginal disutility of a one-dollar increase in that
component of the price. A ratio of one indicates that a cent is weighted
at exactly one hundredth of a dollar. Ratios below one indicate that the
cent component is discounted relative to the dollar component.
Confidence intervals are exact (Fieller). Estimates for GPT-5.6 Terra
and GPT-5.6 Sol are identical. Both LLMs selected the same
cost-minimizing alternative in all 1,014 choice sets.

\subsection{Study 1 - Tool-Lab}\label{study-1---tool-lab}

{\def\LTcaptype{apptbl}
\begin{longtable}[]{@{}
  >{\raggedright\arraybackslash}p{(\linewidth - 2\tabcolsep) * \real{0.1667}}
  >{\raggedright\arraybackslash}p{(\linewidth - 2\tabcolsep) * \real{0.8333}}@{}}
\caption{{\label{apptbl-s1b-prompts}}Prompt manipulation establishing
goal specificity (Studies 1 and 2 Tool-Lab)}\tabularnewline
\toprule\noalign{}
\begin{minipage}[b]{\linewidth}\raggedright
Prompt goal
\end{minipage} & \begin{minipage}[b]{\linewidth}\raggedright
Prompt text
\end{minipage} \\
\midrule\noalign{}
\endfirsthead
\toprule\noalign{}
\begin{minipage}[b]{\linewidth}\raggedright
Prompt goal
\end{minipage} & \begin{minipage}[b]{\linewidth}\raggedright
Prompt text
\end{minipage} \\
\midrule\noalign{}
\endhead
\bottomrule\noalign{}
\endlastfoot
Vague & You have been delegated by a consumer to make purchase decision.
Your goal is to choose an instant coffee for the user. After you choose
a coffee, the user would purchase 5 bags of your chosen coffee. You can
open the cells to get information about the product using the
\emph{inspect\_cell} tool, or you can stop searching and submit your
choice with the \emph{submit\_choice} tool. Every time you use the
\emph{inspect\_cell} tool, the user loses \{inspect\_tool\_cost\}.
\emph{submit\_choice} has no cost. You need to find the \textbf{best
deal} on instant coffee. Which coffee would you choose for the user? \\
Specific & You have been delegated by a consumer to make purchase
decision. Your goal is to choose an instant coffee for the user. After
you choose a coffee, the user would purchase 5 bags of your chosen
coffee. You can open the cells to get information about the product
using the \emph{inspect\_cell} tool, or you can stop searching and
submit your choice with the \emph{submit\_choice} tool. Every time you
use the \emph{inspect\_cell} tool, the user loses
\{inspect\_tool\_cost\}. \emph{submit\_choice} has no cost. You need to
find the coffee with the \textbf{lowest price per ounce}. Which coffee
would you choose for the user? \\
\end{longtable}
}

\emph{Note.} The parameter \{inspect\_tool\_cost\} was dynamically
replaced with the cost (\$0.00 vs.~\$10.00 in Study 1 and \$0.00
vs.~\$50.00 in Study 2) according to the assigned acquisition cost
condition.

{\def\LTcaptype{apptbl}
\begin{longtable}[]{@{}
  >{\raggedright\arraybackslash}p{(\linewidth - 18\tabcolsep) * \real{0.0929}}
  >{\raggedright\arraybackslash}p{(\linewidth - 18\tabcolsep) * \real{0.0500}}
  >{\raggedright\arraybackslash}p{(\linewidth - 18\tabcolsep) * \real{0.1000}}
  >{\raggedright\arraybackslash}p{(\linewidth - 18\tabcolsep) * \real{0.1000}}
  >{\raggedright\arraybackslash}p{(\linewidth - 18\tabcolsep) * \real{0.1000}}
  >{\raggedright\arraybackslash}p{(\linewidth - 18\tabcolsep) * \real{0.1000}}
  >{\raggedright\arraybackslash}p{(\linewidth - 18\tabcolsep) * \real{0.1000}}
  >{\raggedright\arraybackslash}p{(\linewidth - 18\tabcolsep) * \real{0.1000}}
  >{\raggedright\arraybackslash}p{(\linewidth - 18\tabcolsep) * \real{0.1000}}
  >{\raggedright\arraybackslash}p{(\linewidth - 18\tabcolsep) * \real{0.1000}}@{}}
\caption{{\label{apptbl-s1b-descriptives}}Inspecting Cents, Weight,
Dollars, Diagnostic information acquisition and choice optimality (Study
1 Tool-Lab)}\tabularnewline
\toprule\noalign{}
\begin{minipage}[b]{\linewidth}\raggedright
LLM
\end{minipage} & \begin{minipage}[b]{\linewidth}\raggedright
Cost
\end{minipage} &
\multicolumn{4}{>{\raggedright\arraybackslash}p{(\linewidth - 18\tabcolsep) * \real{0.4000} + 6\tabcolsep}}{%
\begin{minipage}[b]{\linewidth}\raggedright
Vague
\end{minipage}} &
\multicolumn{4}{>{\raggedright\arraybackslash}p{(\linewidth - 18\tabcolsep) * \real{0.4000} + 6\tabcolsep}@{}}{%
\begin{minipage}[b]{\linewidth}\raggedright
Specific
\end{minipage}} \\
\begin{minipage}[b]{\linewidth}\raggedright
\end{minipage} & \begin{minipage}[b]{\linewidth}\raggedright
\end{minipage} & \begin{minipage}[b]{\linewidth}\raggedright
Cents
\end{minipage} & \begin{minipage}[b]{\linewidth}\raggedright
Weight
\end{minipage} & \begin{minipage}[b]{\linewidth}\raggedright
Diagnostic
\end{minipage} & \begin{minipage}[b]{\linewidth}\raggedright
Optimality
\end{minipage} & \begin{minipage}[b]{\linewidth}\raggedright
Cents
\end{minipage} & \begin{minipage}[b]{\linewidth}\raggedright
Weight
\end{minipage} & \begin{minipage}[b]{\linewidth}\raggedright
Diagnostic
\end{minipage} & \begin{minipage}[b]{\linewidth}\raggedright
Optimality
\end{minipage} \\
\midrule\noalign{}
\endfirsthead
\toprule\noalign{}
\begin{minipage}[b]{\linewidth}\raggedright
LLM
\end{minipage} & \begin{minipage}[b]{\linewidth}\raggedright
Cost
\end{minipage} &
\multicolumn{4}{>{\raggedright\arraybackslash}p{(\linewidth - 18\tabcolsep) * \real{0.4000} + 6\tabcolsep}}{%
\begin{minipage}[b]{\linewidth}\raggedright
Vague
\end{minipage}} &
\multicolumn{4}{>{\raggedright\arraybackslash}p{(\linewidth - 18\tabcolsep) * \real{0.4000} + 6\tabcolsep}@{}}{%
\begin{minipage}[b]{\linewidth}\raggedright
Specific
\end{minipage}} \\
\begin{minipage}[b]{\linewidth}\raggedright
\end{minipage} & \begin{minipage}[b]{\linewidth}\raggedright
\end{minipage} & \begin{minipage}[b]{\linewidth}\raggedright
Cents
\end{minipage} & \begin{minipage}[b]{\linewidth}\raggedright
Weight
\end{minipage} & \begin{minipage}[b]{\linewidth}\raggedright
Diagnostic
\end{minipage} & \begin{minipage}[b]{\linewidth}\raggedright
Optimality
\end{minipage} & \begin{minipage}[b]{\linewidth}\raggedright
Cents
\end{minipage} & \begin{minipage}[b]{\linewidth}\raggedright
Weight
\end{minipage} & \begin{minipage}[b]{\linewidth}\raggedright
Diagnostic
\end{minipage} & \begin{minipage}[b]{\linewidth}\raggedright
Optimality
\end{minipage} \\
\midrule\noalign{}
\endhead
\bottomrule\noalign{}
\endlastfoot
Flash Lite & \$0 & 2.00 (0.00) & 1.54 (0.85) & 5.54 (0.85) & 0.47 (0.50)
& 2.00 (0.00) & 2.00 (0.00) & 6.00 (0.00) & 0.54 (0.50) \\
Flash Lite & \$10 & 2.00 (0.00) & 1.10 (1.00) & 5.10 (1.00) & 0.35
(0.48) & 2.00 (0.00) & 2.00 (0.00) & 6.00 (0.00) & 0.75 (0.44) \\
Flash & \$0 & 2.00 (0.00) & 2.00 (0.00) & 6.00 (0.00) & 1.00 (0.00) &
2.00 (0.00) & 2.00 (0.00) & 6.00 (0.00) & 1.00 (0.00) \\
Flash & \$10 & 0.20 (0.60) & 0.96 (0.96) & 3.01 (1.45) & 0.16 (0.37) &
1.58 (0.82) & 2.00 (0.00) & 5.58 (0.82) & 0.79 (0.41) \\
Pro & \$0 & 1.75 (0.64) & 1.68 (0.74) & 5.43 (1.34) & 0.76 (0.43) & 2.00
(0.00) & 2.00 (0.00) & 6.00 (0.00) & 1.00 (0.00) \\
Pro & \$10 & 0.10 (0.44) & 0.14 (0.51) & 2.20 (0.94) & 0.07 (0.26) &
1.98 (0.20) & 2.00 (0.00) & 5.98 (0.20) & 0.99 (0.10) \\
Luna & \$0 & 2.00 (0.00) & 2.00 (0.00) & 6.00 (0.00) & 0.96 (0.20) &
2.00 (0.00) & 2.00 (0.00) & 6.00 (0.00) & 1.00 (0.00) \\
Luna & \$10 & 1.82 (0.58) & 2.00 (0.00) & 5.82 (0.58) & 0.84 (0.37) &
1.98 (0.20) & 2.00 (0.00) & 5.98 (0.20) & 1.00 (0.00) \\
Terra & \$0 & 1.98 (0.14) & 1.94 (0.34) & 5.92 (0.46) & 0.96 (0.20) &
2.00 (0.00) & 2.00 (0.00) & 6.00 (0.00) & 1.00 (0.00) \\
Terra & \$10 & 0.58 (0.79) & 0.45 (0.83) & 2.78 (1.68) & 0.22 (0.42) &
1.95 (0.26) & 2.00 (0.00) & 5.95 (0.26) & 0.99 (0.10) \\
Sol & \$0 & 2.00 (0.00) & 2.00 (0.00) & 6.00 (0.00) & 1.00 (0.00) & 2.00
(0.00) & 2.00 (0.00) & 6.00 (0.00) & 1.00 (0.00) \\
Sol & \$10 & 0.37 (0.77) & 0.90 (1.00) & 3.27 (1.56) & 0.18 (0.39) &
2.00 (0.00) & 2.00 (0.00) & 6.00 (0.00) & 1.00 (0.00) \\
Sonnet & \$0 & 2.00 (0.00) & 2.00 (0.00) & 6.00 (0.00) & 1.00 (0.00) &
2.00 (0.00) & 2.00 (0.00) & 6.00 (0.00) & 1.00 (0.00) \\
Sonnet & \$10 & 1.06 (1.00) & 1.88 (0.48) & 4.94 (1.22) & 0.53 (0.50) &
2.00 (0.00) & 2.00 (0.00) & 6.00 (0.00) & 1.00 (0.00) \\
Opus & \$0 & 2.00 (0.00) & 2.00 (0.00) & 6.00 (0.00) & 1.00 (0.00) &
2.00 (0.00) & 2.00 (0.00) & 6.00 (0.00) & 1.00 (0.00) \\
Opus & \$10 & 0.00 (0.00) & 1.38 (0.93) & 3.38 (0.93) & 0.00 (0.00) &
0.31 (0.71) & 2.00 (0.00) & 4.31 (0.71) & 0.14 (0.35) \\
\end{longtable}
}

\emph{Note.} \(N = 3{,}200\) sessions across eight LLMs from three
providers (\(n = 100\) independent Monte Carlo sessions per cell). Means
with standard deviation in parentheses. Diagnostic acquisitions are the
combined count of \emph{price\_dollars}, \emph{price\_cents}, and
\emph{weight} across both alternatives (maximum = 6).\\
\emph{price\_dollars} acquisitions are omitted for brevity. Choice
Optimality is the proportion of sessions selecting the normatively
optimal alternative (\emph{Coffee B}: \$5.00 for 11 oz; 0.455 \$/oz).

{\def\LTcaptype{apptbl}
\begin{longtable}[]{@{}
  >{\raggedright\arraybackslash}p{(\linewidth - 8\tabcolsep) * \real{0.1667}}
  >{\centering\arraybackslash}p{(\linewidth - 8\tabcolsep) * \real{0.2083}}
  >{\centering\arraybackslash}p{(\linewidth - 8\tabcolsep) * \real{0.2083}}
  >{\centering\arraybackslash}p{(\linewidth - 8\tabcolsep) * \real{0.2083}}
  >{\centering\arraybackslash}p{(\linewidth - 8\tabcolsep) * \real{0.2083}}@{}}
\caption{{\label{apptbl-s1b-permodel-paths}}Moderated mediation
structural path estimates by LLM (Study 1 Tool-Lab)}\tabularnewline
\toprule\noalign{}
\begin{minipage}[b]{\linewidth}\raggedright
LLM
\end{minipage} & \begin{minipage}[b]{\linewidth}\centering
\(a_1\) (Cost: Unit price)
\end{minipage} & \begin{minipage}[b]{\linewidth}\centering
\(a_3\) (Cost \(\times\) Vague)
\end{minipage} & \begin{minipage}[b]{\linewidth}\centering
\(b_1\) (Diagnostic acquistion)
\end{minipage} & \begin{minipage}[b]{\linewidth}\centering
\(c'\) (Direct effect)
\end{minipage} \\
\midrule\noalign{}
\endfirsthead
\toprule\noalign{}
\begin{minipage}[b]{\linewidth}\raggedright
LLM
\end{minipage} & \begin{minipage}[b]{\linewidth}\centering
\(a_1\) (Cost: Unit price)
\end{minipage} & \begin{minipage}[b]{\linewidth}\centering
\(a_3\) (Cost \(\times\) Vague)
\end{minipage} & \begin{minipage}[b]{\linewidth}\centering
\(b_1\) (Diagnostic acquistion)
\end{minipage} & \begin{minipage}[b]{\linewidth}\centering
\(c'\) (Direct effect)
\end{minipage} \\
\midrule\noalign{}
\endhead
\bottomrule\noalign{}
\endlastfoot
Flash Lite & 0.005 {[}-0.149, 0.159{]} & -0.479 {[}-0.729, -0.225{]}* &
0.297 {[}0.265, 0.328{]}* & 0.106 {[}0.062, 0.151{]}* \\
Flash & -0.423 {[}-0.573, -0.270{]}* & -2.548 {[}-2.790, -2.305{]}* &
0.210 {[}0.191, 0.230{]}* & -0.166 {[}-0.219, -0.113{]}* \\
Pro & -0.023 {[}-0.175, 0.131{]} & -3.233 {[}-3.482, -2.983{]}* & 0.224
{[}0.206, 0.241{]}* & 0.013 {[}-0.038, 0.065{]} \\
Luna & -0.017 {[}-0.170, 0.132{]} & -0.163 {[}-0.402, 0.085{]} & 0.385
{[}0.321, 0.451{]}* & -0.022 {[}-0.066, 0.022{]} \\
Sol & -0.012 {[}-0.166, 0.138{]} & -2.694 {[}-2.937, -2.457{]}* & 0.248
{[}0.228, 0.267{]}* & -0.072 {[}-0.124, -0.022{]}* \\
Terra & -0.059 {[}-0.211, 0.090{]} & -3.068 {[}-3.309, -2.827{]}* &
0.179 {[}0.161, 0.197{]}* & -0.090 {[}-0.143, -0.038{]}* \\
Sonnet & -0.006 {[}-0.158, 0.150{]} & -1.043 {[}-1.285, -0.802{]}* &
0.379 {[}0.347, 0.411{]}* & -0.035 {[}-0.082, 0.012{]} \\
Opus & -1.664 {[}-1.819, -1.507{]}* & -0.960 {[}-1.210, -0.710{]}* &
0.179 {[}0.145, 0.213{]}* & -0.538 {[}-0.624, -0.452{]}* \\
\end{longtable}
}

\emph{Note.} \(N = 3{,}200\) sessions across eight LLMs (400 sessions
per LLM: 100 per prompt framing \(\times\) cost condition). Entries are
posterior means with 95\% equal-tailed credible intervals in brackets
from the multilevel Model 7 linear probability model (12,000 post-warmup
draws across four independent chains).\\
\(a_1\) is the simple slope of acquisition cost under explicit
unit-price instructions (in cells).\\
\(a_3\) is the first-stage moderation interaction of cost by ambiguous
deal framing (in cells).\\
\(b_1\) is the marginal effect of diagnostic search depth on choice
optimality probability.\\
\(c'\) is the unmoderated direct effect of acquisition cost on choice
optimality.\\
* indicates that the 95\% credible interval excludes zero.

{\def\LTcaptype{apptbl}
\begin{longtable}[]{@{}
  >{\raggedright\arraybackslash}p{(\linewidth - 8\tabcolsep) * \real{0.1667}}
  >{\centering\arraybackslash}p{(\linewidth - 8\tabcolsep) * \real{0.2083}}
  >{\centering\arraybackslash}p{(\linewidth - 8\tabcolsep) * \real{0.2083}}
  >{\centering\arraybackslash}p{(\linewidth - 8\tabcolsep) * \real{0.2083}}
  >{\centering\arraybackslash}p{(\linewidth - 8\tabcolsep) * \real{0.2083}}@{}}
\caption{{\label{apptbl-s1b-permodel-indirect}}Conditional indirect
effects and index of moderated mediation by LLM (Study 1 Tool-Lab)}\tabularnewline
\toprule\noalign{}
\begin{minipage}[b]{\linewidth}\raggedright
LLM
\end{minipage} & \begin{minipage}[b]{\linewidth}\centering
Indirect (Unit price)
\end{minipage} & \begin{minipage}[b]{\linewidth}\centering
Indirect (Vague prompt)
\end{minipage} & \begin{minipage}[b]{\linewidth}\centering
Index of Moderated Mediation (IMM)
\end{minipage} & \begin{minipage}[b]{\linewidth}\centering
Total effect (Vague)
\end{minipage} \\
\midrule\noalign{}
\endfirsthead
\toprule\noalign{}
\begin{minipage}[b]{\linewidth}\raggedright
LLM
\end{minipage} & \begin{minipage}[b]{\linewidth}\centering
Indirect (Unit price)
\end{minipage} & \begin{minipage}[b]{\linewidth}\centering
Indirect (Vague prompt)
\end{minipage} & \begin{minipage}[b]{\linewidth}\centering
Index of Moderated Mediation (IMM)
\end{minipage} & \begin{minipage}[b]{\linewidth}\centering
Total effect (Vague)
\end{minipage} \\
\midrule\noalign{}
\endhead
\bottomrule\noalign{}
\endlastfoot
Flash Lite & 0.001 {[}-0.044, 0.048{]} & -0.141 {[}-0.201, -0.081{]}* &
-0.142 {[}-0.218, -0.066{]}* & -0.035 {[}-0.108, 0.038{]} \\
Flash & -0.089 {[}-0.122, -0.056{]}* & -0.624 {[}-0.695, -0.556{]}* &
-0.535 {[}-0.607, -0.467{]}* & -0.791 {[}-0.855, -0.727{]}* \\
Pro & -0.005 {[}-0.040, 0.029{]} & -0.730 {[}-0.802, -0.657{]}* & -0.725
{[}-0.805, -0.646{]}* & -0.716 {[}-0.785, -0.649{]}* \\
Luna & -0.006 {[}-0.066, 0.051{]} & -0.069 {[}-0.145, 0.005{]} & -0.063
{[}-0.158, 0.033{]} & -0.091 {[}-0.176, -0.005{]}* \\
Terra & -0.011 {[}-0.038, 0.016{]} & -0.561 {[}-0.626, -0.494{]}* &
-0.550 {[}-0.622, -0.481{]}* & -0.650 {[}-0.712, -0.589{]}* \\
Sol & -0.003 {[}-0.041, 0.035{]} & -0.671 {[}-0.743, -0.601{]}* & -0.668
{[}-0.750, -0.590{]}* & -0.742 {[}-0.813, -0.674{]}* \\
Sonnet & -0.002 {[}-0.060, 0.057{]} & -0.397 {[}-0.481, -0.320{]}* &
-0.395 {[}-0.496, -0.300{]}* & -0.432 {[}-0.520, -0.348{]}* \\
Opus & -0.298 {[}-0.363, -0.236{]}* & -0.470 {[}-0.568, -0.374{]}* &
-0.172 {[}-0.230, -0.119{]}* & -1.008 {[}-1.067, -0.952{]}* \\
\end{longtable}
}

\emph{Note.} \(N = 3{,}200\) sessions across eight LLMs (400 sessions
per LLM). Entries are posterior means with 95\% equal-tailed credible
intervals in brackets. All values are expressed in probability units.
Indirect (Unit Price) is \(a_1 \times b_1\). Indirect (Deal Prompt) is
\((a_1 + a_3) \times b_1\). The Index of Moderated Mediation (IMM) is
the difference in indirect effects (\(a_3 \times b_1\)), quantifying the
amplification of the search-mediated penalty under ambiguous framing.
Total Effect (Deal) is \(c' + (a_1 + a_3)b_1\). Because the second stage
is a linear probability model, effects are not bounded between \(-1\)
and \(1\). For example, the total effect for Opus (\(-1.008\)) reflects
this.\\
* indicates that the 95\% credible interval excludes zero.

{\def\LTcaptype{apptbl}
\begin{longtable}[]{@{}
  >{\raggedright\arraybackslash}p{(\linewidth - 2\tabcolsep) * \real{0.5000}}
  >{\raggedright\arraybackslash}p{(\linewidth - 2\tabcolsep) * \real{0.5000}}@{}}
\caption{{\label{apptbl-s1b-prompt-variants}}Prompts used in the prompt
variation study (Study 1 Tool-Lab)}\tabularnewline
\toprule\noalign{}
\begin{minipage}[b]{\linewidth}\raggedright
Variant
\end{minipage} & \begin{minipage}[b]{\linewidth}\raggedright
Prompt text
\end{minipage} \\
\midrule\noalign{}
\endfirsthead
\toprule\noalign{}
\begin{minipage}[b]{\linewidth}\raggedright
Variant
\end{minipage} & \begin{minipage}[b]{\linewidth}\raggedright
Prompt text
\end{minipage} \\
\midrule\noalign{}
\endhead
\bottomrule\noalign{}
\endlastfoot
Original & You have been delegated by a consumer to make purchase
decision. Your goal is to choose an instant coffee for the user. After
you choose a coffee, the user would purchase 5 bags of your chosen
coffee. You can open the cells to get information about the product
using the \emph{inspect\_cell} tool, or you can stop searching and
submit your choice with the \emph{submit\_choice} tool. Every time you
use the \emph{inspect\_cell} tool, the user loses
\{inspect\_tool\_cost\}. \emph{submit\_choice} has no cost.
\textbf{\{goal\_prompt\}} Which coffee would you choose for the user? \\
Sentence reorder & You have been delegated by a consumer to make
purchase decision. Your goal is to choose an instant coffee for the
user. \textbf{\{goal\_prompt\}} After you choose a coffee, the user
would purchase 5 bags of your chosen coffee. You can open the cells to
get information about the product using the \emph{inspect\_cell} tool,
or you can stop searching and submit your choice with the
\emph{submit\_choice} tool. Every time you use the \emph{inspect\_cell}
tool, the user loses \{inspect\_tool\_cost\}. \emph{submit\_choice} has
no cost. Which coffee would you choose for the user? \\
Cost paraphrase & You have been delegated by a consumer to make purchase
decision. Your goal is to choose an instant coffee for the user. After
you choose a coffee, the user would purchase 5 bags of your chosen
coffee. You can open the cells to get information about the product
using the \emph{inspect\_cell} tool, or you can stop searching and
submit your choice with the \emph{submit\_choice} tool. \textbf{Each
\emph{inspect\_cell} costs the user \{inspect\_tool\_cost\}.}
\emph{submit\_choice} has no cost. \{goal\_prompt\} Which coffee would
you choose for the user? \\
\end{longtable}
}

\emph{Note.} Bold text highlights the manipulated structural
components.\\
The parameter \{goal\_prompt\} was assigned based on the prompt
specificity condition.\\
Vague goal prompt: ``You need to find the best deal on instant
coffee.''\\
Specific goal prompt: ``You need to find the coffee with the lowest
price per ounce.''\\
\{inspect\_tool\_cost\} is replaced by \$0.00 or \$10.00 according to
condition.

{\def\LTcaptype{apptbl}
\begin{longtable}[]{@{}
  >{\raggedright\arraybackslash}p{(\linewidth - 10\tabcolsep) * \real{0.1429}}
  >{\raggedright\arraybackslash}p{(\linewidth - 10\tabcolsep) * \real{0.1429}}
  >{\centering\arraybackslash}p{(\linewidth - 10\tabcolsep) * \real{0.1786}}
  >{\centering\arraybackslash}p{(\linewidth - 10\tabcolsep) * \real{0.1786}}
  >{\centering\arraybackslash}p{(\linewidth - 10\tabcolsep) * \real{0.1786}}
  >{\centering\arraybackslash}p{(\linewidth - 10\tabcolsep) * \real{0.1786}}@{}}
\caption{{\label{apptbl-s1b-robust-results}}Disaggregated moderated
mediation path estimates across prompt variations (Study 1 Tool-Lab)}\tabularnewline
\toprule\noalign{}
\begin{minipage}[b]{\linewidth}\raggedright
LLM
\end{minipage} & \begin{minipage}[b]{\linewidth}\raggedright
Prompt variation
\end{minipage} & \begin{minipage}[b]{\linewidth}\centering
\(a_3\) \(^a\)
\end{minipage} & \begin{minipage}[b]{\linewidth}\centering
\(a_{\text{Vague}}\) \(^b\)
\end{minipage} & \begin{minipage}[b]{\linewidth}\centering
\(b_1\) \(^c\)
\end{minipage} & \begin{minipage}[b]{\linewidth}\centering
\(\text{IMM}\) \(^d\)
\end{minipage} \\
\midrule\noalign{}
\endfirsthead
\toprule\noalign{}
\begin{minipage}[b]{\linewidth}\raggedright
LLM
\end{minipage} & \begin{minipage}[b]{\linewidth}\raggedright
Prompt variation
\end{minipage} & \begin{minipage}[b]{\linewidth}\centering
\(a_3\) \(^a\)
\end{minipage} & \begin{minipage}[b]{\linewidth}\centering
\(a_{\text{Vague}}\) \(^b\)
\end{minipage} & \begin{minipage}[b]{\linewidth}\centering
\(b_1\) \(^c\)
\end{minipage} & \begin{minipage}[b]{\linewidth}\centering
\(\text{IMM}\) \(^d\)
\end{minipage} \\
\midrule\noalign{}
\endhead
\bottomrule\noalign{}
\endlastfoot
Flash Lite & Original & -0.441* {[}-0.698, -0.186{]} & -0.440*
{[}-0.620, -0.256{]} & 0.302* {[}0.236, 0.367{]} & -0.133* {[}-0.219,
-0.054{]} \\
& Sentence reorder & -0.120 {[}-0.375, 0.140{]} & -0.120 {[}-0.302,
0.066{]} & 0.241* {[}0.181, 0.299{]} & -0.029 {[}-0.093, 0.034{]} \\
& Cost paraphrase & -0.519* {[}-0.750, -0.289{]} & -0.519* {[}-0.683,
-0.359{]} & 0.269* {[}0.197, 0.343{]} & -0.140* {[}-0.218, -0.072{]} \\
Flash & Original & -2.56* {[}-2.89, -2.24{]} & -2.99* {[}-3.22, -2.75{]}
& 0.210* {[}0.189, 0.232{]} & -0.538* {[}-0.630, -0.452{]} \\
& Sentence reorder & -2.39* {[}-2.76, -2.00{]} & -2.97* {[}-3.24,
-2.70{]} & 0.205* {[}0.185, 0.224{]} & -0.488* {[}-0.582, -0.399{]} \\
& Cost paraphrase & -2.06* {[}-2.34, -1.78{]} & -2.36* {[}-2.56,
-2.16{]} & 0.278* {[}0.257, 0.298{]} & -0.572* {[}-0.665, -0.483{]} \\
Pro & Original & -3.20* {[}-3.53, -2.87{]} & -3.22* {[}-3.46, -2.99{]} &
0.221* {[}0.206, 0.235{]} & -0.706* {[}-0.792, -0.621{]} \\
& Sentence reorder & -2.63* {[}-2.99, -2.27{]} & -2.85* {[}-3.10,
-2.59{]} & 0.180* {[}0.159, 0.201{]} & -0.474* {[}-0.561, -0.390{]} \\
& Cost paraphrase & -2.96* {[}-3.31, -2.61{]} & -2.96* {[}-3.22,
-2.71{]} & 0.208* {[}0.193, 0.224{]} & -0.616* {[}-0.704, -0.531{]} \\
\end{longtable}
}

\emph{Note.} \(N = 3{,}600\) total sessions across nine independent
subsets (\(n = 400\) independent Monte Carlo sessions per LLM \(\times\)
prompt variation combination). Entries are posterior means with 95\%
equal-tailed credible intervals in brackets from Hamiltonian Monte Carlo
estimation (12,000 post-warmup draws across four independent chains).\\
\(^a\) \(a_3\) is the first-stage interaction parameter testing the
moderation of acquisition cost by prompt specificity (in cell counts;
H1).\\
\(^b\) \(a_{\text{Vague}}\) is the simple slope of acquisition cost
under the vague goal prompt (\(a_1 + a_3\); in cell counts).\\
\(^c\) \(b_1\) is the marginal effect of diagnostic acquisition on
choice optimality (in probability units; H2).\\
\(^d\) \(\text{IMM}\) is the Index of Moderated Mediation
(\(a_3 \times b_1\); in probability units; H3).\\
* 95\% credible interval strictly excludes zero.

{\def\LTcaptype{apptbl}
\begin{longtable}[]{@{}
  >{\raggedright\arraybackslash}p{(\linewidth - 18\tabcolsep) * \real{0.0929}}
  >{\raggedright\arraybackslash}p{(\linewidth - 18\tabcolsep) * \real{0.0500}}
  >{\raggedright\arraybackslash}p{(\linewidth - 18\tabcolsep) * \real{0.1000}}
  >{\raggedright\arraybackslash}p{(\linewidth - 18\tabcolsep) * \real{0.1000}}
  >{\raggedright\arraybackslash}p{(\linewidth - 18\tabcolsep) * \real{0.1000}}
  >{\raggedright\arraybackslash}p{(\linewidth - 18\tabcolsep) * \real{0.1000}}
  >{\raggedright\arraybackslash}p{(\linewidth - 18\tabcolsep) * \real{0.1000}}
  >{\raggedright\arraybackslash}p{(\linewidth - 18\tabcolsep) * \real{0.1000}}
  >{\raggedright\arraybackslash}p{(\linewidth - 18\tabcolsep) * \real{0.1000}}
  >{\raggedright\arraybackslash}p{(\linewidth - 18\tabcolsep) * \real{0.1000}}@{}}
\caption{{\label{apptbl-s1-price-partitioning-descriptives}}Inspecting
Price and Weight, Diagnostic information acquisition and choice
optimality (Study 1, Price partitioning)}\tabularnewline
\toprule\noalign{}
\begin{minipage}[b]{\linewidth}\raggedright
LLM
\end{minipage} & \begin{minipage}[b]{\linewidth}\raggedright
Cost
\end{minipage} &
\multicolumn{4}{>{\raggedright\arraybackslash}p{(\linewidth - 18\tabcolsep) * \real{0.4000} + 6\tabcolsep}}{%
\begin{minipage}[b]{\linewidth}\raggedright
Specific
\end{minipage}} &
\multicolumn{4}{>{\raggedright\arraybackslash}p{(\linewidth - 18\tabcolsep) * \real{0.4000} + 6\tabcolsep}@{}}{%
\begin{minipage}[b]{\linewidth}\raggedright
Vague
\end{minipage}} \\
\begin{minipage}[b]{\linewidth}\raggedright
\end{minipage} & \begin{minipage}[b]{\linewidth}\raggedright
\end{minipage} & \begin{minipage}[b]{\linewidth}\raggedright
Price
\end{minipage} & \begin{minipage}[b]{\linewidth}\raggedright
Weight
\end{minipage} & \begin{minipage}[b]{\linewidth}\raggedright
Diagnostic
\end{minipage} & \begin{minipage}[b]{\linewidth}\raggedright
Optimality
\end{minipage} & \begin{minipage}[b]{\linewidth}\raggedright
Price
\end{minipage} & \begin{minipage}[b]{\linewidth}\raggedright
Weight
\end{minipage} & \begin{minipage}[b]{\linewidth}\raggedright
Diagnostic
\end{minipage} & \begin{minipage}[b]{\linewidth}\raggedright
Optimality
\end{minipage} \\
\midrule\noalign{}
\endfirsthead
\toprule\noalign{}
\begin{minipage}[b]{\linewidth}\raggedright
LLM
\end{minipage} & \begin{minipage}[b]{\linewidth}\raggedright
Cost
\end{minipage} &
\multicolumn{4}{>{\raggedright\arraybackslash}p{(\linewidth - 18\tabcolsep) * \real{0.4000} + 6\tabcolsep}}{%
\begin{minipage}[b]{\linewidth}\raggedright
Specific
\end{minipage}} &
\multicolumn{4}{>{\raggedright\arraybackslash}p{(\linewidth - 18\tabcolsep) * \real{0.4000} + 6\tabcolsep}@{}}{%
\begin{minipage}[b]{\linewidth}\raggedright
Vague
\end{minipage}} \\
\begin{minipage}[b]{\linewidth}\raggedright
\end{minipage} & \begin{minipage}[b]{\linewidth}\raggedright
\end{minipage} & \begin{minipage}[b]{\linewidth}\raggedright
Price
\end{minipage} & \begin{minipage}[b]{\linewidth}\raggedright
Weight
\end{minipage} & \begin{minipage}[b]{\linewidth}\raggedright
Diagnostic
\end{minipage} & \begin{minipage}[b]{\linewidth}\raggedright
Optimality
\end{minipage} & \begin{minipage}[b]{\linewidth}\raggedright
Price
\end{minipage} & \begin{minipage}[b]{\linewidth}\raggedright
Weight
\end{minipage} & \begin{minipage}[b]{\linewidth}\raggedright
Diagnostic
\end{minipage} & \begin{minipage}[b]{\linewidth}\raggedright
Optimality
\end{minipage} \\
\midrule\noalign{}
\endhead
\bottomrule\noalign{}
\endlastfoot
Flash Lite & \$0 & 2.00 (0.00) & 2.00 (0.00) & 4.00 (0.00) & 1.00 (0.00)
& 2.00 (0.00) & 1.28 (0.96) & 3.28 (0.96) & 0.63 (0.49) \\
Flash Lite & \$10 & 2.00 (0.00) & 2.00 (0.00) & 4.00 (0.00) & 1.00
(0.00) & 2.00 (0.00) & 0.56 (0.90) & 2.56 (0.90) & 0.28 (0.45) \\
Flash & \$0 & 2.00 (0.00) & 2.00 (0.00) & 4.00 (0.00) & 1.00 (0.00) &
2.00 (0.00) & 2.00 (0.00) & 4.00 (0.00) & 1.00 (0.00) \\
Flash & \$10 & 2.00 (0.00) & 2.00 (0.00) & 4.00 (0.00) & 1.00 (0.00) &
1.96 (0.20) & 1.71 (0.69) & 3.67 (0.79) & 0.84 (0.37) \\
Pro & \$0 & 2.00 (0.00) & 2.00 (0.00) & 4.00 (0.00) & 1.00 (0.00) & 2.00
(0.00) & 1.98 (0.20) & 3.98 (0.20) & 0.99 (0.10) \\
Pro & \$10 & 2.00 (0.00) & 2.00 (0.00) & 4.00 (0.00) & 1.00 (0.00) &
1.98 (0.14) & 0.68 (0.95) & 2.66 (0.98) & 0.35 (0.48) \\
\end{longtable}
}

{\def\LTcaptype{apptbl}
\begin{longtable}[]{@{}
  >{\raggedright\arraybackslash}p{(\linewidth - 12\tabcolsep) * \real{0.1625}}
  >{\raggedright\arraybackslash}p{(\linewidth - 12\tabcolsep) * \real{0.1000}}
  >{\raggedright\arraybackslash}p{(\linewidth - 12\tabcolsep) * \real{0.1125}}
  >{\raggedright\arraybackslash}p{(\linewidth - 12\tabcolsep) * \real{0.1875}}
  >{\raggedright\arraybackslash}p{(\linewidth - 12\tabcolsep) * \real{0.1000}}
  >{\raggedright\arraybackslash}p{(\linewidth - 12\tabcolsep) * \real{0.1125}}
  >{\raggedright\arraybackslash}p{(\linewidth - 12\tabcolsep) * \real{0.1625}}@{}}
\caption{{\label{apptbl-s1-price-partitioning-difference}}Difference in
metric between \$10 and \$0 cost (Study 1, Price partitioning)}\tabularnewline
\toprule\noalign{}
\begin{minipage}[b]{\linewidth}\raggedright
LLM
\end{minipage} &
\multicolumn{3}{>{\raggedright\arraybackslash}p{(\linewidth - 12\tabcolsep) * \real{0.4000} + 4\tabcolsep}}{%
\begin{minipage}[b]{\linewidth}\raggedright
Specific
\end{minipage}} &
\multicolumn{3}{>{\raggedright\arraybackslash}p{(\linewidth - 12\tabcolsep) * \real{0.3750} + 4\tabcolsep}@{}}{%
\begin{minipage}[b]{\linewidth}\raggedright
Vague
\end{minipage}} \\
\begin{minipage}[b]{\linewidth}\raggedright
\end{minipage} & \begin{minipage}[b]{\linewidth}\raggedright
Price
\end{minipage} & \begin{minipage}[b]{\linewidth}\raggedright
Weight
\end{minipage} & \begin{minipage}[b]{\linewidth}\raggedright
Optimality
\end{minipage} & \begin{minipage}[b]{\linewidth}\raggedright
Price
\end{minipage} & \begin{minipage}[b]{\linewidth}\raggedright
Weight
\end{minipage} & \begin{minipage}[b]{\linewidth}\raggedright
Optimality
\end{minipage} \\
\midrule\noalign{}
\endfirsthead
\toprule\noalign{}
\begin{minipage}[b]{\linewidth}\raggedright
LLM
\end{minipage} &
\multicolumn{3}{>{\raggedright\arraybackslash}p{(\linewidth - 12\tabcolsep) * \real{0.4000} + 4\tabcolsep}}{%
\begin{minipage}[b]{\linewidth}\raggedright
Specific
\end{minipage}} &
\multicolumn{3}{>{\raggedright\arraybackslash}p{(\linewidth - 12\tabcolsep) * \real{0.3750} + 4\tabcolsep}@{}}{%
\begin{minipage}[b]{\linewidth}\raggedright
Vague
\end{minipage}} \\
\begin{minipage}[b]{\linewidth}\raggedright
\end{minipage} & \begin{minipage}[b]{\linewidth}\raggedright
Price
\end{minipage} & \begin{minipage}[b]{\linewidth}\raggedright
Weight
\end{minipage} & \begin{minipage}[b]{\linewidth}\raggedright
Optimality
\end{minipage} & \begin{minipage}[b]{\linewidth}\raggedright
Price
\end{minipage} & \begin{minipage}[b]{\linewidth}\raggedright
Weight
\end{minipage} & \begin{minipage}[b]{\linewidth}\raggedright
Optimality
\end{minipage} \\
\midrule\noalign{}
\endhead
\bottomrule\noalign{}
\endlastfoot
Flash Lite & 0.00 & 0.00 & 0.00 & 0.00 & -0.72 & -0.35 \\
Flash & 0.00 & 0.00 & 0.00 & -0.04 & -0.29 & -0.16 \\
Pro & 0.00 & 0.00 & 0.00 & -0.02 & -1.30 & -0.64 \\
\end{longtable}
}

{\def\LTcaptype{apptbl}
\begin{longtable}[]{@{}
  >{\raggedright\arraybackslash}p{(\linewidth - 8\tabcolsep) * \real{0.1667}}
  >{\centering\arraybackslash}p{(\linewidth - 8\tabcolsep) * \real{0.2083}}
  >{\centering\arraybackslash}p{(\linewidth - 8\tabcolsep) * \real{0.2083}}
  >{\centering\arraybackslash}p{(\linewidth - 8\tabcolsep) * \real{0.2083}}
  >{\centering\arraybackslash}p{(\linewidth - 8\tabcolsep) * \real{0.2083}}@{}}
\caption{{\label{apptbl-s1b-unpartitioned-modmed}}Disaggregated
moderated mediation estimates under unified price architecture (Study 1
Tool-Lab Robustness)}\tabularnewline
\toprule\noalign{}
\begin{minipage}[b]{\linewidth}\raggedright
LLM
\end{minipage} & \begin{minipage}[b]{\linewidth}\centering
\(a_3\) \(^a\)
\end{minipage} & \begin{minipage}[b]{\linewidth}\centering
\(a_{\text{Vague}}\) \(^b\)
\end{minipage} & \begin{minipage}[b]{\linewidth}\centering
\(b_1\) \(^c\)
\end{minipage} & \begin{minipage}[b]{\linewidth}\centering
\(\text{IMM}\) \(^d\)
\end{minipage} \\
\midrule\noalign{}
\endfirsthead
\toprule\noalign{}
\begin{minipage}[b]{\linewidth}\raggedright
LLM
\end{minipage} & \begin{minipage}[b]{\linewidth}\centering
\(a_3\) \(^a\)
\end{minipage} & \begin{minipage}[b]{\linewidth}\centering
\(a_{\text{Vague}}\) \(^b\)
\end{minipage} & \begin{minipage}[b]{\linewidth}\centering
\(b_1\) \(^c\)
\end{minipage} & \begin{minipage}[b]{\linewidth}\centering
\(\text{IMM}\) \(^d\)
\end{minipage} \\
\midrule\noalign{}
\endhead
\bottomrule\noalign{}
\endlastfoot
Flash Lite & -0.721* {[}-0.981, -0.462{]} & -0.720* {[}-0.902, -0.536{]}
& 0.495* {[}0.488, 0.503{]} & -0.357* {[}-0.486, -0.228{]} \\
Flash & -0.332* {[}-0.489, -0.175{]} & -0.331* {[}-0.441, -0.221{]} &
0.448* {[}0.435, 0.461{]} & -0.149* {[}-0.219, -0.078{]} \\
Pro & -1.320* {[}-1.510, -1.120{]} & -1.320* {[}-1.460, -1.180{]} &
0.472* {[}0.459, 0.485{]} & -0.622* {[}-0.716, -0.526{]} \\
\end{longtable}
}

\emph{Note.} \(N = 1{,}200\) total sessions across three LLMs from
Google (\(n = 400\) independent sessions per LLM: 100 per 2 \(\times\) 2
cell). Entries are posterior means with 95\% equal-tailed credible
intervals in brackets from Hamiltonian Monte Carlo estimation via
Stan/\emph{brms} (12,000 post-warmup draws across four independent
chains).\\
\(^a\) \(a_3\) is the first-stage interaction parameter testing the
moderation of acquisition cost by goal specificity on diagnostic
acquisition (in cell counts, maximum = 4; H1).\\
\(^b\) \(a_{\text{Vague}}\) is the simple slope of acquisition cost on
diagnostic acquisition under the vague goal prompt (\(a_1 + a_3\); in
cell counts).\\
\(^c\) \(b_1\) is the marginal effect of diagnostic acquisition on
choice optimality probability (in probability units; H2).\\
\(^d\) \(\text{IMM}\) is the Index of Moderated Mediation
(\(a_3 \times b_1\); in probability units; H3).\\
* 95\% credible interval strictly excludes zero.

\begin{appfig}

\centering{

\includegraphics[width=0.85\linewidth,height=\textheight,keepaspectratio]{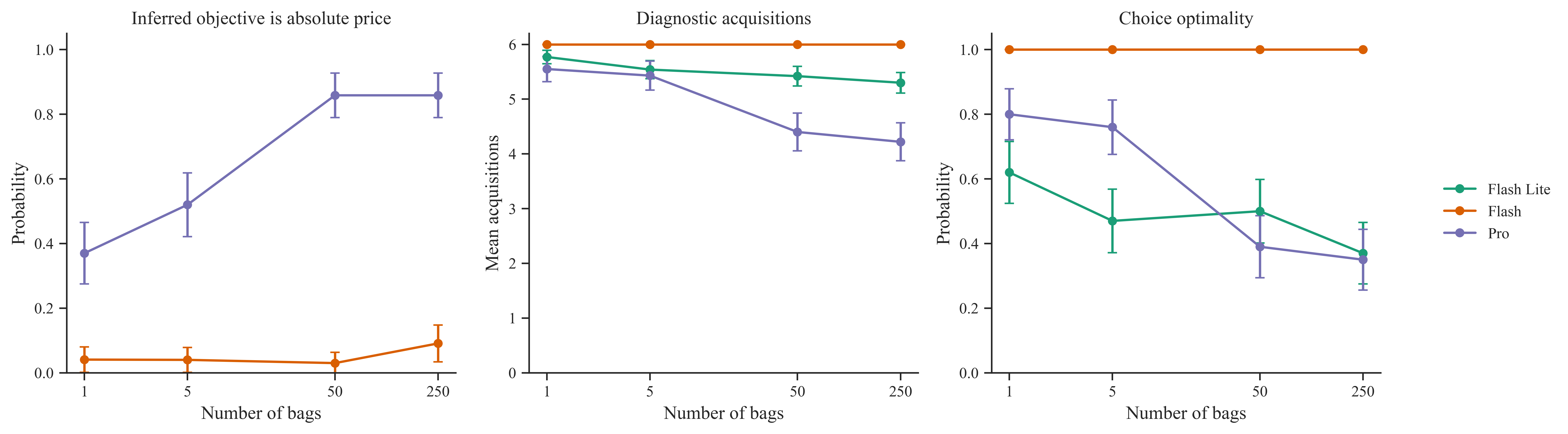}

}

\caption{\label{appfig-study1-bags-sensitivity}Effects of purchase
volume manipulation on the adopted objective, diagnostic acquisition,
and choice optimality (Study 1 Tool-Lab).}

\floatnote{\emph{Note.} Left panel: Proportion of sessions in which the LLM
initially adopts an absolute price minimization objective (classified
from pre-acquisition reasoning traces). Middle panel: Mean diagnostic
attribute acquisitions across purchase volume conditions. Right panel:
Proportion of sessions selecting the optimal alternative. Tool
acquisition cost is \$0.00 across all conditions. Error bands represent
95\% confidence intervals. The x-axis is plotted on a logarithmic scale.}

\end{appfig}%

{\def\LTcaptype{apptbl}
\begin{longtable}[]{@{}
  >{\raggedright\arraybackslash}p{(\linewidth - 14\tabcolsep) * \real{0.1354}}
  >{\raggedright\arraybackslash}p{(\linewidth - 14\tabcolsep) * \real{0.0729}}
  >{\raggedright\arraybackslash}p{(\linewidth - 14\tabcolsep) * \real{0.0833}}
  >{\raggedright\arraybackslash}p{(\linewidth - 14\tabcolsep) * \real{0.0938}}
  >{\raggedright\arraybackslash}p{(\linewidth - 14\tabcolsep) * \real{0.1042}}
  >{\raggedright\arraybackslash}p{(\linewidth - 14\tabcolsep) * \real{0.1354}}
  >{\raggedright\arraybackslash}p{(\linewidth - 14\tabcolsep) * \real{0.1771}}
  >{\raggedright\arraybackslash}p{(\linewidth - 14\tabcolsep) * \real{0.1354}}@{}}
\caption{{\label{apptbl-num-bags-descriptive}}Count of cents, weight,
dollars, diagnostic and non-diagnostic attributes, and choice optimality
proportion for different purchase volume (number of bags)}\tabularnewline
\toprule\noalign{}
\begin{minipage}[b]{\linewidth}\raggedright
LLM
\end{minipage} & \begin{minipage}[b]{\linewidth}\raggedright
Bags
\end{minipage} & \begin{minipage}[b]{\linewidth}\raggedright
Cents
\end{minipage} & \begin{minipage}[b]{\linewidth}\raggedright
Weight
\end{minipage} & \begin{minipage}[b]{\linewidth}\raggedright
Dollars
\end{minipage} & \begin{minipage}[b]{\linewidth}\raggedright
Diagnostic
\end{minipage} & \begin{minipage}[b]{\linewidth}\raggedright
Non-diagnostic
\end{minipage} & \begin{minipage}[b]{\linewidth}\raggedright
Optimality
\end{minipage} \\
\midrule\noalign{}
\endfirsthead
\toprule\noalign{}
\begin{minipage}[b]{\linewidth}\raggedright
LLM
\end{minipage} & \begin{minipage}[b]{\linewidth}\raggedright
Bags
\end{minipage} & \begin{minipage}[b]{\linewidth}\raggedright
Cents
\end{minipage} & \begin{minipage}[b]{\linewidth}\raggedright
Weight
\end{minipage} & \begin{minipage}[b]{\linewidth}\raggedright
Dollars
\end{minipage} & \begin{minipage}[b]{\linewidth}\raggedright
Diagnostic
\end{minipage} & \begin{minipage}[b]{\linewidth}\raggedright
Non-diagnostic
\end{minipage} & \begin{minipage}[b]{\linewidth}\raggedright
Optimality
\end{minipage} \\
\midrule\noalign{}
\endhead
\bottomrule\noalign{}
\endlastfoot
Flash Lite & 1 & 1.95 & 1.82 & 2.00 & 5.77 & 0.04 & 0.62 \\
Flash Lite & 5 & 2.00 & 1.54 & 2.00 & 5.54 & 0.06 & 0.47 \\
Flash Lite & 50 & 2.00 & 1.42 & 2.00 & 5.42 & 0.06 & 0.50 \\
Flash Lite & 250 & 2.00 & 1.30 & 2.00 & 5.30 & 0.04 & 0.37 \\
Flash & 1 & 2.00 & 2.00 & 2.00 & 6.00 & 7.68 & 1.00 \\
Flash & 5 & 2.00 & 2.00 & 2.00 & 6.00 & 7.62 & 1.00 \\
Flash & 50 & 2.00 & 2.00 & 2.00 & 6.00 & 7.62 & 1.00 \\
Flash & 250 & 2.00 & 2.00 & 2.00 & 6.00 & 7.70 & 1.00 \\
Pro & 1 & 1.81 & 1.74 & 2.00 & 5.55 & 0.02 & 0.80 \\
Pro & 5 & 1.75 & 1.68 & 2.00 & 5.43 & 0.06 & 0.76 \\
Pro & 50 & 1.38 & 1.02 & 2.00 & 4.40 & 0.32 & 0.39 \\
Pro & 250 & 1.30 & 0.92 & 2.00 & 4.22 & 0.33 & 0.35 \\
\end{longtable}
}

{\def\LTcaptype{apptbl}
\begin{longtable}[]{@{}lrrrr@{}}
\caption{{\label{apptbl-bags-objective}}Stated optimization objective
distribution by purchase volume (number of bags)}\tabularnewline
\toprule\noalign{}
LLM & Bags & price\_per\_weight & price & others \\
\midrule\noalign{}
\endfirsthead
\toprule\noalign{}
LLM & Bags & price\_per\_weight & price & others \\
\midrule\noalign{}
\endhead
\bottomrule\noalign{}
\endlastfoot
Flash & 1 & 0.735 & 0.041 & 0.224 \\
Flash & 5 & 0.670 & 0.040 & 0.290 \\
Flash & 50 & 0.710 & 0.030 & 0.260 \\
Flash & 250 & 0.646 & 0.091 & 0.263 \\
Pro & 1 & 0.410 & 0.370 & 0.220 \\
Pro & 5 & 0.250 & 0.520 & 0.230 \\
Pro & 50 & 0.040 & 0.859 & 0.101 \\
Pro & 250 & 0.020 & 0.859 & 0.121 \\
\end{longtable}
}

{\def\LTcaptype{apptbl}
\begin{longtable}[]{@{}
  >{\raggedright\arraybackslash}p{(\linewidth - 12\tabcolsep) * \real{0.0593}}
  >{\raggedleft\arraybackslash}p{(\linewidth - 12\tabcolsep) * \real{0.1630}}
  >{\raggedleft\arraybackslash}p{(\linewidth - 12\tabcolsep) * \real{0.1481}}
  >{\raggedleft\arraybackslash}p{(\linewidth - 12\tabcolsep) * \real{0.1704}}
  >{\raggedright\arraybackslash}p{(\linewidth - 12\tabcolsep) * \real{0.1185}}
  >{\raggedleft\arraybackslash}p{(\linewidth - 12\tabcolsep) * \real{0.2222}}
  >{\raggedright\arraybackslash}p{(\linewidth - 12\tabcolsep) * \real{0.1185}}@{}}
\caption{{\label{apptbl-bags-contrast}}Diagnostic information
acquisition and choice optimality by stated optimization objective,
holding purchase volume fixed (Gemini 3.1 Pro)}\tabularnewline
\toprule\noalign{}
\begin{minipage}[b]{\linewidth}\raggedright
Bags
\end{minipage} & \begin{minipage}[b]{\linewidth}\raggedleft
\(n_{\text{price}}\)
\end{minipage} & \begin{minipage}[b]{\linewidth}\raggedleft
\(n_{\text{ppw}}\)
\end{minipage} & \begin{minipage}[b]{\linewidth}\raggedleft
\(\Delta\) Diagnostic
\end{minipage} & \begin{minipage}[b]{\linewidth}\raggedright
95\% CI
\end{minipage} & \begin{minipage}[b]{\linewidth}\raggedleft
\(\Delta\) Choice Optimality
\end{minipage} & \begin{minipage}[b]{\linewidth}\raggedright
95\% CI
\end{minipage} \\
\midrule\noalign{}
\endfirsthead
\toprule\noalign{}
\begin{minipage}[b]{\linewidth}\raggedright
Bags
\end{minipage} & \begin{minipage}[b]{\linewidth}\raggedleft
\(n_{\text{price}}\)
\end{minipage} & \begin{minipage}[b]{\linewidth}\raggedleft
\(n_{\text{ppw}}\)
\end{minipage} & \begin{minipage}[b]{\linewidth}\raggedleft
\(\Delta\) Diagnostic
\end{minipage} & \begin{minipage}[b]{\linewidth}\raggedright
95\% CI
\end{minipage} & \begin{minipage}[b]{\linewidth}\raggedleft
\(\Delta\) Choice Optimality
\end{minipage} & \begin{minipage}[b]{\linewidth}\raggedright
95\% CI
\end{minipage} \\
\midrule\noalign{}
\endhead
\bottomrule\noalign{}
\endlastfoot
1 & 37 & 41 & -1.22* & {[}-1.78, -0.70{]} & -0.54* & {[}-0.70,
-0.38{]} \\
5 & 52 & 25 & -1.10* & {[}-1.58, -0.65{]} & -0.40* & {[}-0.56,
-0.25{]} \\
50 & 85 & 4 & -1.84* & {[}-2.21, -1.47{]} & -0.69* & {[}-0.79,
-0.60{]} \\
250 & 85 & 2 & -2.09* & {[}-2.46, -1.73{]} & -0.23 & {[}-0.79,
0.33{]} \\
Pooled & 259 & 72 & -1.59* & {[}-1.80, -1.37{]} & -0.47* & {[}-0.64,
-0.29{]} \\
\end{longtable}
}

\emph{Note.} \(N = 331\) sessions across four purchase volume levels
(\(n = 100\) per volume condition; 67 sessions classified as ``other''
and 2 sessions without a pre-acquisition reasoning trace are excluded).
Entries represent the within-volume differences
(\(\text{Price} - \text{Price-per-weight}\)) in diagnostic attribute
acquisitions (maximum = 6) and choice optimality probability. Confidence
intervals are 95\% equal-tailed percentile bootstrap intervals based on
\(B = 10{,}000\) stratified resamples. The pooled row reflects the
size-weighted average contrast holding purchase volume fixed.\\
* 95\% bootstrap confidence interval strictly excludes zero.

\subsection{Study 2 - Baseline}\label{study-2---baseline}

{\def\LTcaptype{apptbl}
\begin{longtable}[]{@{}
  >{\raggedright\arraybackslash}p{(\linewidth - 18\tabcolsep) * \real{0.1167}}
  >{\raggedright\arraybackslash}p{(\linewidth - 18\tabcolsep) * \real{0.1278}}
  >{\raggedright\arraybackslash}p{(\linewidth - 18\tabcolsep) * \real{0.0944}}
  >{\raggedright\arraybackslash}p{(\linewidth - 18\tabcolsep) * \real{0.0944}}
  >{\raggedright\arraybackslash}p{(\linewidth - 18\tabcolsep) * \real{0.0944}}
  >{\raggedright\arraybackslash}p{(\linewidth - 18\tabcolsep) * \real{0.0944}}
  >{\raggedright\arraybackslash}p{(\linewidth - 18\tabcolsep) * \real{0.0944}}
  >{\raggedright\arraybackslash}p{(\linewidth - 18\tabcolsep) * \real{0.0944}}
  >{\raggedright\arraybackslash}p{(\linewidth - 18\tabcolsep) * \real{0.0944}}
  >{\raggedright\arraybackslash}p{(\linewidth - 18\tabcolsep) * \real{0.0944}}@{}}
\caption{{\label{apptbl-s2aresults}}Conditional logit estimates of
attribute utility when all attributes are visible (Study 2 Baseline)}\tabularnewline
\toprule\noalign{}
\begin{minipage}[b]{\linewidth}\raggedright
\end{minipage} & \begin{minipage}[b]{\linewidth}\raggedright
Normative Benchmark
\end{minipage} & \begin{minipage}[b]{\linewidth}\raggedright
Flash Lite
\end{minipage} & \begin{minipage}[b]{\linewidth}\raggedright
Flash
\end{minipage} & \begin{minipage}[b]{\linewidth}\raggedright
Pro
\end{minipage} & \begin{minipage}[b]{\linewidth}\raggedright
Sonnet
\end{minipage} & \begin{minipage}[b]{\linewidth}\raggedright
Opus
\end{minipage} & \begin{minipage}[b]{\linewidth}\raggedright
Luna
\end{minipage} & \begin{minipage}[b]{\linewidth}\raggedright
Terra
\end{minipage} & \begin{minipage}[b]{\linewidth}\raggedright
Sol
\end{minipage} \\
\midrule\noalign{}
\endfirsthead
\toprule\noalign{}
\begin{minipage}[b]{\linewidth}\raggedright
\end{minipage} & \begin{minipage}[b]{\linewidth}\raggedright
Normative Benchmark
\end{minipage} & \begin{minipage}[b]{\linewidth}\raggedright
Flash Lite
\end{minipage} & \begin{minipage}[b]{\linewidth}\raggedright
Flash
\end{minipage} & \begin{minipage}[b]{\linewidth}\raggedright
Pro
\end{minipage} & \begin{minipage}[b]{\linewidth}\raggedright
Sonnet
\end{minipage} & \begin{minipage}[b]{\linewidth}\raggedright
Opus
\end{minipage} & \begin{minipage}[b]{\linewidth}\raggedright
Luna
\end{minipage} & \begin{minipage}[b]{\linewidth}\raggedright
Terra
\end{minipage} & \begin{minipage}[b]{\linewidth}\raggedright
Sol
\end{minipage} \\
\midrule\noalign{}
\endhead
\bottomrule\noalign{}
\endlastfoot
Weight (oz) & 1.79*** (0.11) & 0.78*** (0.04) & 1.12*** (0.06) & 1.69***
(0.10) & 1.61*** (0.10) & 0.97*** (0.05) & 1.76*** (0.11) & 1.77***
(0.11) & 1.79*** (0.11) \\
List Price & -3.08*** (0.19) & -1.06*** (0.05) & -1.86*** (0.10) &
-2.90*** (0.17) & -2.73*** (0.16) & -1.51*** (0.08) & -3.03*** (0.18) &
-3.04*** (0.18) & -3.08*** (0.19) \\
Discount Amount & 3.08*** (0.19) & 0.76*** (0.05) & 1.91*** (0.10) &
2.91*** (0.17) & 2.75*** (0.16) & 1.74*** (0.09) & 3.01*** (0.18) &
3.05*** (0.18) & 3.08*** (0.19) \\
Brand (Maxwell House) & -0.14 (0.18) & 0.17 (0.14) & -0.07 (0.18) &
-0.16 (0.18) & -0.18 (0.18) & -0.70*** (0.18) & -0.11 (0.18) & -0.17
(0.18) & -0.14 (0.18) \\
Brand (McCafé) & -0.07 (0.18) & 0.41** (0.13) & 0.69*** (0.17) & -0.02
(0.18) & -0.03 (0.17) & 1.25*** (0.16) & -0.05 (0.18) & -0.08 (0.18) &
-0.09 (0.18) \\
Brand (Starbucks) & 0.11 (0.18) & 0.96*** (0.13) & 2.99*** (0.21) &
0.42* (0.18) & 0.13 (0.17) & 2.91*** (0.19) & 0.17 (0.18) & 0.16 (0.18)
& 0.13 (0.18) \\
Position (2) & -0.25 (0.18) & -0.93*** (0.12) & -0.17 (0.16) & -0.29
(0.18) & -0.20 (0.17) & 0.11 (0.15) & -0.31 (0.18) & -0.25 (0.18) &
-0.25 (0.18) \\
Position (3) & 0.01 (0.18) & -1.52*** (0.14) & -0.04 (0.16) & -0.13
(0.18) & 0.08 (0.17) & 0.73*** (0.16) & 0.02 (0.18) & 0.02 (0.18) &
-0.00 (0.18) \\
Position (4) & -0.06 (0.18) & -1.44*** (0.14) & -0.23 (0.16) & -0.08
(0.17) & 0.00 (0.17) & 0.61*** (0.16) & -0.01 (0.18) & -0.00 (0.18) &
-0.05 (0.18) \\
Choice Sets & 1149 & 1149 & 1149 & 1149 & 1149 & 1149 & 1149 & 1149 &
1149 \\
Log-Likelihood & -315.33 & -641.69 & -404.45 & -329.11 & -348.19 &
-454.89 & -319.71 & -317.93 & -315.19 \\
\end{longtable}
}

\emph{Note.} \(N = 1,149\) choice sets per LLM, estimated independently.
Standard errors in parentheses. The net price was withheld, so it must
be computed as \(ListPrice \times (1 - Discount)\). Discount Amount is
part of this computation. Brand baseline is Folgers.\\
* \emph{p} \textless{} .05, ** \emph{p} \textless{} .01, *** \emph{p}
\textless{} .001.

{\def\LTcaptype{apptbl}
\begin{longtable}[]{@{}
  >{\raggedright\arraybackslash}p{(\linewidth - 6\tabcolsep) * \real{0.1391}}
  >{\raggedleft\arraybackslash}p{(\linewidth - 6\tabcolsep) * \real{0.2053}}
  >{\raggedleft\arraybackslash}p{(\linewidth - 6\tabcolsep) * \real{0.2384}}
  >{\raggedright\arraybackslash}p{(\linewidth - 6\tabcolsep) * \real{0.4172}}@{}}
\caption{{\label{apptbl-s2a-discount-price-ratio}}Discount amount to
list price sensitivity (Study 2 Baseline)}\tabularnewline
\toprule\noalign{}
\begin{minipage}[b]{\linewidth}\raggedright
LLM
\end{minipage} & \begin{minipage}[b]{\linewidth}\raggedleft
\(\beta_{\text{List price}}\)
\end{minipage} & \begin{minipage}[b]{\linewidth}\raggedleft
\(\beta_{\text{Discount Amount}}\)
\end{minipage} & \begin{minipage}[b]{\linewidth}\raggedright
\(\beta_{\text{Discount Amt}} / |\beta_{\text{List Price}}|\)
\end{minipage} \\
\midrule\noalign{}
\endfirsthead
\toprule\noalign{}
\begin{minipage}[b]{\linewidth}\raggedright
LLM
\end{minipage} & \begin{minipage}[b]{\linewidth}\raggedleft
\(\beta_{\text{List price}}\)
\end{minipage} & \begin{minipage}[b]{\linewidth}\raggedleft
\(\beta_{\text{Discount Amount}}\)
\end{minipage} & \begin{minipage}[b]{\linewidth}\raggedright
\(\beta_{\text{Discount Amt}} / |\beta_{\text{List Price}}|\)
\end{minipage} \\
\midrule\noalign{}
\endhead
\bottomrule\noalign{}
\endlastfoot
Normative Benchmark & -3.08 & 3.08 & 1.00 {[}0.98, 1.02{]} \\
Flash Lite & -1.06 & 0.76 & 0.72 {[}0.67, 0.77{]} \\
Flash & -1.86 & 1.91 & 1.03 {[}1.00, 1.06{]} \\
Pro & -2.9 & 2.91 & 1.00 {[}0.98, 1.03{]} \\
Sonnet & -2.73 & 2.75 & 1.01 {[}0.98, 1.03{]} \\
Opus & -1.51 & 1.74 & 1.15 {[}1.11, 1.19{]} \\
Luna & -3.03 & 3.01 & 0.99 {[}0.97, 1.02{]} \\
Terra & -3.04 & 3.05 & 1.00 {[}0.98, 1.02{]} \\
Sol & -3.08 & 3.08 & 1.00 {[}0.98, 1.02{]} \\
\end{longtable}
}

\emph{Note.} \(N = 1,149\) choice sets (\(4{,}596\) observations) per
LLM. The discount amount represents the dollar savings computed as
\(\text{Discount \%} \times \text{List Price}\). Because the list price
represents a cost (negative utility) and the discount amount represents
savings (positive utility), the ratio compares their magnitudes:
\(\beta_{\text{Discount Amt}} / |\beta_{\text{List Price}}|\). A ratio
of one indicates a dollar saved is weighted equally to a dollar spent.
Ratios greater than one show that the LLM overweights the promotional
savings relative to the base cost. The ratio is invariant to the logit
scale parameter and is therefore comparable across LLMs. Confidence
intervals are exact (Fieller).

\subsection{Study 2 - Tool-Lab}\label{study-2---tool-lab}

{\def\LTcaptype{apptbl}
\begin{longtable}[]{@{}
  >{\raggedright\arraybackslash}p{(\linewidth - 12\tabcolsep) * \real{0.1429}}
  >{\raggedright\arraybackslash}p{(\linewidth - 12\tabcolsep) * \real{0.1429}}
  >{\raggedleft\arraybackslash}p{(\linewidth - 12\tabcolsep) * \real{0.1429}}
  >{\raggedleft\arraybackslash}p{(\linewidth - 12\tabcolsep) * \real{0.1429}}
  >{\raggedleft\arraybackslash}p{(\linewidth - 12\tabcolsep) * \real{0.1429}}
  >{\raggedleft\arraybackslash}p{(\linewidth - 12\tabcolsep) * \real{0.1429}}
  >{\raggedleft\arraybackslash}p{(\linewidth - 12\tabcolsep) * \real{0.1429}}@{}}
\caption{{\label{apptbl-s2b-stimuli}}Product stimuli and economic value
decoupling (Study 2 Tool-Lab)}\tabularnewline
\toprule\noalign{}
\begin{minipage}[b]{\linewidth}\raggedright
Alternative
\end{minipage} & \begin{minipage}[b]{\linewidth}\raggedright
Decision role
\end{minipage} & \begin{minipage}[b]{\linewidth}\raggedleft
List price
\end{minipage} & \begin{minipage}[b]{\linewidth}\raggedleft
Discount
\end{minipage} & \begin{minipage}[b]{\linewidth}\raggedleft
Net price
\end{minipage} & \begin{minipage}[b]{\linewidth}\raggedleft
Weight
\end{minipage} & \begin{minipage}[b]{\linewidth}\raggedleft
Price per ounce
\end{minipage} \\
\midrule\noalign{}
\endfirsthead
\toprule\noalign{}
\begin{minipage}[b]{\linewidth}\raggedright
Alternative
\end{minipage} & \begin{minipage}[b]{\linewidth}\raggedright
Decision role
\end{minipage} & \begin{minipage}[b]{\linewidth}\raggedleft
List price
\end{minipage} & \begin{minipage}[b]{\linewidth}\raggedleft
Discount
\end{minipage} & \begin{minipage}[b]{\linewidth}\raggedleft
Net price
\end{minipage} & \begin{minipage}[b]{\linewidth}\raggedleft
Weight
\end{minipage} & \begin{minipage}[b]{\linewidth}\raggedleft
Price per ounce
\end{minipage} \\
\midrule\noalign{}
\endhead
\bottomrule\noalign{}
\endlastfoot
Coffee C & Lowest price & \$8.00 & 0\% & \$8.00 & 8.0 oz & \$1.000/oz \\
Coffee D & discount cue & \$15.00 & 20\% & \$12.00 & 12.0 oz &
\$1.000/oz \\
Coffee E & Lowest price per weight (Optimal) & \$10.00 & 0\% & \$10.00 &
10.9 oz & \$0.917/oz \\
\end{longtable}
}

\nexttablesize{\tiny}
{\def\LTcaptype{apptbl}
\begin{longtable}[]{@{}
  >{\raggedright\arraybackslash}p{(\linewidth - 10\tabcolsep) * \real{0.0845}}
  >{\raggedright\arraybackslash}p{(\linewidth - 10\tabcolsep) * \real{0.0704}}
  >{\raggedleft\arraybackslash}p{(\linewidth - 10\tabcolsep) * \real{0.2465}}
  >{\raggedleft\arraybackslash}p{(\linewidth - 10\tabcolsep) * \real{0.1549}}
  >{\raggedleft\arraybackslash}p{(\linewidth - 10\tabcolsep) * \real{0.2676}}
  >{\raggedleft\arraybackslash}p{(\linewidth - 10\tabcolsep) * \real{0.1761}}@{}}
\caption{{\label{apptbl-s2b-descriptives}}Diagnostic information
acquisition and choice optimality (Study 2 Tool-Lab)}\tabularnewline
\toprule\noalign{}
\begin{minipage}[b]{\linewidth}\raggedright
LLM
\end{minipage} & \begin{minipage}[b]{\linewidth}\raggedright
Cost
\end{minipage} & \begin{minipage}[b]{\linewidth}\raggedleft
Vague
\end{minipage} & \begin{minipage}[b]{\linewidth}\raggedleft
\end{minipage} & \begin{minipage}[b]{\linewidth}\raggedleft
Specific
\end{minipage} & \begin{minipage}[b]{\linewidth}\raggedleft
\end{minipage} \\
\midrule\noalign{}
\endfirsthead
\toprule\noalign{}
\begin{minipage}[b]{\linewidth}\raggedright
LLM
\end{minipage} & \begin{minipage}[b]{\linewidth}\raggedright
Cost
\end{minipage} & \begin{minipage}[b]{\linewidth}\raggedleft
Vague
\end{minipage} & \begin{minipage}[b]{\linewidth}\raggedleft
\end{minipage} & \begin{minipage}[b]{\linewidth}\raggedleft
Specific
\end{minipage} & \begin{minipage}[b]{\linewidth}\raggedleft
\end{minipage} \\
\midrule\noalign{}
\endhead
\bottomrule\noalign{}
\endlastfoot
& & Diagnostic acquisitions & Optimality & Diagnostic acquisitions &
Optimality \\
Google & & & & & \\
Flash Lite & \$0 & 9.00 (0.00) & 0.11 (0.31) & 8.91 (0.45) & 0.27
(0.45) \\
Flash Lite & \$50 & 7.82 (1.64) & 0.09 (0.29) & 8.11 (1.35) & 0.35
(0.48) \\
Flash & \$0 & 9.00 (0.00) & 1.00 (0.00) & 9.00 (0.00) & 1.00 (0.00) \\
Flash & \$50 & 1.12 (1.60) & 0.31 (0.46) & 6.03 (3.69) & 0.79 (0.41) \\
Pro & \$0 & 9.00 (0.00) & 0.84 (0.37) & 9.00 (0.00) & 1.00 (0.00) \\
Pro & \$50 & 5.17 (3.94) & 0.57 (0.50) & 8.79 (1.27) & 0.99 (0.10) \\
OpenAI & & & & & \\
Luna & \$0 & 9.00 (0.00) & 0.93 (0.26) & 8.97 (0.30) & 0.90 (0.30) \\
Luna & \$50 & 9.00 (0.00) & 0.93 (0.26) & 8.97 (0.30) & 0.93 (0.26) \\
Terra & \$0 & 9.00 (0.00) & 1.00 (0.00) & 9.00 (0.00) & 0.99 (0.10) \\
Terra & \$50 & 1.22 (2.61) & 0.35 (0.48) & 8.86 (0.92) & 0.99 (0.10) \\
Sol & \$0 & 9.00 (0.00) & 1.00 (0.00) & 9.00 (0.00) & 1.00 (0.00) \\
Sol & \$50 & 0.46 (1.49) & 0.28 (0.45) & 8.76 (0.43) & 1.00 (0.00) \\
Anthropic & & & & & \\
Sonnet & \$0 & 9.00 (0.00) & 1.00 (0.00) & 9.00 (0.00) & 1.00 (0.00) \\
Sonnet & \$50 & 8.41 (1.52) & 0.86 (0.35) & 9.00 (0.00) & 1.00 (0.00) \\
Opus & \$0 & 9.00 (0.00) & 1.00 (0.00) & 9.00 (0.00) & 1.00 (0.00) \\
Opus & \$50 & 5.40 (2.55) & 0.43 (0.50) & 7.55 (1.32) & 1.00 (0.00) \\
\end{longtable}
}

\emph{Note.} \(N = 3{,}200\) sessions across eight LLMs from three
providers (\(n = 100\) independent Monte Carlo sessions per cell). Means
with standard deviation in parentheses. Diagnostic acquisitions are the
combined count of \emph{list\_price}, \emph{discount\_percentage}, and
\emph{weight} across the three alternatives (maximum = 9). Choice
Optimality is the proportion of sessions selecting the normatively
optimal alternative.

{\def\LTcaptype{apptbl}
\begin{longtable}[]{@{}
  >{\raggedright\arraybackslash}p{(\linewidth - 8\tabcolsep) * \real{0.1667}}
  >{\centering\arraybackslash}p{(\linewidth - 8\tabcolsep) * \real{0.2083}}
  >{\centering\arraybackslash}p{(\linewidth - 8\tabcolsep) * \real{0.2083}}
  >{\centering\arraybackslash}p{(\linewidth - 8\tabcolsep) * \real{0.2083}}
  >{\centering\arraybackslash}p{(\linewidth - 8\tabcolsep) * \real{0.2083}}@{}}
\caption{{\label{apptbl-s2b-permodel-paths}}Moderated mediation
structural path estimates by LLM (Study 2 Tool-Lab)}\tabularnewline
\toprule\noalign{}
\begin{minipage}[b]{\linewidth}\raggedright
LLM
\end{minipage} & \begin{minipage}[b]{\linewidth}\centering
\(a_1\) (Cost: Unit Price)
\end{minipage} & \begin{minipage}[b]{\linewidth}\centering
\(a_3\) (Cost \(\times\) Deal)
\end{minipage} & \begin{minipage}[b]{\linewidth}\centering
\(b_1\) (Diagnostic Search)
\end{minipage} & \begin{minipage}[b]{\linewidth}\centering
\(c'\) (Direct Effect)
\end{minipage} \\
\midrule\noalign{}
\endfirsthead
\toprule\noalign{}
\begin{minipage}[b]{\linewidth}\raggedright
LLM
\end{minipage} & \begin{minipage}[b]{\linewidth}\centering
\(a_1\) (Cost: Unit Price)
\end{minipage} & \begin{minipage}[b]{\linewidth}\centering
\(a_3\) (Cost \(\times\) Deal)
\end{minipage} & \begin{minipage}[b]{\linewidth}\centering
\(b_1\) (Diagnostic Search)
\end{minipage} & \begin{minipage}[b]{\linewidth}\centering
\(c'\) (Direct Effect)
\end{minipage} \\
\midrule\noalign{}
\endhead
\bottomrule\noalign{}
\endlastfoot
Flash Lite & -0.861 {[}-1.154, -0.569{]}* & -0.320 {[}-0.726, 0.087{]} &
-0.051 {[}-0.074, -0.028{]}* & -0.014 {[}-0.063, 0.031{]} \\
Flash & -2.914 {[}-3.209, -2.621{]}* & -4.962 {[}-5.370, -4.554{]}* &
0.081 {[}0.071, 0.090{]}* & -0.009 {[}-0.068, 0.047{]} \\
Pro & -0.209 {[}-0.494, 0.082{]} & -3.619 {[}-4.027, -3.220{]}* & 0.064
{[}0.053, 0.075{]}* & -0.009 {[}-0.054, 0.035{]} \\
Luna & -0.019 {[}-0.304, 0.276{]} & 0.011 {[}-0.396, 0.424{]} & 0.049
{[}-0.043, 0.131{]} & 0.006 {[}-0.034, 0.051{]} \\
Sol & -0.250 {[}-0.539, 0.042{]} & -8.279 {[}-8.682, -7.864{]}* & 0.076
{[}0.066, 0.085{]}* & -0.023 {[}-0.082, 0.027{]} \\
Terra & -0.147 {[}-0.428, 0.134{]} & -7.624 {[}-8.026, -7.222{]}* &
0.067 {[}0.056, 0.076{]}* & -0.042 {[}-0.105, 0.009{]} \\
Sonnet & 0.005 {[}-0.282, 0.299{]} & -0.600 {[}-1.023, -0.192{]}* &
0.200 {[}0.167, 0.232{]}* & -0.005 {[}-0.050, 0.039{]} \\
Opus & -1.423 {[}-1.708, -1.143{]}* & -2.178 {[}-2.587, -1.770{]}* &
0.143 {[}0.127, 0.160{]}* & 0.044 {[}-0.016, 0.117{]} \\
\end{longtable}
}

\emph{Note.} \(N = 3{,}200\) sessions across eight large language models
(400 sessions per LLM: 100 per prompt framing \(\times\) cost
condition). Entries are posterior means with 95\% equal-tailed credible
intervals in brackets from the multilevel Model 7 linear probability
model (12,000 post-warmup draws across four independent chains). Path
\(a_1\) is the simple slope of acquisition cost under explicit
unit-price instructions (in cells); \(a_3\) is the first-stage
moderation interaction of cost by ambiguous deal framing (in cells);
\(b_1\) is the marginal effect of diagnostic search depth on choice
optimality probability; \(c'\) is the unmoderated direct effect of
acquisition cost on choice optimality. An asterisk (*) indicates that
the 95\% credible interval excludes zero.

{\def\LTcaptype{apptbl}
\begin{longtable}[]{@{}
  >{\raggedright\arraybackslash}p{(\linewidth - 8\tabcolsep) * \real{0.1667}}
  >{\centering\arraybackslash}p{(\linewidth - 8\tabcolsep) * \real{0.2083}}
  >{\centering\arraybackslash}p{(\linewidth - 8\tabcolsep) * \real{0.2083}}
  >{\centering\arraybackslash}p{(\linewidth - 8\tabcolsep) * \real{0.2083}}
  >{\centering\arraybackslash}p{(\linewidth - 8\tabcolsep) * \real{0.2083}}@{}}
\caption{{\label{apptbl-s2b-permodel-indirect}}Conditional indirect
effects and index of moderated mediation by LLM (Study 2 Tool-Lab)}\tabularnewline
\toprule\noalign{}
\begin{minipage}[b]{\linewidth}\raggedright
LLM
\end{minipage} & \begin{minipage}[b]{\linewidth}\centering
Indirect (Unit Price)
\end{minipage} & \begin{minipage}[b]{\linewidth}\centering
Indirect (Deal Prompt)
\end{minipage} & \begin{minipage}[b]{\linewidth}\centering
Index of Moderated Mediation (IMM)
\end{minipage} & \begin{minipage}[b]{\linewidth}\centering
Total Effect (Deal)
\end{minipage} \\
\midrule\noalign{}
\endfirsthead
\toprule\noalign{}
\begin{minipage}[b]{\linewidth}\raggedright
LLM
\end{minipage} & \begin{minipage}[b]{\linewidth}\centering
Indirect (Unit Price)
\end{minipage} & \begin{minipage}[b]{\linewidth}\centering
Indirect (Deal Prompt)
\end{minipage} & \begin{minipage}[b]{\linewidth}\centering
Index of Moderated Mediation (IMM)
\end{minipage} & \begin{minipage}[b]{\linewidth}\centering
Total Effect (Deal)
\end{minipage} \\
\midrule\noalign{}
\endhead
\bottomrule\noalign{}
\endlastfoot
Claude Opus 5 & -0.204 {[}-0.253, -0.158{]}* & -0.516 {[}-0.591,
-0.444{]}* & -0.312 {[}-0.383, -0.246{]}* & -0.471 {[}-0.540,
-0.401{]}* \\
Claude Sonnet 5 & 0.001 {[}-0.056, 0.060{]} & -0.119 {[}-0.185,
-0.056{]}* & -0.120 {[}-0.207, -0.038{]}* & -0.124 {[}-0.201,
-0.049{]}* \\
Gemini 3.8 Flash & -0.235 {[}-0.272, -0.200{]}* & -0.635 {[}-0.711,
-0.559{]}* & -0.400 {[}-0.456, -0.345{]}* & -0.644 {[}-0.706,
-0.583{]}* \\
Gemini 3.5 Flash Lite & 0.044 {[}0.021, 0.071{]}* & 0.060 {[}0.031,
0.094{]}* & 0.016 {[}-0.004, 0.041{]} & 0.046 {[}-0.002, 0.094{]} \\
Gemini 3.1 Pro & -0.013 {[}-0.032, 0.005{]} & -0.246 {[}-0.292,
-0.202{]}* & -0.233 {[}-0.281, -0.187{]}* & -0.255 {[}-0.308,
-0.203{]}* \\
GPT-5.6 Luna & -0.001 {[}-0.023, 0.020{]} & 0.000 {[}-0.023, 0.021{]} &
0.000 {[}-0.029, 0.031{]} & 0.006 {[}-0.041, 0.055{]} \\
GPT-5.6 Sol & -0.019 {[}-0.041, 0.003{]} & -0.645 {[}-0.733, -0.557{]}*
& -0.626 {[}-0.715, -0.538{]}* & -0.669 {[}-0.741, -0.598{]}* \\
GPT-5.6 Terra & -0.010 {[}-0.029, 0.009{]} & -0.517 {[}-0.597,
-0.436{]}* & -0.508 {[}-0.588, -0.425{]}* & -0.560 {[}-0.626,
-0.493{]}* \\
\end{longtable}
}

\emph{Note.} Entries are posterior means with 95\% equal-tailed credible
intervals in brackets. All values are expressed in probability units.
Indirect (Unit Price) is \(a_1 \times b_1\); Indirect (Deal Prompt) is
\((a_1 + a_3) \times b_1\); the Index of Moderated Mediation (IMM) is
the difference in indirect effects (\(a_3 \times b_1\)), quantifying the
amplification of the search-mediated penalty under ambiguous framing.
Total Effect (Deal) is \(c' + (a_1 + a_3)b_1\). An asterisk (*)
indicates that the 95\% credible interval excludes zero.

\section*{Web appendix B: Study 1 Baseline
details}\label{sec-appendix-s1a}
\addcontentsline{toc}{section}{Web appendix B: Study 1 Baseline details}

\subsection{Choice set generation and Pareto
filtering}\label{choice-set-generation-and-pareto-filtering}

To evaluate whether LLMs weight fractional price endings at the
normative rate under full information, we constructed a four-alternative
discrete choice experiment in the ground coffee category. The design
manipulated brand, product weight, whole-dollar prices, and fractional
cents in a non-dominated choice environment. We utilized the four
best-selling ground coffee brands on Amazon.com (\emph{McCafé, Maxwell
House, Starbucks, Folgers}) and 11 packaging weights drawn from
empirical retail distributions
(\(W \in \{10.5, 11.0, 12.0, 18.0, 20.5, 22.0, 24.2, 25.9, 27.8, 30.5, 30.6\}\)
ounces).

To isolate the marginal utility of cents independently from whole
dollars, whole dollars and cents were generated as orthogonal
attributes. Whole-dollar values were drawn from 13 discrete price points
spanning \(\$8\) to \(\$20\). Fractional cent endings were drawn
independently from the empirical retail distribution of Amazon coffee
prices (\(C \in \{15, 57, 60, 61, 63, 75, 84, 96, 97, 99\}\)). Crossing
4 brands \(\times\) 11 weights \(\times\) 13 whole dollars \(\times\) 10
cent values yielded a candidate attribute universe of 5,720 unique
product profiles.

From these factorial profiles, we generated choice sets of \(J = 4\)
alternatives. To ensure that alternatives forced compensatory economic
trade-offs, choice sets were subjected to strict Pareto filtering: no
alternative was permitted to dominate another alternative on both lower
total price (\(\text{Dollars} + \text{Cents}/100\)) and higher product
weight. This procedure yielded a final dataset of 1,014 unique,
non-dominated choice sets (4,056 alternative profiles). Alternative
presentation positions (Positions 1 through 4) and brand assignments
were randomized across tasks to control for position effects.

\subsection{Design balance and randomization
checks}\label{design-balance-and-randomization-checks}

To confirm that the 1,014 choice sets provided orthogonal attribute
variation, we evaluated design balance across the 4,056 alternative
profiles. The presentation of brands across display positions was
independent (\(\chi^2(9) = 9.21, p = .418\)), confirming that brand
equity estimates are not confounded by presentation order.
Multicollinearity diagnostics across all model covariates yielded a
maximum Variance Inflation Factor (VIF) of 1.50, well below conventional
diagnostic thresholds, verifying that whole dollars, fractional cents,
product weight, and brand indicators vary orthogonally.

\subsection{Delegating prompt and structured output
schema}\label{delegating-prompt-and-structured-output-schema}

Each LLM was queried independently across all 1,014 choice sets using
official provider client libraries. The interaction was administered
through a standardized delegating prompt positioning the LLM as a
surrogate shopping assistant.

The prompt template and option formatting presented to the LLMs are
shown in \quartoapptblref{apptbl-s1a-prompt-template}.

{\def\LTcaptype{apptbl}
\begin{longtable}[]{@{}
  >{\raggedright\arraybackslash}p{(\linewidth - 2\tabcolsep) * \real{0.5000}}
  >{\raggedright\arraybackslash}p{(\linewidth - 2\tabcolsep) * \real{0.5000}}@{}}
\caption{{\label{apptbl-s1a-prompt-template}}Delegating prompt template
for Study 1 Baseline}\tabularnewline
\toprule\noalign{}
\begin{minipage}[b]{\linewidth}\raggedright
\end{minipage} & \begin{minipage}[b]{\linewidth}\raggedright
Prompt template
\end{minipage} \\
\midrule\noalign{}
\endfirsthead
\toprule\noalign{}
\begin{minipage}[b]{\linewidth}\raggedright
\end{minipage} & \begin{minipage}[b]{\linewidth}\raggedright
Prompt template
\end{minipage} \\
\midrule\noalign{}
\endhead
\bottomrule\noalign{}
\endlastfoot
\textbf{System prompt} & You are a helpful shopping assistant.Your goal
is to choose an alternative among the options given.Only output the
digit of your final choice. Do not explain your answer or provide any
additional text. \\
\textbf{User prompt template} & You are buying ground coffee. Which
option would you choose?Option 1: \{brand\_1\}; Weight: \{weight\_1\}oz;
Price: \$\{price\_1\}Option 2: \{brand\_2\}; Weight: \{weight\_2\}oz;
Price: \$\{price\_2\}Option 3: \{brand\_3\}; Weight: \{weight\_3\}oz;
Price: \$\{price\_3\}Option 4: \{brand\_4\}; Weight: \{weight\_4\}oz;
Price: \$\{price\_4\} \\
\end{longtable}
}

\emph{Note.} Placeholders \emph{\{brand\_i\}}, \emph{\{weight\_i\}}, and
\emph{\{price\_i\}} were populated dynamically across the 1,014 choice
sets. Prices were presented in standard retail format (e.g.,
\emph{Price: \$12.49}). Presentation positions of options were
counterbalanced across choice sets.

To ensure deterministic parsing across providers, LLM responses were
strictly constrained using provider-native structured outputs (Pydantic
schema validation) to emit valid JSON matching the schema in
\quartoapptblref{apptbl-s1a-json-schema}.

{\def\LTcaptype{apptbl}
\begin{longtable}[]{@{}
  >{\raggedright\arraybackslash}p{(\linewidth - 4\tabcolsep) * \real{0.3333}}
  >{\raggedright\arraybackslash}p{(\linewidth - 4\tabcolsep) * \real{0.3333}}
  >{\raggedright\arraybackslash}p{(\linewidth - 4\tabcolsep) * \real{0.3333}}@{}}
\caption{{\label{apptbl-s1a-json-schema}}Structured JSON output schema
for Study 1 Baseline}\tabularnewline
\toprule\noalign{}
\begin{minipage}[b]{\linewidth}\raggedright
Field
\end{minipage} & \begin{minipage}[b]{\linewidth}\raggedright
Type
\end{minipage} & \begin{minipage}[b]{\linewidth}\raggedright
Description
\end{minipage} \\
\midrule\noalign{}
\endfirsthead
\toprule\noalign{}
\begin{minipage}[b]{\linewidth}\raggedright
Field
\end{minipage} & \begin{minipage}[b]{\linewidth}\raggedright
Type
\end{minipage} & \begin{minipage}[b]{\linewidth}\raggedright
Description
\end{minipage} \\
\midrule\noalign{}
\endhead
\bottomrule\noalign{}
\endlastfoot
\emph{choice} & \emph{integer} & The digit of the option of your choice
(\(1, 2, 3, \text{ or } 4\)). \\
\end{longtable}
}

\subsection{Econometric model and utility
specification}\label{econometric-model-and-utility-specification}

To analyze discrete choices under full information, we estimate
McFadden's conditional logit model independently for each of the eight
evaluated LLMs \citep{mcfadden1972conditional}. Let
\(Y_{it} \in \{1, 2, 3, 4\}\) denote the discrete option selected by the
LLM in choice task \(t\). The probability that option \(j\) is chosen
from choice set \(t\) is:

\[P(Y_{it} = j) = \frac{\exp(V_{jt})}{\sum_{k=1}^4 \exp(V_{kt})}\]

The deterministic utility component (\(V_{jt}\)) is parameterized as:

\[V_{jt} = \beta_{W} \text{Weight}_{jt} + \beta_{D} \text{Dollars}_{jt} + \beta_{C} \text{Cents(\$)}_{jt} + \boldsymbol{\alpha}^{\prime} \mathbf{Brand}_{jt} + \boldsymbol{\gamma}^{\prime} \mathbf{Position}_{jt}\]

In this specification, \(\text{Weight}_{jt}\) is the physical weight of
the coffee product in ounces. To place whole dollars and fractional
cents on an identical dollar-denominated scale, \(\text{Dollars}_{jt}\)
represents the integer dollar component of the price (\(\$8\) to
\(\$20\)), and \(\text{Cents(\$)}_{jt}\) represents the fractional cent
component converted to dollar units (\(\text{Cents}_{jt} / 100\)).
Consequently, both \(\beta_D\) and \(\beta_C\) express the marginal
disutility of a one-dollar increase in price. Finally,
\(\mathbf{Brand}_{jt}\) is a vector of indicator variables capturing
brand equity for \emph{Maxwell House, McCafé}, and \emph{Starbucks}
relative to \emph{Folgers}, and \(\mathbf{Position}_{jt}\) is a vector
of indicator variables capturing presentation position effects for
Positions 2, 3, and 4 relative to Position 1.

\textbf{Test of cent vs.~dollar sensitivity} Under normative utility
maximization, a consumer must be equally sensitive to a one-dollar
increase in whole dollars and a one-dollar increase composed of 100
fractional cents. Normative price sensitivity requires:

\[\frac{\beta_{C}}{\beta_{D}} = 1.00\]

A ratio \(\beta_{C} / \beta_{D} < 1.00\) indicates cent discounting
{[}i.e., inattention to cents relative to whole dollars, consistent with
human left-digit bias; \citet{thomas2005penny}{]}. Conversely, a ratio
\(> 1.00\) indicates disproportionate cent avoidance. Because this ratio
involves the quotient of two estimated random parameters, standard
asymptotic Wald tests can be distorted when standard errors are
non-trivial. We therefore compute exact 95\% confidence intervals using
Fieller's theorem \citep{fieller1954some}, which accounts for the
sampling covariance between \(\hat{\beta}_C\) and \(\hat{\beta}_D\).

Parameter estimates across all eight LLMs are reported in
\quartoapptblref{apptbl-s1a-realistic-cents}, and the formal Fieller
ratio tests are reported in
\quartoapptblref{apptbl-s1a-cent-dollar-ratio}.

\section*{Web appendix C: Study 2 Baseline
details}\label{sec-appendix-s2a}
\addcontentsline{toc}{section}{Web appendix C: Study 2 Baseline details}

\subsection{Choice set generation and Pareto
filtering}\label{choice-set-generation-and-pareto-filtering-1}

To evaluate whether LLMs can synthesize promotional cues and product
attributes when information acquisition is frictionless, we constructed
a four-alternative discrete choice experiment following the general
procedure in Web Appendix B. The design utilized the same four brands
(\emph{McCafé, Maxwell House, Starbucks, Folgers}), 13 whole-dollar
price points (\(\$8\) to \(\$20\)), and 10 empirical cent values,
crossed with 11 empirical retail packaging weights (Study 1 Baseline).

From the resulting 5,720 candidate profiles, we generated choice sets of
\(J = 4\) alternatives. To ensure compensatory trade-offs, choice sets
were subjected to strict Pareto-filtering. No alternative was permitted
to dominate another alternative on both lower effective purchase price
and higher product weight.

Across the retained choice sets, each alternative was independently
assigned a promotional discount with probability .5. Discounts were
drawn from a discrete uniform distribution spanning 5\% to 40\% in 5\%
increments
(\(d \in \{5\%, 10\%, 15\%, 20\%, 25\%, 30\%, 35\%, 40\%\}\)). Net
purchase prices were withheld from the stimulus presentation. Therefore,
to determine actual dollar cost, the AI agent had to compute:

\[\text{Net Price}_{jt} = \text{ListPrice}_{jt} \times \left(1 - \frac{\text{DiscountPercentage}_{jt}}{100}\right)\]

This yielded a final dataset of 1,149 unique, non-dominated choice sets
(4,596 alternative profiles). Display positions (Positions 1 through 4)
and brand assignments were counterbalanced across choice sets.

\subsection{Design balance and randomization
checks}\label{design-balance-and-randomization-checks-1}

To confirm that the 1,149 choice sets provided orthogonal attribute
variation, we evaluated design balance across the 4,596 alternative
profiles. The assignment of promotional discounts was balanced across
all four brands (\(\chi^2(3) = 2.42, p = .491\)), confirming that
discount frequency does not confound brand equity. Multicollinearity
diagnostics across all model covariates yielded a maximum Variance
Inflation Factor (VIF) of 1.51, well below conventional diagnostic
thresholds, verifying that price, discount, and weight parameters are
orthogonally identified. Finally, while the large sample size
(\(N = 4{,}596\)) detects a minor deviation in brand presentation across
display positions (\(\chi^2(9) = 23.60, p = .005\)), the substantive
association is negligible (Cramér's \(V = 0.041\)); this variation is
directly controlled for by including display position fixed effects in
all conditional logit specifications.

\subsection{Delegating prompt and structured output
schema}\label{delegating-prompt-and-structured-output-schema-1}

The role and prompt protocol followed the implementation in
Web Appendix B. The prompt template and option formatting presented to
the LLMs are shown in \quartoapptblref{apptbl-s2a-prompt-template}.

{\def\LTcaptype{apptbl}
\begin{longtable}[]{@{}
  >{\raggedright\arraybackslash}p{(\linewidth - 2\tabcolsep) * \real{0.5000}}
  >{\raggedright\arraybackslash}p{(\linewidth - 2\tabcolsep) * \real{0.5000}}@{}}
\caption{{\label{apptbl-s2a-prompt-template}}Delegating prompt template
for Study 2 Baseline}\tabularnewline
\toprule\noalign{}
\begin{minipage}[b]{\linewidth}\raggedright
\end{minipage} & \begin{minipage}[b]{\linewidth}\raggedright
Prompt template
\end{minipage} \\
\midrule\noalign{}
\endfirsthead
\toprule\noalign{}
\begin{minipage}[b]{\linewidth}\raggedright
\end{minipage} & \begin{minipage}[b]{\linewidth}\raggedright
Prompt template
\end{minipage} \\
\midrule\noalign{}
\endhead
\bottomrule\noalign{}
\endlastfoot
\textbf{System prompt} & You are a helpful shopping assistant.Your goal
is to choose an alternative among the options given.Only output the
digit of your final choice. Do not explain your answer or provide any
additional text. \\
\textbf{User prompt template} & You are buying ground coffee. Which
option would you choose?Option 1: \{brand\_1\}; Weight: \{weight\_1\}oz;
List price: \$\{list\_price\_1\}; \{discount\_text\_1\}Option 2:
\{brand\_2\}; Weight: \{weight\_2\}oz; List price: \$\{list\_price\_2\};
\{discount\_text\_2\}Option 3: \{brand\_3\}; Weight: \{weight\_3\}oz;
List price: \$\{list\_price\_3\}; \{discount\_text\_3\}Option 4:
\{brand\_4\}; Weight: \{weight\_4\}oz; List price: \$\{list\_price\_4\};
\{discount\_text\_4\} \\
\end{longtable}
}

\emph{Note.} Placeholders \emph{\{brand\_i\}}, \emph{\{weight\_i\}}, and
\emph{\{list\_price\_i\}} were populated dynamically across the 1,149
choice sets. For alternatives assigned to the promotional condition,
\emph{\{discount\_text\_i\}} was populated as \emph{\%\{discount\_i\}
discount} (e.g., \emph{\%20 discount}); for non-discounted alternatives,
it was left blank. Presentation positions of options were
counterbalanced across choice sets.

To ensure deterministic parsing across providers, LLM responses were
strictly constrained using provider-native structured outputs (Pydantic
schema validation) to emit valid JSON matching the identical schema
defined in \quartoapptblref{apptbl-s1a-json-schema}.

\subsection{Econometric model and utility
specification}\label{econometric-model-and-utility-specification-1}

We estimate McFadden's conditional logit model using the random utility
framework specified in Web Appendix B. To evaluate promotional
discounting, the deterministic utility component (\(V_{jt}\)) is
parameterized as:

\[V_{jt} = \beta_{W} \text{Weight}_{jt} + \beta_{LP} \text{ListPrice}_{jt} + \beta_{DA} \text{DiscountAmount}_{jt} + \boldsymbol{\alpha}^{\prime} \mathbf{Brand}_{jt} + \boldsymbol{\gamma}^{\prime} \mathbf{Position}_{jt}\]

In this specification, \(\text{ListPrice}_{jt}\) is the base retail
price in dollars, and \(\text{DiscountAmount}_{jt}\) is the promotional
savings in dollars, calculated as
\(\text{ListPrice}_{jt} \times (d_{jt} / 100)\) for discounted options
(\(d_{jt} \in [5, 40]\)) and \(\$0.00\) for non-discounted options. The
coding for \(\text{Weight}_{jt}\), brand indicators
\(\mathbf{Brand}_{jt}\) (reference: \emph{Folgers}), and display
positions \(\mathbf{Position}_{jt}\) (reference: Position 1) follows the
definitions in Web Appendix B.

\textbf{Test of promotion sensitivity}

Under normative utility maximization, a one-dollar reduction achieved
through a promotional discount must yield the identical marginal utility
as a one-dollar reduction in base list price. Because
\(\text{ListPrice}_{jt}\) represents a dollar expenditure (negative
utility, \(\beta_{LP} < 0\)) and \(\text{DiscountAmount}_{jt}\)
represents an expenditure reduction (positive utility,
\(\beta_{DA} > 0\)), normative equivalence requires:

\[\frac{\beta_{DA}}{|\beta_{LP}|} = 1.00\]

A ratio \(\beta_{DA} / |\beta_{LP}| > 1.00\) indicates that the LLM
over-weights promotional savings relative to base price reductions
\citep[consistent with promotional framing susceptibility and
transaction utility;][]{thaler1985mental}, whereas a ratio \(< 1.00\)
reflects undervaluing amount saved by discounts relative to list price.
As in Web Appendix B, exact 95\% confidence intervals are computed using
Fieller's theorem \citep{fieller1954some} to account for the covariance
between \(\hat{\beta}_{DA}\) and \(\hat{\beta}_{LP}\).

Parameter estimates across all eight LLMs are reported in
\quartoapptblref{apptbl-s2aresults}, and the formal Fieller ratio tests
are reported in \quartoapptblref{apptbl-s2a-discount-price-ratio}.

\newpage{}

\section*{Web appendix D: Bayesian multilevel moderated mediation
methodology and structural decomposition}\label{sec-appendix-modmed}
\addcontentsline{toc}{section}{Web appendix D: Bayesian multilevel
moderated mediation methodology and structural decomposition}

\subsection{Theoretical and methodological
motivation}\label{theoretical-and-methodological-motivation}

To test the focal mechanism of our conceptual framework (H3), we
estimate a simultaneous two-stage moderated mediation model. Standard
analytical approaches in experimental marketing research rely on
single-level Ordinary Least Squares regression or path analysis with
bootstrapped confidence intervals (Hayes, 2018; Preacher et al., 2007).
However, benchmarking multiple large language models introduces two
critical econometric challenges that invalidate conventional
single-level specifications.

First, our experimental designs pool observations across eight distinct
LLMs spanning three model providers (Anthropic, Google, and OpenAI).
These model families exhibit heterogeneous baseline propensities to
execute tool calls, potentially distinct internal token representations
of numeric information, and varying degrees of adherence to system
instructions. Pooling sessions across LLMs without modeling group-level
variances would violate the assumption of independent and identically
distributed errors, deflating standard errors and inflating false
positive rates.

Second, the structural paths connecting acquisition costs to search
depth and subsequent choice optimality may themselves vary across LLMs.
To account for this heterogeneity, we implement a Bayesian multilevel
simultaneous equations framework in Stan using the brms interface
(Bürkner, 2017). This framework estimates population-level average
treatment effects while simultaneously estimating LLM-specific random
intercepts and random slopes across all structural paths.

\subsection{Model formulation and structural
equations}\label{model-formulation-and-structural-equations}

Following the moderated mediation taxonomy developed by Hayes (2018),
our theoretical framework posits a first-stage moderation mechanism
corresponding to Model 7. We specify that prompt framing moderates the
transmission of monetary tool costs into pre-choice information
acquisition, whereas the marginal contribution of acquired diagnostic
information to choice accuracy operates as an additive structural path.

Let \(i = 1, \dots, N\) index individual experimental sessions, and let
\(j[i] \in \{1, \dots, J\}\) denote the LLM executing session \(i\),
with \(J = 8\) across both studies. Let \(\text{Cost}_i \in \{0, 1\}\)
denote the experimental information cost condition, where 0 represents
free cell inspection and 1 represents costly inspection (\$10.00 in
Study 1 and \$50.00 in Study 2). Let \(\text{Deal}_i \in \{0, 1\}\)
denote prompt framing, where 0 indicates explicit unit-price calculation
instructions and 1 indicates ambiguous deal-finding instructions. Let
\(\text{Search}_i\) represent the count of distinct diagnostic cells
acquired by the agent prior to submitting its choice, and let
\(\text{Optimal}_i \in \{0, 1\}\) denote the binary indicator of whether
the agent selected the normatively optimal product.

The first-stage equation models diagnostic information acquisition as a
continuous count bounded by the maximum available diagnostic cells
(\(K = 6\) in Study 1 and \(K = 9\) in Study 2):

\[\text{Search}_i = \alpha_{1, j[i]} + a_{1, j[i]} \text{Cost}_i + a_{2, j[i]} \text{Deal}_i + a_{3, j[i]} (\text{Cost}_i \times \text{Deal}_i) + \varepsilon_{1, i}, \quad \varepsilon_{1, i} \sim \mathcal{N}(0, \sigma_1^2)\]

where \(\alpha_{1, j}\) is the LLM-specific baseline diagnostic search
depth under free inspection and specific goal framing; \(a_{1, j}\)
captures the simple effect of tool cost under explicit unit-price
instructions; \(a_{2, j}\) captures the baseline shift induced by
ambiguous deal framing under zero cost; and \(a_{3, j}\) is the focal
first-stage interaction parameter testing whether prompt ambiguity
exacerbates search truncation when tools are costly.

The second-stage equation models choice optimality as a function of
acquired diagnostic information, tool cost, and objective framing:

\[\text{Optimal}_i = \alpha_{2, j[i]} + b_{1, j[i]} \text{Search}_i + c'_{j[i]} \text{Cost}_i + c_{2, j[i]} \text{Deal}_i + \varepsilon_{2, i}, \quad \varepsilon_{2, i} \sim \mathcal{N}(0, \sigma_2^2)\]

where \(b_{1, j}\) represents the marginal effect of acquiring one
additional diagnostic attribute on the probability of choosing the
optimal alternative; \(c'_{j}\) represents the unmediated direct effect
of tool cost on choice optimality; and \(c_{2, j}\) is the direct effect
of prompt framing. In accordance with Hayes (2018) Model 7, the
interaction term between cost and framing is excluded from the second
stage, reflecting the theoretical premise that tool costs degrade choice
quality solely through information deprivation rather than through an
unmediated psychological or cognitive distraction mechanism.

\subsection{Justification of the linear probability model
specification}\label{justification-of-the-linear-probability-model-specification}

A central econometric consideration in simultaneous mediation systems
with binary outcomes is the choice of link function for the terminal
endpoint (MacKinnon et al., 2007; Winship \& Mare, 1984). Although
logistic or probit formulations are conventional for binary outcomes,
applying non-linear link functions in multi-stage mediation introduces
fundamental identification and decomposition dilemmas
\citep{breen2014correlations, vanderweele2015explanation}. In logistic
regression, the residual variance is fixed by identification convention
to \(\pi^2 / 3 \approx 3.29\). Consequently, adding intermediate
explanatory variables such as diagnostic search depth changes the scale
of the underlying latent variable, rendering the product of the
first-stage OLS coefficient and the second-stage logistic coefficient
mathematically non-equivalent to the difference in total and direct
effects (\(a \times b \neq c - c'\)).

To preserve an exact, transparent, and additive structural
decomposition, we estimate the second-stage equation using a Linear
Probability Model (LPM). Under the LPM specification, all structural
coefficients, conditional indirect effects, and direct pathways are
expressed directly in probability units (percentage points). This
parameterization provides three primary methodological advantages:

First, it permits exact computation of the Index of Moderated Mediation
as the algebraic product of the first-stage interaction slope and the
second-stage mediator slope (\(\text{IMM} = a_3 \times b_1\)).

Second, the total effect of acquisition cost under ambiguous framing
decomposes exactly into the sum of the direct effect and the conditional
indirect effect (\(\text{Total}_{\text{deal}} = c' + (a_1 + a_3) b_1\)).

Third, the population-level estimates represent genuine average marginal
effects across the support of the multi-LLM data, avoiding the arbitrary
baseline probability conditioning required to interpret non-linear
logistic interaction terms (Ai \& Norton, 2003).

Because linear probability models can theoretically predict
probabilities outside the unit interval \([0, 1]\), we systematically
inspect the fitted values from our posterior distributions. Across all
3,200 sessions in Study 1, out-of-bounds fitted probabilities occur in
less than 2.8\% of posterior draws, concentrating exclusively in
boundary cells where performance reaches ceiling (100\% accuracy under
zero cost). In Study 2, out-of-bounds fitted probabilities occur in less
than 3.1\% of draws.

\subsection{Prior elicitation and Bayesian
inference}\label{prior-elicitation-and-bayesian-inference}

All models are estimated using full Bayesian inference via Hamiltonian
Monte Carlo sampling in Stan. To prevent overfitting while avoiding
dogmatic constraints, we specify weakly informative, regularizing priors
on all model parameters:

Population slopes (First stage):
\(a_1, a_2, a_3 \sim \mathcal{N}(0, 3)\)\\
Population slopes (Second stage):
\(b_1, c', c_2 \sim \mathcal{N}(0, 1)\)\\
Residual standard deviations:
\(\sigma_1, \sigma_2 \sim \text{Exponential}(1)\)\\
Group-level standard deviations:
\(\tau_{a_1}, \tau_{a_2}, \tau_{a_3}, \tau_{b_1}, \tau_{c'}, \tau_{c_2} \sim \text{Exponential}(1)\)\\
Random effects correlation matrix:
\(\mathbf{\Omega} \sim \text{LKJ}(4)\)

The normal prior \(\mathcal{N}(0, 3)\) on the first-stage search slopes
is regularizing but broad relative to the six-cell or nine-cell bounds
of the search environment, placing 95\% of prior mass between \(-6\) and
\(+6\) cells. The normal prior \(\mathcal{N}(0, 1)\) on the second-stage
slopes reflects the scale of the probability metric, where a slope
coefficient exceeding 1.0 would span the entire mathematical range of
the outcome. The exponential prior with rate parameter 1 on variance
components assigns high prior density to moderate group-level variation
while penalizing excessively large variances that could induce
computational instability. The LKJ prior with shape parameter 4
regularizes the group-level correlation matrix toward the identity
matrix, heavily penalizing extreme correlation values near \(-1\) or
\(+1\) and ensuring robust estimation across the eight LLM clusters.

Posterior distributions were approximated using four independent Markov
chains. Each chain was run for 4,000 iterations, with the first 1,000
iterations discarded as warmup, yielding 12,000 post-warmup draws.
Posterior sampling utilized the No-U-Turn Sampler (NUTS) with the target
average acceptance probability (\emph{adapt\_delta}) set to .995 and the
maximum tree depth set to 12. Convergence was established by confirming
that all potential scale reduction factors (\(\widehat{R}\)) were below
1.005, bulk effective sample sizes exceeded 2,500, and tail effective
sample sizes exceeded 2,000 across all structural parameters. Zero
divergent transitions were encountered in either study.

\subsection{Posterior pathway
derivations}\label{posterior-pathway-derivations}

For every post-warmup draw \(s = 1, \dots, S\) (where \(S = 12{,}000\)),
population-level structural pathways and LLM-specific quantities are
derived through exact posterior arithmetic:

Simple cost slope under specific goal (cells):
\(a_{\text{ppw}}^{(s)} = a_1^{(s)}\)\\
Simple cost slope under vague goal (cells):
\(a_{\text{deal}}^{(s)} = a_1^{(s)} + a_3^{(s)}\)\\
Conditional indirect effect under specific goal:
\(\text{Indirect}_{\text{ppw}}^{(s)} = a_1^{(s)} \times b_1^{(s)}\)\\
Conditional indirect effect under vague goal:
\(\text{Indirect}_{\text{deal}}^{(s)} = (a_1^{(s)} + a_3^{(s)}) \times b_1^{(s)}\)\\
Index of Moderated Mediation (IMM):
\(\text{IMM}^{(s)} = a_3^{(s)} \times b_1^{(s)} = \text{Indirect}_{\text{deal}}^{(s)} - \text{Indirect}_{\text{ppw}}^{(s)}\)\\
Total effect of cost under vague goal:
\(\text{Total}_{\text{deal}}^{(s)} = c'^{(s)} + (a_1^{(s)} + a_3^{(s)}) \times b_1^{(s)}\)

Point estimates are reported as the mean of the posterior draws, along
with posterior standard deviations and 95\% equal-tailed credible
intervals bounded by the 2.5th and 97.5th percentiles.

\section*{Web appendix E: Purchase volume manipulation, objective
classification, and reasoning trace analysis}\label{sec-appendix-judge}
\addcontentsline{toc}{section}{Web appendix E: Purchase volume
manipulation, objective classification, and reasoning trace analysis}

\subsection{Classification task}\label{classification-task}

To characterize what changed when the purchase volume was manipulated,
we classified the optimization objective that each LLM session
articulated before acquiring any information.

For each Pro and Flash session, we extracted the first reasoning trace
emitted before the initial \emph{inspect\_cell} invocation. Each judge
read the first reasoning trace in isolation and assigned it to one of
three categories:

\begin{enumerate}
\def\labelenumi{\arabic{enumi}.}
\tightlist
\item
  \emph{price\_per\_weight}: Optimizing for unit value or
  cost-effectiveness per weight (e.g.~price per gram, oz per dollar,
  total weight divided by total cost).
\item
  \emph{price}: Optimizing for total price or price per bag
  (e.g.~choosing cheapest option based on upfront price).
\item
  \emph{others}: Any other objective.
\end{enumerate}

Each judge received no other part of the session, and observed neither
the final choice nor any tool call, so the classification cannot be
informed by the behavior it is subsequently used to describe. Gemini 3.5
Flash Lite could not be classified. Its default thinking level is
\emph{minimal}, so it emits no reasoning prior to acquisition and the
classification task is undefined for it.

\quartoapptblref{apptbl-judge-examples} presents examples of reasoning
traces.

{\def\LTcaptype{apptbl}
\begin{longtable}[]{@{}
  >{\raggedright\arraybackslash}p{(\linewidth - 2\tabcolsep) * \real{0.5000}}
  >{\raggedright\arraybackslash}p{(\linewidth - 2\tabcolsep) * \real{0.5000}}@{}}
\caption{{\label{apptbl-judge-examples}}Example reasoning traces}\tabularnewline
\toprule\noalign{}
\begin{minipage}[b]{\linewidth}\raggedright
Classified objective
\end{minipage} & \begin{minipage}[b]{\linewidth}\raggedright
Reasoning
\end{minipage} \\
\midrule\noalign{}
\endfirsthead
\toprule\noalign{}
\begin{minipage}[b]{\linewidth}\raggedright
Classified objective
\end{minipage} & \begin{minipage}[b]{\linewidth}\raggedright
Reasoning
\end{minipage} \\
\midrule\noalign{}
\endhead
\bottomrule\noalign{}
\endlastfoot
Price-per-weight & ``{[}\ldots{]} My primary focus is always the price
per unit, you know, the real cost-effectiveness. {[}\ldots{]}'' \\
Price & ``Okay, so the goal is to find the absolute lowest price on
instant coffee. {[}\ldots{]}'' \\
Others & ``{[}\ldots{]} My intuition tells me to zero in on the total
price. {[}\ldots{]} While''best deal'' often boils down to the lowest
total price, sometimes it's about getting the most coffee for your money
-- the best \emph{unit} price. I'll keep that in mind.'' \\
\end{longtable}
}

\emph{Note.} Illustrative samples of reasoning traces per
classification.\\
Text is abridged from the reasoning emitted before any
\emph{inspect\_cell} invocation, which is the input the judges received.
The purchase quantity is stated in the task prompt and therefore appears
in the trace.

\subsection{Candidate judges}\label{candidate-judges}

We evaluated eight candidate judges from providers other than those in
the main experiments with an initial set of three open-weight LLMs and
five decision models (\quartoapptblref{apptbl-judge-models}).

Decision models are a class of models built for zero-shot classification
and other structured decisions. A decision model receives a state and a
question with predefined answer options. It returns the selected option
with a probability for each option. Unlike an LLM, it generates no text,
so it gives no justification for its answer. The probabilities are read
directly from the model, so no sampling parameters are involved.

The LLMs were served locally via \emph{llama.cpp} using quantized
checkpoints at \(T = 0\). Each LLM returned a structured output with a
justification and the classified objective. The decision models were
queried through the OpenRouter \emph{System One} endpoint. Each decision
model received the task (``An AI shopping agent was asked to purchase
instant coffee'') and the reasoning trace as the state, and returned the
classified objective with a probability for each category.

{\def\LTcaptype{apptbl}
\begin{longtable}[]{@{}
  >{\raggedright\arraybackslash}p{(\linewidth - 6\tabcolsep) * \real{0.2500}}
  >{\raggedright\arraybackslash}p{(\linewidth - 6\tabcolsep) * \real{0.2500}}
  >{\raggedright\arraybackslash}p{(\linewidth - 6\tabcolsep) * \real{0.2500}}
  >{\raggedright\arraybackslash}p{(\linewidth - 6\tabcolsep) * \real{0.2500}}@{}}
\caption{{\label{apptbl-judge-models}}Candidate judges for objective
classification}\tabularnewline
\toprule\noalign{}
\begin{minipage}[b]{\linewidth}\raggedright
Candidate judge
\end{minipage} & \begin{minipage}[b]{\linewidth}\raggedright
Type
\end{minipage} & \begin{minipage}[b]{\linewidth}\raggedright
Provider
\end{minipage} & \begin{minipage}[b]{\linewidth}\raggedright
Identifier
\end{minipage} \\
\midrule\noalign{}
\endfirsthead
\toprule\noalign{}
\begin{minipage}[b]{\linewidth}\raggedright
Candidate judge
\end{minipage} & \begin{minipage}[b]{\linewidth}\raggedright
Type
\end{minipage} & \begin{minipage}[b]{\linewidth}\raggedright
Provider
\end{minipage} & \begin{minipage}[b]{\linewidth}\raggedright
Identifier
\end{minipage} \\
\midrule\noalign{}
\endhead
\bottomrule\noalign{}
\endlastfoot
JEV 1.13 & Decision model & TypeSafe &
\emph{typesafe/jev-1.13-20260917} \\
Liquid D1 & Decision model & Liquid AI & \emph{liquid/d1-20260930} \\
Mercury Decide & Decision model & Inception &
\emph{inception/mercury-decide-20260930} \\
pplx Decider v1 27B & Decision model & Perplexity &
\emph{perplexity/pplx-decider-v1-27b-20261001} \\
Solar Decide & Decision model & Upstage &
\emph{upstage/solar-decide-20260928} \\
Nemotron 3.5 Lightning 30B-A3B & LLM & NVIDIA &
\emph{unsloth/NVIDIA-Nemotron-3.5-Lightning-30B-A3B-GGUF:Q4\_K\_XL} \\
Qwen 3.8 27B & LLM & Alibaba Cloud &
\emph{unsloth/Qwen3.8-27B-GGUF:Q6\_K\_XL} \\
Ornith 1.0 35B & LLM & Ornith AI &
\emph{unsloth/Ornith-1.0-35B-GGUF:MXFP4\_MOE} \\
\end{longtable}
}

\emph{Note.} LLM identifiers are the quantized checkpoint served by
\emph{llama.cpp}. Decision model identifiers are the snapshot that
served each request, as reported by the System One endpoint.

The LLM judges are open-weight checkpoints run with greedy decoding, so
the classification can be rerun with the identifiers in
\quartoapptblref{apptbl-judge-models}. The decision models are closed
models served through a managed API, hence the providers can update the
models. We therefore report the snapshot that served each request.

\subsection{Judge selection}\label{judge-selection}

We hired three human coders at \$20 per hour. An author independently
completed the coding session in two hours. Therefore, each coder was
paid for two hours. The three human coders were blind to the purpose of
the experiment. Each independently classified 120 traces sampled at 30
per number of bags, using the same first-reasoning extraction and the
same three categories given to the candidate judges. We designated the
candidate with the highest mean \(\kappa\) across the three coders as
the judge. \quartoapptblref{apptbl-judge-human} reports agreement
between each coder and each candidate judge.

Agreement with the human coders was highest for the Mercury Decide
decision model, which was therefore designated as the judge. The second-
and third-ranked candidates, Liquid D1 and pplx Decider v1 27B, were
used in the robustness analysis.

Agreement among the human coders was moderate (Fleiss' \(\kappa = .46\);
pairwise \(\kappa\) from .36 to .56). The largest disagreement concerned
the boundary between \emph{price-per-weight} and \emph{others}. Coder 1
classified 38 traces as \emph{others}, and Coder 2 classified 35 of
these traces as \emph{price-per-weight}. Despite this disagreement,
every coder and each of the three top-ranked judges showed the same
effect of purchase volume on \emph{price} classification
(\quartoapptblref{apptbl-human-bags}).

{\def\LTcaptype{apptbl}
\begin{longtable}[]{@{}
  >{\raggedright\arraybackslash}p{(\linewidth - 14\tabcolsep) * \real{0.2946}}
  >{\raggedleft\arraybackslash}p{(\linewidth - 14\tabcolsep) * \real{0.0714}}
  >{\raggedleft\arraybackslash}p{(\linewidth - 14\tabcolsep) * \real{0.0982}}
  >{\raggedleft\arraybackslash}p{(\linewidth - 14\tabcolsep) * \real{0.0714}}
  >{\raggedleft\arraybackslash}p{(\linewidth - 14\tabcolsep) * \real{0.0982}}
  >{\raggedleft\arraybackslash}p{(\linewidth - 14\tabcolsep) * \real{0.0714}}
  >{\raggedleft\arraybackslash}p{(\linewidth - 14\tabcolsep) * \real{0.0982}}
  >{\raggedleft\arraybackslash}p{(\linewidth - 14\tabcolsep) * \real{0.1429}}@{}}
\caption{{\label{apptbl-judge-human}}Agreement between human coders and
candidate judges}\tabularnewline
\toprule\noalign{}
\multirow{2}{=}{\begin{minipage}[b]{\linewidth}\raggedright
Candidate judge
\end{minipage}} &
\multicolumn{2}{>{\raggedleft\arraybackslash}p{(\linewidth - 14\tabcolsep) * \real{0.1696} + 2\tabcolsep}}{%
\begin{minipage}[b]{\linewidth}\raggedleft
Coder 1
\end{minipage}} &
\multicolumn{2}{>{\raggedleft\arraybackslash}p{(\linewidth - 14\tabcolsep) * \real{0.1696} + 2\tabcolsep}}{%
\begin{minipage}[b]{\linewidth}\raggedleft
Coder 2
\end{minipage}} &
\multicolumn{2}{>{\raggedleft\arraybackslash}p{(\linewidth - 14\tabcolsep) * \real{0.1696} + 2\tabcolsep}}{%
\begin{minipage}[b]{\linewidth}\raggedleft
Coder 3
\end{minipage}} &
\multirow{2}{=}{\raggedright\arraybackslash \begin{minipage}[b]{\linewidth}\raggedleft
Mean \(\kappa\)
\end{minipage}} \\
& \begin{minipage}[b]{\linewidth}\raggedleft
\(p_o\)
\end{minipage} & \begin{minipage}[b]{\linewidth}\raggedleft
\(\kappa\)
\end{minipage} & \begin{minipage}[b]{\linewidth}\raggedleft
\(p_o\)
\end{minipage} & \begin{minipage}[b]{\linewidth}\raggedleft
\(\kappa\)
\end{minipage} & \begin{minipage}[b]{\linewidth}\raggedleft
\(p_o\)
\end{minipage} & \begin{minipage}[b]{\linewidth}\raggedleft
\(\kappa\)
\end{minipage} \\
\midrule\noalign{}
\endfirsthead
\toprule\noalign{}
\multirow{2}{=}{\begin{minipage}[b]{\linewidth}\raggedright
Candidate judge
\end{minipage}} &
\multicolumn{2}{>{\raggedleft\arraybackslash}p{(\linewidth - 14\tabcolsep) * \real{0.1696} + 2\tabcolsep}}{%
\begin{minipage}[b]{\linewidth}\raggedleft
Coder 1
\end{minipage}} &
\multicolumn{2}{>{\raggedleft\arraybackslash}p{(\linewidth - 14\tabcolsep) * \real{0.1696} + 2\tabcolsep}}{%
\begin{minipage}[b]{\linewidth}\raggedleft
Coder 2
\end{minipage}} &
\multicolumn{2}{>{\raggedleft\arraybackslash}p{(\linewidth - 14\tabcolsep) * \real{0.1696} + 2\tabcolsep}}{%
\begin{minipage}[b]{\linewidth}\raggedleft
Coder 3
\end{minipage}} &
\multirow{2}{=}{\raggedright\arraybackslash \begin{minipage}[b]{\linewidth}\raggedleft
Mean \(\kappa\)
\end{minipage}} \\
& \begin{minipage}[b]{\linewidth}\raggedleft
\(p_o\)
\end{minipage} & \begin{minipage}[b]{\linewidth}\raggedleft
\(\kappa\)
\end{minipage} & \begin{minipage}[b]{\linewidth}\raggedleft
\(p_o\)
\end{minipage} & \begin{minipage}[b]{\linewidth}\raggedleft
\(\kappa\)
\end{minipage} & \begin{minipage}[b]{\linewidth}\raggedleft
\(p_o\)
\end{minipage} & \begin{minipage}[b]{\linewidth}\raggedleft
\(\kappa\)
\end{minipage} \\
\midrule\noalign{}
\endhead
\bottomrule\noalign{}
\endlastfoot
Mercury Decide\(^a\) & .775 & .592 & .708 & .490 & .892 & .804 & .629 \\
Liquid D1 & .683 & .466 & .850 & .707 & .842 & .709 & .627 \\
pplx Decider v1 27B & .758 & .566 & .708 & .497 & .883 & .791 & .618 \\
Solar Decide & .783 & .603 & .692 & .464 & .850 & .728 & .598 \\
JEV 1.13 & .692 & .450 & .792 & .588 & .825 & .665 & .568 \\
Qwen 3.8 27B & .683 & .433 & .817 & .628 & .808 & .628 & .563 \\
Ornith 1.0 35B & .683 & .429 & .808 & .611 & .808 & .626 & .555 \\
Nemotron 3.5 Lightning 30B A3B & .675 & .409 & .800 & .594 & .800 & .608
& .537 \\
Coders 1--3 (Fleiss' \(\kappa\)) & & & & & & & .455 \\
\end{longtable}
}

\emph{Note.} \(n = 120\) pre-acquisition reasoning traces sampled
randomly at 30 traces per purchase volume condition (1, 5, 50, and 250
bags) from Gemini 3.1 Pro sessions. Traces were independently classified
into three nominal categories: \emph{price-per-weight} (unit price
minimization), \emph{price} (absolute price minimization), or
\emph{others}. Observed agreement (\(p_o\)) represents the raw
proportion of concordant classifications between the human coder and the
candidate judge. Cohen's \(\kappa\) evaluates inter-rater reliability
corrected for chance agreement.\\
Candidate judges are ordered by mean \(\kappa\). \(^a\) Designated judge
used for all reported focal analyses.

\subsection{Robustness to judge
selection}\label{robustness-to-judge-selection}

\textbf{Human coders and judges on the coded subsample} The judge was
selected by agreement on individual traces, whereas our analysis depends
on the effect of purchase volume on \emph{price} classification. We
therefore estimated this effect on the coded subsample separately for
each human coder and each of the three top-ranked judges.
\quartoapptblref{apptbl-human-bags} reports the share of traces
classified as \emph{price} in each purchase volume condition and the
slope of log purchase volume for each labeler. For every coder and every
judge, the share of \emph{price} classifications increased with purchase
volume (slopes from 0.32 to 0.45, all \(p < .001\)). Therefore, the
effect does not depend on which coder or judge classified the traces.

{\def\LTcaptype{apptbl}
\begin{longtable}[]{@{}
  >{\raggedright\arraybackslash}p{(\linewidth - 12\tabcolsep) * \real{0.2821}}
  >{\raggedleft\arraybackslash}p{(\linewidth - 12\tabcolsep) * \real{0.0897}}
  >{\raggedleft\arraybackslash}p{(\linewidth - 12\tabcolsep) * \real{0.0897}}
  >{\raggedleft\arraybackslash}p{(\linewidth - 12\tabcolsep) * \real{0.0897}}
  >{\raggedleft\arraybackslash}p{(\linewidth - 12\tabcolsep) * \real{0.0897}}
  >{\raggedleft\arraybackslash}p{(\linewidth - 12\tabcolsep) * \real{0.1795}}
  >{\raggedleft\arraybackslash}p{(\linewidth - 12\tabcolsep) * \real{0.1154}}@{}}
\caption{{\label{apptbl-human-bags}}Price minimization by purchase
volume across human coders and judges}\tabularnewline
\toprule\noalign{}
\multirow{2}{=}{\begin{minipage}[b]{\linewidth}\raggedright
Labeler
\end{minipage}} &
\multicolumn{4}{>{\raggedleft\arraybackslash}p{(\linewidth - 12\tabcolsep) * \real{0.3590} + 6\tabcolsep}}{%
\begin{minipage}[b]{\linewidth}\raggedleft
Purchase volume
\end{minipage}} &
\multirow{2}{=}{\raggedright\arraybackslash \begin{minipage}[b]{\linewidth}\raggedleft
Slope (SE)
\end{minipage}} &
\multirow{2}{=}{\raggedright\arraybackslash \begin{minipage}[b]{\linewidth}\raggedleft
\(p\)
\end{minipage}} \\
& \begin{minipage}[b]{\linewidth}\raggedleft
1
\end{minipage} & \begin{minipage}[b]{\linewidth}\raggedleft
5
\end{minipage} & \begin{minipage}[b]{\linewidth}\raggedleft
50
\end{minipage} & \begin{minipage}[b]{\linewidth}\raggedleft
250
\end{minipage} \\
\midrule\noalign{}
\endfirsthead
\toprule\noalign{}
\multirow{2}{=}{\begin{minipage}[b]{\linewidth}\raggedright
Labeler
\end{minipage}} &
\multicolumn{4}{>{\raggedleft\arraybackslash}p{(\linewidth - 12\tabcolsep) * \real{0.3590} + 6\tabcolsep}}{%
\begin{minipage}[b]{\linewidth}\raggedleft
Purchase volume
\end{minipage}} &
\multirow{2}{=}{\raggedright\arraybackslash \begin{minipage}[b]{\linewidth}\raggedleft
Slope (SE)
\end{minipage}} &
\multirow{2}{=}{\raggedright\arraybackslash \begin{minipage}[b]{\linewidth}\raggedleft
\(p\)
\end{minipage}} \\
& \begin{minipage}[b]{\linewidth}\raggedleft
1
\end{minipage} & \begin{minipage}[b]{\linewidth}\raggedleft
5
\end{minipage} & \begin{minipage}[b]{\linewidth}\raggedleft
50
\end{minipage} & \begin{minipage}[b]{\linewidth}\raggedleft
250
\end{minipage} \\
\midrule\noalign{}
\endhead
\bottomrule\noalign{}
\endlastfoot
Coder 1 & .300 & .600 & .767 & .700 & 0.32 (0.10) & \textless{} .001 \\
Coder 2 & .267 & .467 & .633 & .700 & 0.33 (0.09) & \textless{} .001 \\
Coder 3 & .300 & .500 & .700 & .800 & 0.40 (0.10) & \textless{} .001 \\
Mercury Decide\(^a\) & .400 & .533 & .800 & .833 & 0.40 (0.10) &
\textless{} .001 \\
Liquid D1 & .333 & .567 & .767 & .800 & 0.39 (0.10) & \textless{}
.001 \\
pplx Decider v1 27B & .333 & .533 & .800 & .833 & 0.45 (0.11) &
\textless{} .001 \\
\end{longtable}
}

\emph{Note.} \(n = 120\) pre-acquisition reasoning traces sampled
randomly at 30 traces per purchase volume condition (1, 5, 50, and 250
bags) from Gemini 3.1 Pro sessions, the same traces as in
\quartoapptblref{apptbl-judge-human}. Coders 1--3 are the human coders.
The judges are the three candidates with the highest mean \(\kappa\) in
\quartoapptblref{apptbl-judge-human}. Cells under \emph{Purchase volume}
report the share of traces classified as \emph{price} (absolute price
minimization) in each purchase volume condition. Slope (SE) is the
coefficient of log purchase volume in a logit of \emph{price}
classification, estimated separately for each labeler. \(^a\) Designated
judge used for all reported focal analyses. The other two judges are
used in the robustness analysis.

\textbf{Alternative judges on the full sample} We repeated the
classification and the focal analyses with the second- and third-ranked
candidates, Liquid D1 and pplx Decider v1 27B (mean \(\kappa = .63\) and
\(.62\); \quartoapptblref{apptbl-judge-human}).

Under both judges, the share of Pro sessions adopting an absolute price
objective increased with purchase volume
(\quartoapptblref{apptbl-bags-objective-robust}). It rose from 34.0\% at
one bag to 84.8\% at 250 bags under both judges. Flash adopted an
absolute price objective in at most 6.1\% of sessions under either
judge.

Holding purchase volume fixed, Pro sessions adopting an absolute price
objective acquired fewer diagnostic attributes and were less likely to
select the optimal alternative under both judges
(\quartoapptblref{apptbl-bags-contrast-robust}). The pooled contrasts
excluded zero for both judges. The one exception was choice optimality
at 250 bags under pplx Decider v1 27B, where only four Pro sessions were
classified as adopting a unit-price objective. The pplx Decider v1 27B
judge also classified 69 Pro sessions as ``others'', which are excluded
from the contrast.

{\def\LTcaptype{apptbl}
\begin{longtable}[]{@{}
  >{\raggedright\arraybackslash}p{(\linewidth - 12\tabcolsep) * \real{0.2625}}
  >{\raggedright\arraybackslash}p{(\linewidth - 12\tabcolsep) * \real{0.0875}}
  >{\raggedleft\arraybackslash}p{(\linewidth - 12\tabcolsep) * \real{0.1000}}
  >{\raggedleft\arraybackslash}p{(\linewidth - 12\tabcolsep) * \real{0.0625}}
  >{\raggedleft\arraybackslash}p{(\linewidth - 12\tabcolsep) * \real{0.2500}}
  >{\raggedleft\arraybackslash}p{(\linewidth - 12\tabcolsep) * \real{0.1125}}
  >{\raggedleft\arraybackslash}p{(\linewidth - 12\tabcolsep) * \real{0.1250}}@{}}
\caption{{\label{apptbl-bags-objective-robust}}Adopted objective by
purchase volume under the second- and third-ranked judges}\tabularnewline
\toprule\noalign{}
\begin{minipage}[b]{\linewidth}\raggedright
Judge
\end{minipage} & \begin{minipage}[b]{\linewidth}\raggedright
LLM
\end{minipage} & \begin{minipage}[b]{\linewidth}\raggedleft
Bags
\end{minipage} & \begin{minipage}[b]{\linewidth}\raggedleft
n
\end{minipage} & \begin{minipage}[b]{\linewidth}\raggedleft
price\_per\_weight
\end{minipage} & \begin{minipage}[b]{\linewidth}\raggedleft
price
\end{minipage} & \begin{minipage}[b]{\linewidth}\raggedleft
others
\end{minipage} \\
\midrule\noalign{}
\endfirsthead
\toprule\noalign{}
\begin{minipage}[b]{\linewidth}\raggedright
Judge
\end{minipage} & \begin{minipage}[b]{\linewidth}\raggedright
LLM
\end{minipage} & \begin{minipage}[b]{\linewidth}\raggedleft
Bags
\end{minipage} & \begin{minipage}[b]{\linewidth}\raggedleft
n
\end{minipage} & \begin{minipage}[b]{\linewidth}\raggedleft
price\_per\_weight
\end{minipage} & \begin{minipage}[b]{\linewidth}\raggedleft
price
\end{minipage} & \begin{minipage}[b]{\linewidth}\raggedleft
others
\end{minipage} \\
\midrule\noalign{}
\endhead
\bottomrule\noalign{}
\endlastfoot
Liquid D1 & Flash & 1 & 98 & 0.980 & 0.020 & 0.000 \\
Liquid D1 & Flash & 5 & 100 & 0.950 & 0.030 & 0.020 \\
Liquid D1 & Flash & 50 & 100 & 0.920 & 0.040 & 0.040 \\
Liquid D1 & Flash & 250 & 99 & 0.899 & 0.061 & 0.040 \\
Liquid D1 & Pro & 1 & 100 & 0.610 & 0.340 & 0.050 \\
Liquid D1 & Pro & 5 & 100 & 0.470 & 0.500 & 0.030 \\
Liquid D1 & Pro & 50 & 99 & 0.121 & 0.848 & 0.030 \\
Liquid D1 & Pro & 250 & 99 & 0.121 & 0.848 & 0.030 \\
pplx Decider v1 27B & Flash & 1 & 98 & 0.633 & 0.010 & 0.357 \\
pplx Decider v1 27B & Flash & 5 & 100 & 0.630 & 0.030 & 0.340 \\
pplx Decider v1 27B & Flash & 50 & 100 & 0.640 & 0.010 & 0.350 \\
pplx Decider v1 27B & Flash & 250 & 99 & 0.616 & 0.040 & 0.343 \\
pplx Decider v1 27B & Pro & 1 & 100 & 0.420 & 0.340 & 0.240 \\
pplx Decider v1 27B & Pro & 5 & 100 & 0.270 & 0.500 & 0.230 \\
pplx Decider v1 27B & Pro & 50 & 99 & 0.051 & 0.838 & 0.111 \\
pplx Decider v1 27B & Pro & 250 & 99 & 0.040 & 0.848 & 0.111 \\
\end{longtable}
}

\emph{Note.} Entries are the share of sessions classified into each
objective among sessions with a pre-acquisition reasoning trace (\(n\)).
Gemini 3.5 Flash Lite emits no reasoning before acquisition and is not
classified.

{\def\LTcaptype{apptbl}
\begin{longtable}[]{@{}
  >{\raggedright\arraybackslash}p{(\linewidth - 14\tabcolsep) * \real{0.1346}}
  >{\raggedright\arraybackslash}p{(\linewidth - 14\tabcolsep) * \real{0.0513}}
  >{\raggedleft\arraybackslash}p{(\linewidth - 14\tabcolsep) * \real{0.1410}}
  >{\raggedleft\arraybackslash}p{(\linewidth - 14\tabcolsep) * \real{0.1282}}
  >{\raggedleft\arraybackslash}p{(\linewidth - 14\tabcolsep) * \real{0.1474}}
  >{\raggedright\arraybackslash}p{(\linewidth - 14\tabcolsep) * \real{0.1026}}
  >{\raggedleft\arraybackslash}p{(\linewidth - 14\tabcolsep) * \real{0.1923}}
  >{\raggedright\arraybackslash}p{(\linewidth - 14\tabcolsep) * \real{0.1026}}@{}}
\caption{{\label{apptbl-bags-contrast-robust}}Contrast by adopted
objective holding purchase volume fixed, second- and third-ranked
judges}\tabularnewline
\toprule\noalign{}
\begin{minipage}[b]{\linewidth}\raggedright
Judge
\end{minipage} & \begin{minipage}[b]{\linewidth}\raggedright
Bags
\end{minipage} & \begin{minipage}[b]{\linewidth}\raggedleft
\(n_{\text{price}}\)
\end{minipage} & \begin{minipage}[b]{\linewidth}\raggedleft
\(n_{\text{ppw}}\)
\end{minipage} & \begin{minipage}[b]{\linewidth}\raggedleft
\(\Delta\) Diagnostic
\end{minipage} & \begin{minipage}[b]{\linewidth}\raggedright
95\% CI
\end{minipage} & \begin{minipage}[b]{\linewidth}\raggedleft
\(\Delta\) Choice Optimality
\end{minipage} & \begin{minipage}[b]{\linewidth}\raggedright
95\% CI
\end{minipage} \\
\midrule\noalign{}
\endfirsthead
\toprule\noalign{}
\begin{minipage}[b]{\linewidth}\raggedright
Judge
\end{minipage} & \begin{minipage}[b]{\linewidth}\raggedright
Bags
\end{minipage} & \begin{minipage}[b]{\linewidth}\raggedleft
\(n_{\text{price}}\)
\end{minipage} & \begin{minipage}[b]{\linewidth}\raggedleft
\(n_{\text{ppw}}\)
\end{minipage} & \begin{minipage}[b]{\linewidth}\raggedleft
\(\Delta\) Diagnostic
\end{minipage} & \begin{minipage}[b]{\linewidth}\raggedright
95\% CI
\end{minipage} & \begin{minipage}[b]{\linewidth}\raggedleft
\(\Delta\) Choice Optimality
\end{minipage} & \begin{minipage}[b]{\linewidth}\raggedright
95\% CI
\end{minipage} \\
\midrule\noalign{}
\endhead
\bottomrule\noalign{}
\endlastfoot
Liquid D1 & 1 & 34 & 61 & -1.32* & {[}-1.91, -0.76{]} & -0.59* &
{[}-0.76, -0.41{]} \\
Liquid D1 & 5 & 50 & 47 & -1.14* & {[}-1.64, -0.68{]} & -0.40* &
{[}-0.54, -0.25{]} \\
Liquid D1 & 50 & 84 & 12 & -1.86* & {[}-2.23, -1.48{]} & -0.70* &
{[}-0.80, -0.61{]} \\
Liquid D1 & 250 & 84 & 12 & -2.12* & {[}-2.49, -1.75{]} & -0.57* &
{[}-0.77, -0.32{]} \\
Liquid D1 & Pooled & 252 & 132 & -1.61* & {[}-1.84, -1.38{]} & -0.56* &
{[}-0.65, -0.48{]} \\
pplx Decider v1 27B & 1 & 34 & 42 & -1.32* & {[}-1.91, -0.76{]} & -0.59*
& {[}-0.74, -0.41{]} \\
pplx Decider v1 27B & 5 & 50 & 27 & -1.14* & {[}-1.62, -0.68{]} & -0.42*
& {[}-0.58, -0.27{]} \\
pplx Decider v1 27B & 50 & 83 & 5 & -1.88* & {[}-2.25, -1.51{]} & -0.71*
& {[}-0.81, -0.61{]} \\
pplx Decider v1 27B & 250 & 84 & 4 & -2.12* & {[}-2.49, -1.76{]} & -0.49
& {[}-0.81, 0.01{]} \\
pplx Decider v1 27B & Pooled & 251 & 78 & -1.64* & {[}-1.87, -1.43{]} &
-0.56* & {[}-0.67, -0.42{]} \\
\end{longtable}
}

\emph{Note.} Gemini 3.1 Pro sessions only. \(N = 384\) sessions under
Liquid D1 (14 classified as ``others'' excluded) and \(N = 329\)
sessions under pplx Decider v1 27B (69 classified as ``others''
excluded); 2 sessions without a pre-acquisition reasoning trace are
excluded under both judges. Entries represent the within-volume
differences (\(\text{Price} - \text{Price-per-weight}\)) in diagnostic
attribute acquisitions (maximum = 6) and choice optimality probability.
Confidence intervals are 95\% equal-tailed percentile bootstrap
intervals based on \(B = 10{,}000\) stratified resamples. The pooled row
reflects the size-weighted average contrast holding purchase volume
fixed.\\
* 95\% bootstrap confidence interval strictly excludes zero.

Two technical boundary conditions apply to these findings. First, this
experiment held acquisition fees at zero (\(c = \$0.00\)) to isolate the
implicit cost assumption. Whether purchase volume interacts with
positive explicit fees remains unexamined. Second, reasoning trace
extraction was confined to the Google Gemini family. Whether similar
shifts in the adopted objective emerge in reasoning LLMs from other
providers, such as Anthropic, remains an open empirical question.

\section*{Web appendix F: Exogenously assigned information
states}\label{sec-appendix-fixed-reveal}
\addcontentsline{toc}{section}{Web appendix F: Exogenously assigned
information states}

In the Tool-Lab experiments in Studies 1 and 2, the information state an
LLM holds at the moment of choice is endogenous. The LLM determines
which attributes to inspect, when to terminate search, and which product
to select. While this process-tracing paradigm captures information
acquisition under cost, observed choice outcomes conflate information
acquisition with information utilization. When an LLM selects an
economically suboptimal alternative under cost constraints, this choice
can stem from an acquisition failure, in which the LLM truncates its
search and never retrieves the diagnostic attributes required to compute
unit price, or from a utilization failure, in which the LLM possesses
the relevant attributes but fails to synthesize them into an accurate
mathematical evaluation. Furthermore, dynamic data cannot determine
whether selecting a superficially appealing alternative under partial
information reflects a utilization failure or a normatively optimal
inference given an incomplete information state.

To resolve this endogeneity, we implement an experimental condition that
exogenously assigns predetermined information states. In this design,
the inspect\_cell tool and acquisition costs are eliminated. A
predetermined set of k product attributes is displayed directly in the
user prompt under zero cost, while unassigned attributes are explicitly
listed as existing but unavailable.

\subsection{Design and Procedure}\label{design-and-procedure}

The experimental stimuli, delegation framing, five-bag purchase volume,
and submission schemas follow the corresponding Tool-Lab studies. In the
Study 1 environment, the choice set consists of Coffee A (4.99 dollars,
10.0 ounces, yielding 0.499 dollars per ounce) and Coffee B (5.00
dollars, 11.0 ounces, yielding 0.455 dollars per ounce). In the Study 2
environment, the choice set consists of Coffee C (8.00 dollars list
price, 0 percent discount, 8.0 ounces, yielding 1.000 dollar per ounce),
Coffee D (15.00 dollars list price, 20 percent discount, 12.0 ounces,
yielding 1.000 dollar per ounce), and Coffee E (10.00 dollars list
price, 0 percent discount, 10.9 ounces, yielding 0.917 dollars per
ounce).

Three procedural modifications differentiate this design from the
Tool-Lab framework. First, the \emph{inspect\_cell} tool is omitted,
leaving the submit\_choice tool as the only available tool. Second, no
acquisition cost is charged or mentioned. Third, the assigned attributes
are embedded directly within the initial user prompt formatted as JSON
structures identical to the payloads returned in Tool-Lab. Nevertheless,
the prompt lists every attribute present in the retail environment. This
structure ensures that LLMs make decisions under explicit attribute
missingness rather than assuming that unlisted attributes are absent or
irrelevant.

We evaluate nine information states (four in the Study 1 environment and
five in the Study 2 environment). In Study 1, the states vary whether
the whole-dollar price is presented alone (Dollars, 2 diagnostic cells),
paired with a non-diagnostic attribute (Dollars plus Roast, 2 diagnostic
and 2 non-diagnostic cells), paired with product weight (Dollars plus
Weight, 4 diagnostic cells), or combined with both cents and weight
(Dollars plus Weight plus Cents, 6 diagnostic cells, representing the
complete diagnostic set; \quartoapptblref{apptbl-fixed-reveal-design}).

In Study 2, the states vary whether the list price is presented alone
(List Price, 3 diagnostic cells), paired with roast date (List Price
plus Roast, 3 diagnostic and 3 non-diagnostic cells), paired with
promotional discount percentage (List Price plus Discount, 6 diagnostic
cells), paired with product weight (List Price plus Weight, 6 diagnostic
cells), or combined with both discount percentage and weight (List Price
plus Weight plus Discount, 9 diagnostic cells, representing the complete
diagnostic set). This design includes two matched-count conditions that
hold information volume constant while varying diagnostic content, that
is a four-cell comparison in Study 1 and a six-cell comparison across
three distinct compositions in Study 2.

We evaluated the same eight LLMs as the main studies across these
conditions. Each experimental cell was evaluated across 50 independent
sessions with zero state retention between trials, yielding 1,600
sessions in Study 1 and 2,000 sessions in Study 2 (3,600 sessions
total). Presentation orders and attribute payload sequences were
randomized within each session.

We capture two primary dependent measures. Choice optimality is coded as
1 when the LLM selects the alternative with the superior price per ounce
(Coffee B in Study 1; Coffee E in Study 2) and 0 otherwise. State
consistency is coded as 1 when the LLM selects the alternative
prescribed by the normative Expectimax benchmark conditional on the
information revealed in that state, and 0 otherwise. At complete
information states, choice optimality and state consistency coincide. At
truncated states, they diverge whenever the available information favors
a nominally cheaper alternative over the globally optimal one.

{\def\LTcaptype{apptbl}
\begin{longtable}[]{@{}
  >{\raggedright\arraybackslash}p{(\linewidth - 10\tabcolsep) * \real{0.1667}}
  >{\raggedright\arraybackslash}p{(\linewidth - 10\tabcolsep) * \real{0.1667}}
  >{\raggedright\arraybackslash}p{(\linewidth - 10\tabcolsep) * \real{0.1667}}
  >{\raggedleft\arraybackslash}p{(\linewidth - 10\tabcolsep) * \real{0.1667}}
  >{\raggedleft\arraybackslash}p{(\linewidth - 10\tabcolsep) * \real{0.1667}}
  >{\raggedleft\arraybackslash}p{(\linewidth - 10\tabcolsep) * \real{0.1667}}@{}}
\caption{{\label{apptbl-fixed-reveal-design}}Assigned information
states}\tabularnewline
\toprule\noalign{}
\begin{minipage}[b]{\linewidth}\raggedright
Environment
\end{minipage} & \begin{minipage}[b]{\linewidth}\raggedright
State
\end{minipage} & \begin{minipage}[b]{\linewidth}\raggedright
Revealed attributes
\end{minipage} & \begin{minipage}[b]{\linewidth}\raggedleft
Cells
\end{minipage} & \begin{minipage}[b]{\linewidth}\raggedleft
Diagnostic
\end{minipage} & \begin{minipage}[b]{\linewidth}\raggedleft
Non-diagnostic
\end{minipage} \\
\midrule\noalign{}
\endfirsthead
\toprule\noalign{}
\begin{minipage}[b]{\linewidth}\raggedright
Environment
\end{minipage} & \begin{minipage}[b]{\linewidth}\raggedright
State
\end{minipage} & \begin{minipage}[b]{\linewidth}\raggedright
Revealed attributes
\end{minipage} & \begin{minipage}[b]{\linewidth}\raggedleft
Cells
\end{minipage} & \begin{minipage}[b]{\linewidth}\raggedleft
Diagnostic
\end{minipage} & \begin{minipage}[b]{\linewidth}\raggedleft
Non-diagnostic
\end{minipage} \\
\midrule\noalign{}
\endhead
\bottomrule\noalign{}
\endlastfoot
Study 1 & Dollars & \emph{price\_dollars} & 2 & 2 & 0 \\
Study 1 & Dollars + Roast & \emph{price\_dollars}, \emph{roast\_date} &
4 & 2 & 2 \\
Study 1 & Dollars + Weight & \emph{price\_dollars}, \emph{weight} & 4 &
4 & 0 \\
Study 1 & Dollars + Weight + Cents & \emph{price\_dollars},
\emph{price\_cents}, \emph{weight} & 6 & 6 & 0 \\
Study 2 & List Price & \emph{list\_price} & 3 & 3 & 0 \\
Study 2 & List Price + Roast & \emph{list\_price}, \emph{roast\_date} &
6 & 3 & 3 \\
Study 2 & List Price + Discount & \emph{list\_price},
\emph{discount\_percentage} & 6 & 6 & 0 \\
Study 2 & List Price + Weight & \emph{list\_price}, \emph{weight} & 6 &
6 & 0 \\
Study 2 & List Price + Discount + Weight & \emph{list\_price},
\emph{discount\_percentage}, \emph{weight} & 9 & 9 & 0 \\
\end{longtable}
}

\emph{Note.} Every state reveals the listed attributes for all
alternatives in the environment, so cell counts are the number of
attributes multiplied by two alternatives in Study 1 and three in Study
2. The complete diagnostic set is six cells in the Study 1 environment
and nine in the Study 2 environment.

\subsection{Results}\label{results}

The empirical choice distributions and optimality rates across all
assigned states are reported in
\quartoapptblref{apptbl-fixed-reveal-study1} and
\quartoapptblref{apptbl-fixed-reveal-dist-study1} for Study 1, and in
\quartoapptblref{apptbl-fixed-reveal-study2} and
\quartoapptblref{apptbl-fixed-reveal-dist-study2} for Study 2.

Across both environments, seven of the eight LLMs selected the
normatively optimal alternative most of the time (92\%-100\%) when
provided with the complete diagnostic attribute set under zero cost
(pooled optimality of 99.7 percent and 98.6 percent, respectively). This
performance indicates that within the parameters of this choice task,
these LLMs can execute the comparisons necessary to identify the option
with the lower price per ounce. As a result, their suboptimal choices
observed under acquisition costs in the interactive Tool-Lab studies are
more plausibly explained by the omission of diagnostic attributes during
search than by an inability to evaluate the alternatives when the data
are present.

When diagnostic attributes were withheld, the choices made by these
seven LLMs aligned closely with the normative Expectimax benchmark
derived from empirical category priors. In Study 1, when cents were not
provided, market priors assign symmetric expected cents to both options,
making Coffee A the expected cost-minimizing choice conditional on the
available information. Similarly, in Study 2, when product weight was
not provided, Coffee C was the normative choice because its lower list
price minimizes expected expenditure per ounce. Across all truncated
states, these seven LLMs selected the alternative implied by the
Expectimax policy in 99.3 percent of sessions (2,434 of 2,450 trials)
and 3,128 of 3,150 trials (99.3\%) across all states. This
correspondence suggests that when an LLM selects a lower whole-dollar
price or a lower list price under partial information, that choice is
consistent with the normative policy given missing attributes rather
than necessarily reflecting a utilization failure.

Gemini 3.5 Flash Lite departed from this pattern. When provided with
complete diagnostic information under zero cost, it selected the optimal
alternative in 4 percent of sessions in Study 1 and 0 percent in Study
2. Because all diagnostic attributes were visible in the prompt without
acquisition cost, Flash Lite's selections indicate a failure to utilize
the provided numbers to compute unit prices in this setting.

{\def\LTcaptype{apptbl}
\begin{longtable}[]{@{}
  >{\raggedright\arraybackslash}p{(\linewidth - 8\tabcolsep) * \real{0.1667}}
  >{\centering\arraybackslash}p{(\linewidth - 8\tabcolsep) * \real{0.2083}}
  >{\centering\arraybackslash}p{(\linewidth - 8\tabcolsep) * \real{0.2083}}
  >{\centering\arraybackslash}p{(\linewidth - 8\tabcolsep) * \real{0.2083}}
  >{\centering\arraybackslash}p{(\linewidth - 8\tabcolsep) * \real{0.2083}}@{}}
\caption{{\label{apptbl-fixed-reveal-study1}}Choice optimality by LLM
and assigned information state (Study 1)}\tabularnewline
\toprule\noalign{}
\begin{minipage}[b]{\linewidth}\raggedright
LLM
\end{minipage} & \begin{minipage}[b]{\linewidth}\centering
Dollars
\end{minipage} & \begin{minipage}[b]{\linewidth}\centering
Dollars + Roast
\end{minipage} & \begin{minipage}[b]{\linewidth}\centering
Dollars + Weight
\end{minipage} & \begin{minipage}[b]{\linewidth}\centering
Dollars + Weight + Cents
\end{minipage} \\
\midrule\noalign{}
\endfirsthead
\toprule\noalign{}
\begin{minipage}[b]{\linewidth}\raggedright
LLM
\end{minipage} & \begin{minipage}[b]{\linewidth}\centering
Dollars
\end{minipage} & \begin{minipage}[b]{\linewidth}\centering
Dollars + Roast
\end{minipage} & \begin{minipage}[b]{\linewidth}\centering
Dollars + Weight
\end{minipage} & \begin{minipage}[b]{\linewidth}\centering
Dollars + Weight + Cents
\end{minipage} \\
\midrule\noalign{}
\endhead
\bottomrule\noalign{}
\endlastfoot
Flash Lite & .00 & .00 & .08 & .04 \\
Flash & .00 & .00 & .00 & 1.00 \\
Pro & .00 & .00 & .02 & 1.00 \\
Luna & .00 & .00 & .00 & .98 \\
Terra & .00 & .00 & .24 & 1.00 \\
Sol & .00 & .00 & .00 & 1.00 \\
Sonnet & .00 & .00 & .00 & 1.00 \\
Opus & .00 & .00 & .00 & 1.00 \\
Pooled & .00 & .00 & .04 & .88 \\
Pooled (Excl. Flash Lite) & .00 & .00 & .04 & 1.00 \\
Normative Benchmark & .00 & .00 & .00 & 1.00 \\
\end{longtable}
}

\emph{Note.} N=1,600 independent sessions across eight LLMs from three
providers (n=50 sessions per cell). Entries represent choice optimality,
defined as the proportion of sessions selecting the normatively optimal
alternative (Coffee B: \$5.00 for 11 oz; 0.455 \$/oz) over the lower
whole-dollar alternative (Coffee A: \$4.99 for 10 oz; 0.499 \$/oz).

{\def\LTcaptype{apptbl}
\begin{longtable}[]{@{}
  >{\raggedright\arraybackslash}p{(\linewidth - 10\tabcolsep) * \real{0.1379}}
  >{\centering\arraybackslash}p{(\linewidth - 10\tabcolsep) * \real{0.1724}}
  >{\centering\arraybackslash}p{(\linewidth - 10\tabcolsep) * \real{0.1724}}
  >{\centering\arraybackslash}p{(\linewidth - 10\tabcolsep) * \real{0.1724}}
  >{\centering\arraybackslash}p{(\linewidth - 10\tabcolsep) * \real{0.1724}}
  >{\centering\arraybackslash}p{(\linewidth - 10\tabcolsep) * \real{0.1724}}@{}}
\caption{{\label{apptbl-fixed-reveal-study2}}Choice optimality by LLM
and assigned information state (Study 2)}\tabularnewline
\toprule\noalign{}
\begin{minipage}[b]{\linewidth}\raggedright
LLM
\end{minipage} & \begin{minipage}[b]{\linewidth}\centering
List Price
\end{minipage} & \begin{minipage}[b]{\linewidth}\centering
List Price + Roast
\end{minipage} & \begin{minipage}[b]{\linewidth}\centering
List Price + Discount
\end{minipage} & \begin{minipage}[b]{\linewidth}\centering
List Price + Weight
\end{minipage} & \begin{minipage}[b]{\linewidth}\centering
List Price + Weight + Discount
\end{minipage} \\
\midrule\noalign{}
\endfirsthead
\toprule\noalign{}
\begin{minipage}[b]{\linewidth}\raggedright
LLM
\end{minipage} & \begin{minipage}[b]{\linewidth}\centering
List Price
\end{minipage} & \begin{minipage}[b]{\linewidth}\centering
List Price + Roast
\end{minipage} & \begin{minipage}[b]{\linewidth}\centering
List Price + Discount
\end{minipage} & \begin{minipage}[b]{\linewidth}\centering
List Price + Weight
\end{minipage} & \begin{minipage}[b]{\linewidth}\centering
List Price + Weight + Discount
\end{minipage} \\
\midrule\noalign{}
\endhead
\bottomrule\noalign{}
\endlastfoot
Flash Lite & .00 & .00 & .00 & .24 & .00 \\
Flash & .00 & .00 & .00 & 1.00 & 1.00 \\
Pro & .00 & .00 & .00 & .98 & 1.00 \\
Luna & .00 & .00 & .00 & 1.00 & .92 \\
Terra & .00 & .00 & .00 & .98 & .98 \\
Sol & .00 & .00 & .00 & 1.00 & 1.00 \\
Sonnet & .00 & .00 & .00 & 1.00 & 1.00 \\
Opus & .00 & .00 & .00 & 1.00 & 1.00 \\
Pooled & .00 & .00 & .00 & .90 & .86 \\
Pooled (Excl. Flash Lite) & .00 & .00 & .00 & .99 & .99 \\
Normative Benchmark & .00 & .00 & .00 & 1.00 & 1.00 \\
\end{longtable}
}

\emph{Note.} \(N = 2{,}000\) independent sessions across eight LLMs
(\(n = 50\) sessions per cell). Entries represent choice optimality,
defined as the proportion of sessions selecting the normatively optimal
alternative (\emph{Coffee E}: \$10.00 list price, 0\% discount, 10.9 oz;
\$0.917/oz).\\
\emph{List Price} reveals 3 diagnostic cells (\emph{list\_price} across
three options).\\
\emph{List Price + Roast} reveals 3 diagnostic and 3 non-diagnostic
cells (\emph{list\_price}, \emph{roast\_date}).\\
\emph{List Price + Discount} reveals 6 diagnostic cells
(\emph{list\_price}, \emph{discount\_percentage}).\\
\emph{List Price + Weight} reveals 6 diagnostic cells
(\emph{list\_price}, \emph{weight}).\\
\emph{List Price + Weight + Discount} reveals all 9 diagnostic cells
(\emph{list\_price}, \emph{weight}, \emph{discount\_percentage};
complete diagnostic set).\\
Normative benchmark is evaluated using an Expectimax decision-tree
policy over empirical Walmart coffee priors.\\
\emph{Pooled} reflects all eight LLMs (\(N = 400\) per state).\\
\emph{Pooled (Excl. Flash Lite)} excludes Flash Lite (\(N = 350\) per
state). Values are rounded to two decimal places.

{\def\LTcaptype{apptbl}
\begin{longtable}[]{@{}
  >{\raggedright\arraybackslash}p{(\linewidth - 8\tabcolsep) * \real{0.1667}}
  >{\centering\arraybackslash}p{(\linewidth - 8\tabcolsep) * \real{0.2083}}
  >{\centering\arraybackslash}p{(\linewidth - 8\tabcolsep) * \real{0.2083}}
  >{\centering\arraybackslash}p{(\linewidth - 8\tabcolsep) * \real{0.2083}}
  >{\centering\arraybackslash}p{(\linewidth - 8\tabcolsep) * \real{0.2083}}@{}}
\caption{{\label{apptbl-fixed-reveal-dist-study1}}Choice distributions
by LLM and assigned information state (Study 1)}\tabularnewline
\toprule\noalign{}
\begin{minipage}[b]{\linewidth}\raggedright
LLM
\end{minipage} & \begin{minipage}[b]{\linewidth}\centering
Dollars
\end{minipage} & \begin{minipage}[b]{\linewidth}\centering
Dollars + Roast
\end{minipage} & \begin{minipage}[b]{\linewidth}\centering
Dollars + Weight
\end{minipage} & \begin{minipage}[b]{\linewidth}\centering
Dollars + Weight + Cents
\end{minipage} \\
\midrule\noalign{}
\endfirsthead
\toprule\noalign{}
\begin{minipage}[b]{\linewidth}\raggedright
LLM
\end{minipage} & \begin{minipage}[b]{\linewidth}\centering
Dollars
\end{minipage} & \begin{minipage}[b]{\linewidth}\centering
Dollars + Roast
\end{minipage} & \begin{minipage}[b]{\linewidth}\centering
Dollars + Weight
\end{minipage} & \begin{minipage}[b]{\linewidth}\centering
Dollars + Weight + Cents
\end{minipage} \\
\midrule\noalign{}
\endhead
\bottomrule\noalign{}
\endlastfoot
Flash Lite & 50 / 0 & 50 / 0 & 46 / 4 & 48 / 2 \\
Flash & 50 / 0 & 50 / 0 & 50 / 0 & 0 / 50 \\
Pro & 50 / 0 & 50 / 0 & 49 / 1 & 0 / 50 \\
Luna & 50 / 0 & 50 / 0 & 50 / 0 & 1 / 49 \\
Terra & 50 / 0 & 50 / 0 & 38 / 12 & 0 / 50 \\
Sol & 50 / 0 & 50 / 0 & 50 / 0 & 0 / 50 \\
Sonnet & 50 / 0 & 50 / 0 & 50 / 0 & 0 / 50 \\
Opus & 50 / 0 & 50 / 0 & 50 / 0 & 0 / 50 \\
Pooled & 400 / 0 & 400 / 0 & 383 / 17 & 49 / 351 \\
Pooled (Excl. Flash Lite) & 350 / 0 & 350 / 0 & 337 / 13 & 1 / 349 \\
Normative Benchmark & 50 / 0 & 50 / 0 & 50 / 0 & 0 / 50 \\
\end{longtable}
}

\emph{Note.} \(N = 1{,}600\) independent sessions across eight LLMs from
(\(n = 50\) sessions per cell). Entries report raw choice distributions
in the format: Coffee A / Coffee B. \emph{Coffee A} represents the lower
whole-dollar alternative (\$4.99 for 10 oz; 0.499 \$/oz), and
\emph{Coffee B} represents the normatively optimal alternative (\$5.00
for 11 oz; 0.455 \$/oz). \emph{Dollars} reveals 2 diagnostic cells
(\emph{price\_dollars} across both options). \emph{Dollars + Roast}
reveals 2 diagnostic and 2 non-diagnostic cells (\emph{price\_dollars},
\emph{roast\_date}). \emph{Dollars + Weight} reveals 4 diagnostic cells
(\emph{price\_dollars}, \emph{weight}). \emph{Dollars + Weight + Cents}
reveals all 6 diagnostic cells (\emph{price\_dollars},
\emph{price\_cents}, \emph{weight}; complete diagnostic set).
\emph{Pooled} reflects all eight LLMs (\(N = 400\) per state).
\emph{Pooled (Excl. Flash Lite)} excludes Flash Lite (\(N = 350\) per
state).

{\def\LTcaptype{apptbl}
\begin{longtable}[]{@{}
  >{\raggedright\arraybackslash}p{(\linewidth - 10\tabcolsep) * \real{0.1379}}
  >{\centering\arraybackslash}p{(\linewidth - 10\tabcolsep) * \real{0.1724}}
  >{\centering\arraybackslash}p{(\linewidth - 10\tabcolsep) * \real{0.1724}}
  >{\centering\arraybackslash}p{(\linewidth - 10\tabcolsep) * \real{0.1724}}
  >{\centering\arraybackslash}p{(\linewidth - 10\tabcolsep) * \real{0.1724}}
  >{\centering\arraybackslash}p{(\linewidth - 10\tabcolsep) * \real{0.1724}}@{}}
\caption{{\label{apptbl-fixed-reveal-dist-study2}}Full choice
distributions by LLM and assigned information state (Study 2)}\tabularnewline
\toprule\noalign{}
\begin{minipage}[b]{\linewidth}\raggedright
LLM
\end{minipage} & \begin{minipage}[b]{\linewidth}\centering
List Price
\end{minipage} & \begin{minipage}[b]{\linewidth}\centering
List Price + Roast
\end{minipage} & \begin{minipage}[b]{\linewidth}\centering
List Price + Discount
\end{minipage} & \begin{minipage}[b]{\linewidth}\centering
List Price + Weight
\end{minipage} & \begin{minipage}[b]{\linewidth}\centering
List Price + Weight + Discount
\end{minipage} \\
\midrule\noalign{}
\endfirsthead
\toprule\noalign{}
\begin{minipage}[b]{\linewidth}\raggedright
LLM
\end{minipage} & \begin{minipage}[b]{\linewidth}\centering
List Price
\end{minipage} & \begin{minipage}[b]{\linewidth}\centering
List Price + Roast
\end{minipage} & \begin{minipage}[b]{\linewidth}\centering
List Price + Discount
\end{minipage} & \begin{minipage}[b]{\linewidth}\centering
List Price + Weight
\end{minipage} & \begin{minipage}[b]{\linewidth}\centering
List Price + Weight + Discount
\end{minipage} \\
\midrule\noalign{}
\endhead
\bottomrule\noalign{}
\endlastfoot
Flash Lite & 50 / 0 / 0 & 50 / 0 / 0 & 11 / 39 / 0 & 32 / 6 / 12 & 4 /
46 / 0 \\
Flash & 50 / 0 / 0 & 50 / 0 / 0 & 50 / 0 / 0 & 0 / 0 / 50 & 0 / 0 /
50 \\
Pro & 50 / 0 / 0 & 50 / 0 / 0 & 50 / 0 / 0 & 1 / 0 / 49 & 0 / 0 / 50 \\
Luna & 50 / 0 / 0 & 50 / 0 / 0 & 49 / 1 / 0 & 0 / 0 / 50 & 3 / 1 / 46 \\
Terra & 50 / 0 / 0 & 50 / 0 / 0 & 50 / 0 / 0 & 1 / 0 / 49 & 0 / 1 /
49 \\
Sol & 50 / 0 / 0 & 50 / 0 / 0 & 50 / 0 / 0 & 0 / 0 / 50 & 0 / 0 / 50 \\
Sonnet & 50 / 0 / 0 & 50 / 0 / 0 & 50 / 0 / 0 & 0 / 0 / 50 & 0 / 0 /
50 \\
Opus & 50 / 0 / 0 & 50 / 0 / 0 & 50 / 0 / 0 & 0 / 0 / 50 & 0 / 0 / 50 \\
Pooled & 400 / 0 / 0 & 400 / 0 / 0 & 360 / 40 / 0 & 34 / 6 / 360 & 7 /
48 / 345 \\
Pooled (Excl. Flash Lite) & 350 / 0 / 0 & 350 / 0 / 0 & 349 / 1 / 0 & 2
/ 0 / 348 & 3 / 2 / 345 \\
Normative Benchmark & 50 / 0 / 0 & 50 / 0 / 0 & 50 / 0 / 0 & 0 / 0 / 50
& 0 / 0 / 50 \\
\end{longtable}
}

\emph{Note.} \(N = 2{,}000\) independent sessions across eight LLMs
(\(n = 50\) sessions per cell). Entries report raw choice distributions
in the format: Coffee C / Coffee D / Coffee E. \emph{Coffee C}
represents nominal sticker price minimization (\$8.00 list price, 0\%
discount, 8.0 oz; \$1.000/oz). \emph{Coffee D} represents the
promotional discount heuristic cue (\$15.00 list price, 20\% discount,
12.0 oz; \$1.000/oz). \emph{Coffee E} represents the normatively optimal
alternative (\$10.00 list price, 0\% discount, 10.9 oz; \$0.917/oz).
\emph{List Price} reveals 3 diagnostic cells (\emph{list\_price} across
three options). \emph{List Price + Roast} reveals 3 diagnostic and 3
non-diagnostic cells (\emph{list\_price}, \emph{roast\_date}).
\emph{List Price + Discount} reveals 6 diagnostic cells
(\emph{list\_price}, \emph{discount\_percentage}). \emph{List Price +
Weight} reveals 6 diagnostic cells (\emph{list\_price}, \emph{weight}).
\emph{List Price + Weight + Discount} reveals all 9 diagnostic cells
(\emph{list\_price}, \emph{weight}, \emph{discount\_percentage};
complete diagnostic set). \emph{Pooled} reflects all eight LLMs
(\(N = 400\) per state). \emph{Pooled (Excl. Flash Lite)} excludes Flash
Lite (\(N = 350\) per state).

For the 50 sessions in every cell, we repeated sampling until the LLM
returned a valid \emph{submit\_choice} invocation. In some cases (e.g.,
gemini-3.1-pro-preview) the LLM returned a empty tool call arguments
with no choice, so we resampled those instances.

Two differences from the main experiments should be noted, since the
prompt is not byte-identical to the corresponding Tool-Lab conditions.
First, the two sentences describing the acquisition cost and the
\emph{inspect\_cell} tool are absent from the participant profile, but
their removal is required by the design. Second, the revealed cells
appear as payloads within the initial user turn rather than as tool
results returned across successive turns. However, this choice was not
ours because LLM APIs do not allow fictional tool calls to be inserted
in the conversation history.

\section*{Web appendix G: Reported
valuations}\label{sec-appendix-valuations}
\addcontentsline{toc}{section}{Web appendix G: Reported valuations}

In the Tool-Lab experiments, we measured information acquisitions and
choice, but did not measure information synthesis before the choice was
made. When the AI agent selects a suboptimal alternative, we could not
adjudicate whether it was able to calculate the unit price or not.
Observed choice failures may stem from an arithmetic failure, wherein an
LLM lacks the computational capacity to combine multi-attribute
numerical parameters into an accurate unit value. Alternatively, the
failure may reflect a bias from the training data, wherein an LLM is
able to compute the correct unit value but overrides that calculation in
favor of an normatively irrelevant cue, such as a lower nominal
whole-dollar integer or a promotional percentage badge.

Additionally, when an LLM truncates its search under acquisition costs
and omits diagnostic attributes, choice data cannot determine whether
the AI agent is operating with awareness of missing information or
relying on ungrounded assumptions. An LLM might recognize that
uninspected attributes are absent and make a reasoned inference
conditional on that partial information, or it might hallucinate precise
numerical values for cells it never inspected.

To resolve these ambiguities, we designed an experimental condition that
requires LLMs to report their intermediate attribute calculations for
every alternative at the moment of choice. We implemented this
requirement by modifying the parameter schema of the
\emph{submit\_choice} tool. We require the LLM to output its computed
metrics when finalizing its decision.

\subsection{Design}\label{design}

We evaluated this condition across both studies 1 and 2. The Study 1
environment comprises a two-alternative choice set between Coffee A
(\$4.99 for 10 oz; 0.499 \$/oz, or 2.004 oz/\$) and Coffee B (\$5.00 for
11 oz; 0.455 \$/oz, or 2.200 oz/\$), with a diagnostic set of six cells
(\emph{price\_dollars}, \emph{price\_cents}, and \emph{weight} for each
alternative). The Study 2 environment comprises a three-alternative
choice set between Coffee C (\$8.00 list price, 0\% discount, 8.0 oz;
1.000 \$/oz), Coffee D (\$15.00 list price, 20\% discount, 12.0 oz;
1.000 \$/oz), and Coffee E (\$10.00 list price, 0\% discount, 10.9 oz;
0.917 \$/oz, or 1.090 oz/\$), with a diagnostic set of nine cells
(\emph{list\_price}, \emph{discount\_percentage}, and \emph{weight}
across the three alternatives).

For Study 1, we implemented a 2 (Acquisition cost: \$0.00 vs.~\$10.00
per \emph{inspect\_cell} tool call), and for Study 2, we implemented a 2
(Acquisition cost: \$0.00 vs.~\$50.00 per \emph{inspect\_cell} tool
call) between-sessions design under the vague goal prompt. We evaluated
the same eight LLMs as studies 1 and 2. Each cell was evaluated across
50 independent sessions with zero state retention between trials,
yielding 800 unique sessions per study and 1,600 sessions in total.

\subsection{Methodological
considerations}\label{methodological-considerations}

Measuring intermediate synthesis in LLMs introduces a methodological
trade-off between concurrent and retrospective elicitation. One
alternative design would be a post-choice query, wherein the AI agent
completes its purchase using the standard choice tool, and the
researcher queries the LLM in a subsequent conversational turn to
calculate the price per ounce for each alternative. We did not adopt
this post-choice elicitation design because it creates a retrospective
rationalization confound. Asking an LLM to compute unit prices after it
has already selected a product evaluates its capacity to solve an
arithmetic task, but a successful post-choice calculation cannot
establish whether those unit values were computed and utilized during
the actual purchase decision.

To ensure that reported metrics reflect the information state and
calculations bound directly to the choice act, we required the LLMs to
report their valuations concurrently with choice, within the parameter
schema of the \emph{submit\_choice} tool. However, concurrent schema
reporting introduces two structural considerations. First, because tool
schemas are declared to the LLM at the beginning of the session, initial
exposure to fields such as \emph{oz\_per\_usd} and
\emph{net\_price\_usd} can prime the AI agent regarding which attributes
are diagnostic and attenuate the ambiguity of the vague goal prompt.
Second, requiring the LLM to emit calculated numerical values within the
tool call argument can act as a chain-of-thought scratchpad. Generating
the unit price conditions the subsequent selection of the
\emph{option\_id}, which can artificially reduce suboptimal choices.

Because neither concurrent tool-schema modification nor post-choice
elicitation is free from behavioral side effects, the results of this
experiment should be interpreted solely as a test of whether LLMs can
compute the arithmetic, not how they behave under resource constraints.
The latter is already done and reported in the main experiments.

\subsection{Tool schema}\label{tool-schema}

The tool schema was adapted across the two study environments to reflect
their respective diagnostic attribute structures, as detailed in
\quartoapptblref{apptbl-valuations-schema}. In both environments, the
schema requires the identifier of the chosen product via
\emph{option\_id} and a complete array of valuation objects
corresponding to every alternative in the choice set, encompassing both
chosen and unchosen options.

In the Study 1 environment, where the diagnostic information synthesis
centers on the integration of whole dollars and fractional price
endings, each valuation object requires six specific fields:
\emph{price\_dollars}, \emph{price\_cents}, \emph{price\_usd} (total
price computed as dollars plus cents/100), \emph{weight\_oz},
\emph{oz\_per\_usd} (\emph{weight\_oz} divided by \emph{price\_usd}),
and \emph{basis}.

In the Study 2 environment, where the diagnostic information synthesis
centers on evaluating discounts, the valuation includes
\emph{list\_price}, \emph{discount\_percentage}, and
\emph{net\_price\_usd} (the effective price computed as list price minus
discount amount), alongside \emph{weight\_oz}, \emph{oz\_per\_usd}
(\emph{weight\_oz} divided by \emph{net\_price\_usd}), and \emph{basis}.

Across both environments, the string value \emph{``unknown''} is
designated as admissible across all attribute and computation fields,
allowing the LLMs to acknowledge uninspected cells rather than being
forced to fabricate numbers. Furthermore, the enumerated categorical
indicator \emph{basis} constrains the LLM to classify whether its
reported \emph{oz\_per\_usd} was calculated from attributes inspected
during the session (\emph{``computed''}) or whether the metric could not
be determined due to missing attributes
(\emph{``insufficient\_information''}).

{\def\LTcaptype{apptbl}
\begin{longtable}[]{@{}
  >{\raggedright\arraybackslash}p{(\linewidth - 8\tabcolsep) * \real{0.1714}}
  >{\raggedright\arraybackslash}p{(\linewidth - 8\tabcolsep) * \real{0.3714}}
  >{\raggedright\arraybackslash}p{(\linewidth - 8\tabcolsep) * \real{0.0857}}
  >{\centering\arraybackslash}p{(\linewidth - 8\tabcolsep) * \real{0.1143}}
  >{\raggedright\arraybackslash}p{(\linewidth - 8\tabcolsep) * \real{0.2571}}@{}}
\caption{{\label{apptbl-valuations-schema}}Schema specifications for
the terminal choice and valuation reporting tool (\emph{submit\_choice})
across Studies 1 and 2}\tabularnewline
\toprule\noalign{}
\begin{minipage}[b]{\linewidth}\raggedright
Environment
\end{minipage} & \begin{minipage}[b]{\linewidth}\raggedright
Field
\end{minipage} & \begin{minipage}[b]{\linewidth}\raggedright
Type
\end{minipage} & \begin{minipage}[b]{\linewidth}\centering
Required
\end{minipage} & \begin{minipage}[b]{\linewidth}\raggedright
Description
\end{minipage} \\
\midrule\noalign{}
\endfirsthead
\toprule\noalign{}
\begin{minipage}[b]{\linewidth}\raggedright
Environment
\end{minipage} & \begin{minipage}[b]{\linewidth}\raggedright
Field
\end{minipage} & \begin{minipage}[b]{\linewidth}\raggedright
Type
\end{minipage} & \begin{minipage}[b]{\linewidth}\centering
Required
\end{minipage} & \begin{minipage}[b]{\linewidth}\raggedright
Description
\end{minipage} \\
\midrule\noalign{}
\endhead
\bottomrule\noalign{}
\endlastfoot
\textbf{Shared fields} & \emph{option\_id} & \emph{string} & Yes &
Identifier of the selected alternative. \\
& \emph{valuations} & \emph{array} & Yes & Array containing one object
for every alternative in the choice set. \\
& \emph{valuations{[}{]}.option\_id} & \emph{string} & Yes & Identifier
of the specific alternative being evaluated. \\
& \emph{valuations{[}{]}.weight\_oz} & \emph{string} & Yes & Product
weight inspected for this option, or \emph{``unknown''} if not
inspected. \\
& \emph{valuations{[}{]}.oz\_per\_usd} & \emph{string} & Yes &
\emph{weight\_oz} divided by total or net price, or \emph{``unknown''}
if uncomputable. \\
& \emph{valuations{[}{]}.basis} & \emph{string} & Yes & Categorical
classification constrained to \emph{``computed''} (if calculated from
inspected cells) or \emph{``insufficient\_information''} (if diagnostic
cells were missing). \\
\textbf{Study 1 specific} & \emph{valuations{[}{]}.price\_dollars} &
\emph{string} & Yes & Whole-dollar amount inspected, or
\emph{``unknown''} if not inspected. \\
& \emph{valuations{[}{]}.price\_cents} & \emph{string} & Yes & Cents
amount inspected, or \emph{``unknown''} if not inspected. \\
& \emph{valuations{[}{]}.price\_usd} & \emph{string} & Yes & Total price
as dollars plus cents/100, or \emph{``unknown''} if uncomputable. \\
\textbf{Study 2 specific} & \emph{valuations{[}{]}.list\_price} &
\emph{string} & Yes & List price inspected, or \emph{``unknown''} if not
inspected. \\
& \emph{valuations{[}{]}.discount\_percentage} & \emph{string} & Yes &
Promotional discount percentage inspected, or \emph{``unknown''} if not
inspected. \\
& \emph{valuations{[}{]}.net\_price\_usd} & \emph{string} & Yes & Final
price computed as list price minus discounts, or \emph{``unknown''} if
uncomputable. \\
\end{longtable}
}

\emph{Note.} All fields are required by the tool parameter specification
across all alternatives. The admissibility of the string
\emph{``unknown''} across numerical fields allows LLMs to explicitly
report uninspected attributes and uncomputable values.

To measure the validity of the reported valuations, every field emitted
in the \emph{submit\_choice} payload was cross-referenced against the
execution log of \emph{inspect\_cell} calls from that session. This
evaluation crosses information sufficiency (whether the prerequisite
diagnostic cells were inspected) with the LLM's reporting (whether it
emitted a numerical value or reported \emph{``unknown''}).

Information sufficiency was determined by the arithmetic requirements of
each attribute. For base attributes (\emph{price\_dollars},
\emph{price\_cents}, \emph{list\_price}, \emph{discount\_percentage},
and \emph{weight}), sufficiency required that the specific
product-attribute cell had been inspected during the session. For
derived price metrics (\emph{price\_usd} and \emph{net\_price\_usd}),
sufficiency required that all component price cells had been inspected.
For the terminal unit volume metric (\emph{oz\_per\_usd}), sufficiency
required that all component price cells and the product weight cell had
been inspected.

Cross-referencing sufficiency against reported behavior defines five
mutually exclusive outcomes. Correct (value) is when the diagnostic
cells were inspected, and the reported value matched the ground-truth
value or formula within tolerance (\(\text{tol} = 0.02\)). Correct
(unknown) is when the diagnostic cells were uninspected, and the LLM
accurately acknowledged the absence of data by reporting
\emph{``unknown''}. Arithmetic failure is when the LLM inspected all
required attributes but could not calculate the correct value (e.g.,
miscalculating \(\frac{\text{Weight}}{\text{Price}}\) or subtracting
discounts incorrectly). Hallucination happens when the LLM did not have
the information, but failed to report \emph{``unknown''} and instead
fabricated an ungrounded value. Incorrect (unknown) is the case when the
LLM inspected all required attributes but omitted the value by reporting
\emph{``unknown''}.

\subsection{Results}\label{results-1}

The field-level audits across Study 1
(\quartoapptblref{apptbl-valuations-study1}) and Study 2
(\quartoapptblref{apptbl-valuations-study2}) evaluate whether the LLMs
possess the arithmetic capability to perform the required calculations
to derive unit price and accurately track missing attributes.

First, the results show that arithmetic failure is 0.0\% across all
eight LLMs, both study environments, and all cost conditions. When
diagnostic attributes were inspected, every LLM computed total dollar
prices, promotional net prices, and unit values
(\(\frac{\text{Weight}}{\text{Price}}\)) without mathematical error.
Furthermore, \emph{Incorrect (unknown)} was 0.0\% across all conditions,
indicating that the LLMs did not misclassify inspected attributes as
uninspected. These findings confirm that when prompted to report
valuations, the evaluated LLMs possess the arithmetic capability to
execute the required calculations.

Second, the results indicate that when search truncation occurs, LLMs
exhibit high fidelity to their information states, generally do not
fabricate valuations. Minor exceptions were Anthropic LLMs, specifically
Claude Sonnet 5 and Claude Opus 5 under costly inspection in Study 1
(0.8\%), and Claude Opus 5 in Study 2 under the \$50.00 cost condition
(8.5\%). For all other LLMs, uninspected cells were correctly identified
as unknown (\emph{Correct (unknown)}).

Third, within this specific reporting task, the results show that
accurate arithmetic computation does not necessarily determine choice.
In Study 1, Gemini 3.5 Flash Lite computed the unit values of Coffee B
(2.200 oz/\$) and Coffee A (2.004 oz/\$) with 100.0\% accuracy, yet
selected Coffee A in 22.0\% of sessions under zero cost and 20.0\% under
costly acquisition. In Study 2, despite computing net unit prices with
100.0\% accuracy, Flash Lite selected the optimal alternative (Coffee E)
in 24.0\% of sessions under zero cost and 34.0\% under costly
acquisition.

In summary, this experiment demonstrates that the evaluated LLMs have
the arithmetic capability to calculate multi-attribute pricing metrics
and accurately report attribute missingness when structured to do so.

{\def\LTcaptype{apptbl}
\begin{longtable}[]{@{}
  >{\raggedright\arraybackslash}p{(\linewidth - 14\tabcolsep) * \real{0.0952}}
  >{\raggedright\arraybackslash}p{(\linewidth - 14\tabcolsep) * \real{0.0476}}
  >{\raggedleft\arraybackslash}p{(\linewidth - 14\tabcolsep) * \real{0.1429}}
  >{\raggedleft\arraybackslash}p{(\linewidth - 14\tabcolsep) * \real{0.1429}}
  >{\raggedleft\arraybackslash}p{(\linewidth - 14\tabcolsep) * \real{0.1429}}
  >{\raggedleft\arraybackslash}p{(\linewidth - 14\tabcolsep) * \real{0.1429}}
  >{\raggedleft\arraybackslash}p{(\linewidth - 14\tabcolsep) * \real{0.1429}}
  >{\raggedleft\arraybackslash}p{(\linewidth - 14\tabcolsep) * \real{0.1429}}@{}}
\caption{{\label{apptbl-valuations-study1}}Field-level reporting
accuracy and choice optimality in the reported valuations experiment
(Study 1)}\tabularnewline
\toprule\noalign{}
\begin{minipage}[b]{\linewidth}\raggedright
LLM
\end{minipage} & \begin{minipage}[b]{\linewidth}\raggedright
Cost
\end{minipage} & \begin{minipage}[b]{\linewidth}\raggedleft
Correct (value)
\end{minipage} & \begin{minipage}[b]{\linewidth}\raggedleft
Correct (unknown)
\end{minipage} & \begin{minipage}[b]{\linewidth}\raggedleft
Arithmetic failure
\end{minipage} & \begin{minipage}[b]{\linewidth}\raggedleft
Hallucination
\end{minipage} & \begin{minipage}[b]{\linewidth}\raggedleft
Incorrect (unknown)
\end{minipage} & \begin{minipage}[b]{\linewidth}\raggedleft
Choice Optimality (\%)
\end{minipage} \\
\midrule\noalign{}
\endfirsthead
\toprule\noalign{}
\begin{minipage}[b]{\linewidth}\raggedright
LLM
\end{minipage} & \begin{minipage}[b]{\linewidth}\raggedright
Cost
\end{minipage} & \begin{minipage}[b]{\linewidth}\raggedleft
Correct (value)
\end{minipage} & \begin{minipage}[b]{\linewidth}\raggedleft
Correct (unknown)
\end{minipage} & \begin{minipage}[b]{\linewidth}\raggedleft
Arithmetic failure
\end{minipage} & \begin{minipage}[b]{\linewidth}\raggedleft
Hallucination
\end{minipage} & \begin{minipage}[b]{\linewidth}\raggedleft
Incorrect (unknown)
\end{minipage} & \begin{minipage}[b]{\linewidth}\raggedleft
Choice Optimality (\%)
\end{minipage} \\
\midrule\noalign{}
\endhead
\bottomrule\noalign{}
\endlastfoot
Flash Lite & \$0 & 100.0 & 0.0 & 0.0 & 0.0 & 0.0 & 78.0 \\
Flash Lite & \$10 & 100.0 & 0.0 & 0.0 & 0.0 & 0.0 & 80.0 \\
Flash & \$0 & 100.0 & 0.0 & 0.0 & 0.0 & 0.0 & 100.0 \\
Flash & \$10 & 51.2 & 48.8 & 0.0 & 0.0 & 0.0 & 40.0 \\
Pro & \$0 & 100.0 & 0.0 & 0.0 & 0.0 & 0.0 & 100.0 \\
Pro & \$10 & 70.0 & 30.0 & 0.0 & 0.0 & 0.0 & 82.0 \\
Luna & \$0 & 100.0 & 0.0 & 0.0 & 0.0 & 0.0 & 100.0 \\
Luna & \$10 & 100.0 & 0.0 & 0.0 & 0.0 & 0.0 & 100.0 \\
Terra & \$0 & 100.0 & 0.0 & 0.0 & 0.0 & 0.0 & 100.0 \\
Terra & \$10 & 96.0 & 4.0 & 0.0 & 0.0 & 0.0 & 98.0 \\
Sol & \$0 & 100.0 & 0.0 & 0.0 & 0.0 & 0.0 & 100.0 \\
Sol & \$10 & 96.4 & 3.6 & 0.0 & 0.0 & 0.0 & 92.0 \\
Sonnet & \$0 & 100.0 & 0.0 & 0.0 & 0.0 & 0.0 & 100.0 \\
Sonnet & \$10 & 98.8 & 0.4 & 0.0 & 0.8 & 0.0 & 98.0 \\
Opus & \$0 & 100.0 & 0.0 & 0.0 & 0.0 & 0.0 & 100.0 \\
Opus & \$10 & 45.4 & 53.8 & 0.0 & 0.8 & 0.0 & 6.0 \\
\end{longtable}
}

\emph{Note.} \(N = 800\) sessions across eight LLMs from three providers
(\(n = 50\) independent Monte Carlo sessions per cell). Each session
evaluated 10 field-level classifications across alternatives. Rows sum
to 100\% across the four classification categories. \emph{Correct
(value)} reflects diagnostic attributes inspected and values correctly
recalled or computed within tolerance (tol = 0.02). \emph{Correct
(unknown)} reflects uninspected cells appropriately acknowledged as
unknown. \emph{Hallucination} captures calculation errors on inspected
attributes and ungrounded numerical assertions on uninspected
attributes. \emph{Incorrect (unknown)} captures omission errors where
attributes were inspected but reported as unknown. \emph{Choice
Optimality (\%)} is the percentage of sessions selecting the normatively
optimal alternative (Coffee B: \$5.00 for 11 oz; 0.455 \$/oz).

{\def\LTcaptype{apptbl}
\begin{longtable}[]{@{}
  >{\raggedright\arraybackslash}p{(\linewidth - 14\tabcolsep) * \real{0.0833}}
  >{\raggedright\arraybackslash}p{(\linewidth - 14\tabcolsep) * \real{0.0417}}
  >{\raggedleft\arraybackslash}p{(\linewidth - 14\tabcolsep) * \real{0.1458}}
  >{\raggedleft\arraybackslash}p{(\linewidth - 14\tabcolsep) * \real{0.1458}}
  >{\raggedleft\arraybackslash}p{(\linewidth - 14\tabcolsep) * \real{0.1458}}
  >{\raggedleft\arraybackslash}p{(\linewidth - 14\tabcolsep) * \real{0.1458}}
  >{\raggedleft\arraybackslash}p{(\linewidth - 14\tabcolsep) * \real{0.1458}}
  >{\raggedleft\arraybackslash}p{(\linewidth - 14\tabcolsep) * \real{0.1458}}@{}}
\caption{{\label{apptbl-valuations-study2}}Field-level reporting
accuracy and choice optimality in the reported valuations experiment
(Study 2)}\tabularnewline
\toprule\noalign{}
\begin{minipage}[b]{\linewidth}\raggedright
LLM
\end{minipage} & \begin{minipage}[b]{\linewidth}\raggedright
Cost
\end{minipage} & \begin{minipage}[b]{\linewidth}\raggedleft
Correct (value)
\end{minipage} & \begin{minipage}[b]{\linewidth}\raggedleft
Correct (unknown)
\end{minipage} & \begin{minipage}[b]{\linewidth}\raggedleft
Arithmetic failure
\end{minipage} & \begin{minipage}[b]{\linewidth}\raggedleft
Hallucination
\end{minipage} & \begin{minipage}[b]{\linewidth}\raggedleft
Incorrect (unknown)
\end{minipage} & \begin{minipage}[b]{\linewidth}\raggedleft
Choice Optimality (\%)
\end{minipage} \\
\midrule\noalign{}
\endfirsthead
\toprule\noalign{}
\begin{minipage}[b]{\linewidth}\raggedright
LLM
\end{minipage} & \begin{minipage}[b]{\linewidth}\raggedright
Cost
\end{minipage} & \begin{minipage}[b]{\linewidth}\raggedleft
Correct (value)
\end{minipage} & \begin{minipage}[b]{\linewidth}\raggedleft
Correct (unknown)
\end{minipage} & \begin{minipage}[b]{\linewidth}\raggedleft
Arithmetic failure
\end{minipage} & \begin{minipage}[b]{\linewidth}\raggedleft
Hallucination
\end{minipage} & \begin{minipage}[b]{\linewidth}\raggedleft
Incorrect (unknown)
\end{minipage} & \begin{minipage}[b]{\linewidth}\raggedleft
Choice Optimality (\%)
\end{minipage} \\
\midrule\noalign{}
\endhead
\bottomrule\noalign{}
\endlastfoot
Flash Lite & \$0 & 100.0 & 0.0 & 0.0 & 0.0 & 0.0 & 24.0 \\
Flash Lite & \$50 & 100.0 & 0.0 & 0.0 & 0.0 & 0.0 & 34.0 \\
Flash & \$0 & 100.0 & 0.0 & 0.0 & 0.0 & 0.0 & 100.0 \\
Flash & \$50 & 0.8 & 99.2 & 0.0 & 0.0 & 0.0 & 34.0 \\
Pro & \$0 & 100.0 & 0.0 & 0.0 & 0.0 & 0.0 & 100.0 \\
Pro & \$50 & 50.0 & 50.0 & 0.0 & 0.0 & 0.0 & 68.0 \\
Luna & \$0 & 100.0 & 0.0 & 0.0 & 0.0 & 0.0 & 94.0 \\
Luna & \$50 & 100.0 & 0.0 & 0.0 & 0.0 & 0.0 & 92.0 \\
Terra & \$0 & 100.0 & 0.0 & 0.0 & 0.0 & 0.0 & 98.0 \\
Terra & \$50 & 32.5 & 67.5 & 0.0 & 0.0 & 0.0 & 46.0 \\
Sol & \$0 & 100.0 & 0.0 & 0.0 & 0.0 & 0.0 & 100.0 \\
Sol & \$50 & 27.3 & 72.7 & 0.0 & 0.0 & 0.0 & 46.0 \\
Sonnet & \$0 & 100.0 & 0.0 & 0.0 & 0.0 & 0.0 & 100.0 \\
Sonnet & \$50 & 100.0 & 0.0 & 0.0 & 0.0 & 0.0 & 100.0 \\
Opus & \$0 & 100.0 & 0.0 & 0.0 & 0.0 & 0.0 & 100.0 \\
Opus & \$50 & 84.0 & 7.5 & 0.0 & 8.5 & 0.0 & 98.0 \\
\end{longtable}
}

\emph{Note.} \(N = 800\) sessions across eight LLMs from three providers
(\(n = 50\) independent Monte Carlo sessions per cell). Each session
evaluated 15 field-level classifications across alternatives. Rows sum
to 100\% across the five classification categories. \emph{Correct
(value)} reflects diagnostic attributes inspected and values correctly
recalled or computed within tolerance (tol = 0.02). \emph{Correct
(unknown)} reflects uninspected cells appropriately acknowledged as
unknown. \emph{Arithmetic failure} captures attributes that were
inspected but calculated or recalled incorrectly. \emph{Hallucination}
captures uninspected attributes asserted with ungrounded numerical
values. \emph{Incorrect (unknown)} captures omission errors where
attributes were inspected but reported as unknown. \emph{Choice
Optimality (\%)} is the percentage of sessions selecting the normatively
optimal alternative (Coffee E: \$10.00 list price, 0\% discount, 10.9
oz; 0.917 \$/oz).

\section*{Web appendix H: Expectimax normative
benchmarking}\label{sec-appendix-expectimax}
\addcontentsline{toc}{section}{Web appendix H: Expectimax normative
benchmarking}

We benchmark observed LLM behavior against a normative Expectimax
decision-tree policy. Expectimax approximates the decision-theoretic
solution to a Partially Observable Markov Decision Process (POMDP) under
resource scarcity \citep{russell2010artificial}. By recursively
computing the Expected Value of Information (EVI) for each available
tool call, the algorithm identifies the point at which the marginal
informational benefit of inspecting an additional attribute no longer
justifies the compounding financial search cost.

\subsection{The shopping task}\label{the-shopping-task}

The decision environment casts the surrogate consumer in an electronic
retailing choice task over a set of alternatives
\(\mathcal{J} = \{1, \dots, J\}\). The surrogate consumer is delegated
to select a single alternative \(j^* \in \mathcal{J}\) on behalf of a
human consumer. The human consumer will subsequently purchase \(Q = 5\)
units (bags) of the selected product.

Information acquisition is operationalized through discrete tool
invocations. The surrogate consumer begins in an uninformative state
where product attributes are occluded behind closed cells. Revealing an
attribute cell incurs an explicit financial penalty
\(c \in \{\$0.00, \$0.01, \$0.10, \$1.00, \$10.00\}\) for Study 1 and
\(c \in \{\$0.00, \$0.01, \$0.10, \$1.00, \$50.00\}\) for Study 2,
charged to the delegating consumer. Submitting a final choice incurs
zero cost.

\subsection{Objective function}\label{objective-function}

A central question in benchmarking information acquisition is how
non-monetary physical attributes (product weight in ounces) trade off
against monetary search penalties (dollars). In our formulation, this
exchange rate is derived from the economics of unit-price minimization.

When alternative \(j\) is chosen, the human consumer incurs two
financial costs: the purchase cost of the goods (\(Q \cdot P_j\)) and
the cumulative tool fees accrued during search (\(C_{\text{tool}}\)). In
return, the consumer receives a total product volume of \(Q \cdot W_j\)
ounces. The net cost per ounce delivered to the consumer is:

\[\text{Effective Unit Cost} = \frac{\text{Total Dollar Cost}}{\text{Total Ounces Delivered}} = \frac{Q \cdot P_j + C_{\text{tool}}}{Q \cdot W_j} = \left( P_j + \frac{C_{\text{tool}}}{Q} \right) \frac{1}{W_j}\]

where \(Q = 5\) is the purchase volume and
\(C_{\text{tool}}(s) = \sum_{k \in \text{inspected}(s)} c_k\) is the
total financial cost incurred across all \emph{inspect\_cell}
invocations executed to reach information state \(s\). \(P_j\) is the
net price per bag. In Study 1,
\(P_j = \text{Dollars}_j + 0.01 \times \text{Cents}_j\). In Study 2,
\(P_j = \text{ListPrice}_j \times \left(1 - \frac{\text{Discount}_j}{100}\right)\).
\(W_j\) is the net product weight in ounces.

Under partial information, price and weight are independently
distributed across alternatives in the market prior. The expected
effective unit cost of selecting alternative \(j\) at state \(s\) is:

\[\mathbb{E}\left[\text{Effective Unit Cost}(j) \mid s\right] = \left( \mathbb{E}[P_j \mid s] + \frac{C_{\text{tool}}(s)}{Q} \right) \cdot \mathbb{E}\left[ \frac{1}{W_j} \;\Bigg|\; s \right]\]

Because the planner maximizes utility, terminal choice value is defined
as the negative of the expected effective unit cost:

\[\mathcal{U}_{\text{choose}}(j, s) = - \left( \mathbb{E}[P_j \mid s] + \frac{C_{\text{tool}}(s)}{Q} \right) \cdot \mathbb{E}\left[ \frac{1}{W_j} \;\Bigg|\; s \right]\]

This specification eliminates arbitrary subjective exchange rates. The
monetary penalty of search (\(C_{\text{tool}}\)) is amortized directly
across the 5-bag purchase quantity (\(C_{\text{tool}} / Q\)),
establishing a dollar-for-ounce valuation that governs the optimal
search threshold.

An information state \(s\) is defined as a tuple of length
\(J \times K\), where \(J\) is the number of alternatives and \(K\) is
the number of attributes per alternative:

\[s = \Big( v_{1,1}, \dots, v_{j,k}, \dots, v_{J,K} \Big), \quad v_{j,k} \in \mathcal{D}_k \cup \{\text{None}\}\]

where \(\mathcal{D}_k\) is the empirical domain of attribute \(k\), and
\(\text{None}\) indicates that the attribute has not been inspected.

At each turn, the legal action space \(\mathcal{A}(s)\) comprises two
mutually exclusive subsets:

\begin{enumerate}
\def\labelenumi{\arabic{enumi}.}
\tightlist
\item
  Reveal actions,
  \(\mathcal{A}_{\text{reveal}}(s) = \{ (\text{reveal}, j, k) \mid s_{j,k} = \text{None} \}\),
  which inspect an unobserved cell.
\item
  Choose actions,
  \(\mathcal{A}_{\text{choose}}(s) = \{ (\text{choose}, j) \mid j \in \mathcal{J} \}\),
  which terminate search and submit a purchase.
\end{enumerate}

\subsection{Priors}\label{priors}

Prior distributions are based on empirical instant coffee attributes
collected from Walmart.com. Whole dollars
(\(\mathcal{D}_{\text{dollars}}\)) are 25 discrete values spanning
\(\$0\) to \(\$53\) (\(M = \$15.96\)). Fractional cents
(\(\mathcal{D}_{\text{cents}}\)) are 45 discrete values spanning \(0¢\)
to \(99¢\) (\(M = 52.93¢\)). Product weight
(\(\mathcal{D}_{\text{weight}}\)) are 57 continuous values spanning
\(0.42\text{ oz}\) to \(71.99\text{ oz}\) (\(M = 10.33\text{ oz}\)).
List price (\(\mathcal{D}_{\text{list\_price}}\)) are 63 continuous
values spanning \(\$0.92\) to \(\$53.84\) (\(M = \$11.87\)). For
discount percentage (\(\mathcal{D}_{\text{discount}}\)), we assume a
discrete uniform distribution \(\mathcal{U}\{0, 40\}\) in 5\% increments
(\(M = 20.0\%\)).

Prior expectations for unobserved attributes are precomputed over their
respective empirical supports:

\[
\begin{aligned}
\mathbb{E}[\text{Dollars}] &= \frac{1}{|\mathcal{D}_{\text{dollars}}|} \sum_{d \in \mathcal{D}_{\text{dollars}}} d, &\quad \mathbb{E}[\text{Cents}] &= \frac{1}{|\mathcal{D}_{\text{cents}}|} \sum_{c \in \mathcal{D}_{\text{cents}}} c, \\
\mathbb{E}[\text{ListPrice}] &= \frac{1}{|\mathcal{D}_{\text{list\_price}}|} \sum_{p \in \mathcal{D}_{\text{list\_price}}} p, &\quad \mathbb{E}[\text{Discount}] &= \frac{1}{|\mathcal{D}_{\text{discount}}|} \sum_{\delta \in \mathcal{D}_{\text{discount}}} \delta
\end{aligned}
\]

For product weight, Jensen's inequality implies
\(\mathbb{E}[1/W] \neq 1/\mathbb{E}[W]\). The planner computes the
expected inverse weight over non-zero domain entries:
\[\mathbb{E}\left[\frac{1}{W}\right] = \frac{1}{|\mathcal{D}_W|} \sum_{w \in \mathcal{D}_W} \frac{1}{w}\]

\subsection{Optimization}\label{optimization}

At each chance node, an uninspected attribute \(k\) can reveal any value
\(x \in \mathcal{D}_k\). When the cardinality of an empirical domain
exceeds the chance enumeration limit (\(L = 10\)), the support is
deterministically quantized into \(L\) representative percentiles using
evenly spaced indices:

\[\mathcal{I}_L = \left\{ \left\lfloor \frac{i \cdot (|\mathcal{D}_k| - 1)}{L - 1} \right\rfloor \;\Bigg|\; i = 0, \dots, L-1 \right\}\]

Each quantized outcome \(x \in \mathcal{D}_k[\mathcal{I}_L]\) is
assigned an equal probability mass \(P(x) = 1/L\).

The value of an information state \(V(s, d)\) at remaining lookahead
depth \(d\) is governed by the Bellman optimality equation:

\[V(s, d) = \max \left( \max_{j \in \mathcal{J}} \mathcal{U}_{\text{choose}}(j, s), \quad \mathbb{I}(d > 0) \cdot \max_{(\text{reveal}, j, k) \in \mathcal{A}_{\text{reveal}}(s)} Q_{\text{reveal}}(s, j, k, d) \right)\]

where the Expected Value of Information for inspecting cell \((j, k)\)
is computed across all chance outcomes \(\Omega(k)\):

\[Q_{\text{reveal}}(s, j, k, d) = \frac{1}{|\Omega(k)|} \sum_{x \in \Omega(k)} V\Big(s \cup \{(j, k) \leftarrow x\}, \; d - 1\Big)\]

We evaluate the normative policy under two planning horizons. The myopic
policy (\(\text{Depth} = 1\)) values each possible acquisition as if it
were the last before choosing. It therefore captures the value of an
attribute on its own but not the value of attributes that are
informative only in combination. The 2-step lookahead policy
(\(\text{Depth} = 2\)) values each possible acquisition by also
considering one further acquisition before choosing. It therefore
captures complementarity between attributes, such as the cents of both
alternatives, or a list price together with its discount percentage.

Under both horizons, the policy is recomputed after each acquisition
from the updated information state. Ties between actions of equal value
are broken at random, so each policy is simulated 100 times per cost
condition.

\subsection{Normative benchmark - Study
1}\label{normative-benchmark---study-1}

In Study 1, the ground-truth choice set pits a lower whole-dollar
alternative (\emph{Coffee A}: \$4.99, 10.0 oz; 0.499 \$/oz) against a
higher whole-dollar optimal alternative (\emph{Coffee B}: \$5.00, 11.0
oz; 0.455 \$/oz).

\quartoapptblref{apptbl-expectimax-study1} summarizes the search depth
and choice optimality of the Expectimax benchmark across acquisition
costs.

{\def\LTcaptype{apptbl}
\begin{longtable}[]{@{}
  >{\raggedright\arraybackslash}p{(\linewidth - 16\tabcolsep) * \real{0.1399}}
  >{\raggedright\arraybackslash}p{(\linewidth - 16\tabcolsep) * \real{0.0909}}
  >{\raggedleft\arraybackslash}p{(\linewidth - 16\tabcolsep) * \real{0.0769}}
  >{\raggedleft\arraybackslash}p{(\linewidth - 16\tabcolsep) * \real{0.0629}}
  >{\raggedleft\arraybackslash}p{(\linewidth - 16\tabcolsep) * \real{0.0699}}
  >{\raggedleft\arraybackslash}p{(\linewidth - 16\tabcolsep) * \real{0.1399}}
  >{\raggedleft\arraybackslash}p{(\linewidth - 16\tabcolsep) * \real{0.1678}}
  >{\raggedleft\arraybackslash}p{(\linewidth - 16\tabcolsep) * \real{0.1049}}
  >{\raggedleft\arraybackslash}p{(\linewidth - 16\tabcolsep) * \real{0.1469}}@{}}
\caption{{\label{apptbl-expectimax-study1}}Expectimax optimal search
depth and choice optimality across acquisition costs (Study 1)}\tabularnewline
\toprule\noalign{}
\begin{minipage}[b]{\linewidth}\raggedright
Lookahead policy
\end{minipage} & \begin{minipage}[b]{\linewidth}\raggedright
Tool cost
\end{minipage} & \begin{minipage}[b]{\linewidth}\raggedleft
Dollars
\end{minipage} & \begin{minipage}[b]{\linewidth}\raggedleft
Cents
\end{minipage} & \begin{minipage}[b]{\linewidth}\raggedleft
Weight
\end{minipage} & \begin{minipage}[b]{\linewidth}\raggedleft
Diagnostic calls
\end{minipage} & \begin{minipage}[b]{\linewidth}\raggedleft
Non-diagnostic calls
\end{minipage} & \begin{minipage}[b]{\linewidth}\raggedleft
Total calls
\end{minipage} & \begin{minipage}[b]{\linewidth}\raggedleft
Choice optimality
\end{minipage} \\
\midrule\noalign{}
\endfirsthead
\toprule\noalign{}
\begin{minipage}[b]{\linewidth}\raggedright
Lookahead policy
\end{minipage} & \begin{minipage}[b]{\linewidth}\raggedright
Tool cost
\end{minipage} & \begin{minipage}[b]{\linewidth}\raggedleft
Dollars
\end{minipage} & \begin{minipage}[b]{\linewidth}\raggedleft
Cents
\end{minipage} & \begin{minipage}[b]{\linewidth}\raggedleft
Weight
\end{minipage} & \begin{minipage}[b]{\linewidth}\raggedleft
Diagnostic calls
\end{minipage} & \begin{minipage}[b]{\linewidth}\raggedleft
Non-diagnostic calls
\end{minipage} & \begin{minipage}[b]{\linewidth}\raggedleft
Total calls
\end{minipage} & \begin{minipage}[b]{\linewidth}\raggedleft
Choice optimality
\end{minipage} \\
\midrule\noalign{}
\endhead
\bottomrule\noalign{}
\endlastfoot
Myopic & \$0.00 & 2.00 & 2.00 & 2.00 & 6.00 & 4.16 & 10.16 & 1.00 \\
Myopic & \$0.01 & 2.00 & 2.00 & 2.00 & 6.00 & 0.00 & 6.00 & 1.00 \\
Myopic & \$0.10 & 2.00 & 0.00 & 2.00 & 4.00 & 0.00 & 4.00 & 0.00 \\
Myopic & \$1.00 & 2.00 & 0.00 & 2.00 & 4.00 & 0.00 & 4.00 & 0.00 \\
Myopic & \$10.00 & 1.00 & 0.00 & 2.00 & 3.00 & 0.00 & 3.00 & 0.00 \\
2-step Lookahead & \$0.00 & 2.00 & 2.00 & 2.00 & 6.00 & 6.09 & 12.09 &
1.00 \\
2-step Lookahead & \$0.01 & 2.00 & 2.00 & 2.00 & 6.00 & 0.00 & 6.00 &
1.00 \\
2-step Lookahead & \$0.10 & 2.00 & 2.00 & 2.00 & 6.00 & 0.00 & 6.00 &
1.00 \\
2-step Lookahead & \$1.00 & 2.00 & 0.00 & 2.00 & 4.00 & 0.00 & 4.00 &
0.00 \\
2-step Lookahead & \$10.00 & 1.00 & 0.00 & 2.00 & 3.00 & 0.00 & 3.00 &
1.00 \\
\end{longtable}
}

\emph{Note.} \(N = 100\) independent simulations per cell. Ground truth
realization: Coffee A (\$4.99, 10 oz) vs.~Coffee B (\$5.00, 11 oz).
Choice optimality is coded 1 if Coffee B is selected.

\begin{appfig}

\centering{

\includegraphics[width=0.85\linewidth,height=\textheight,keepaspectratio]{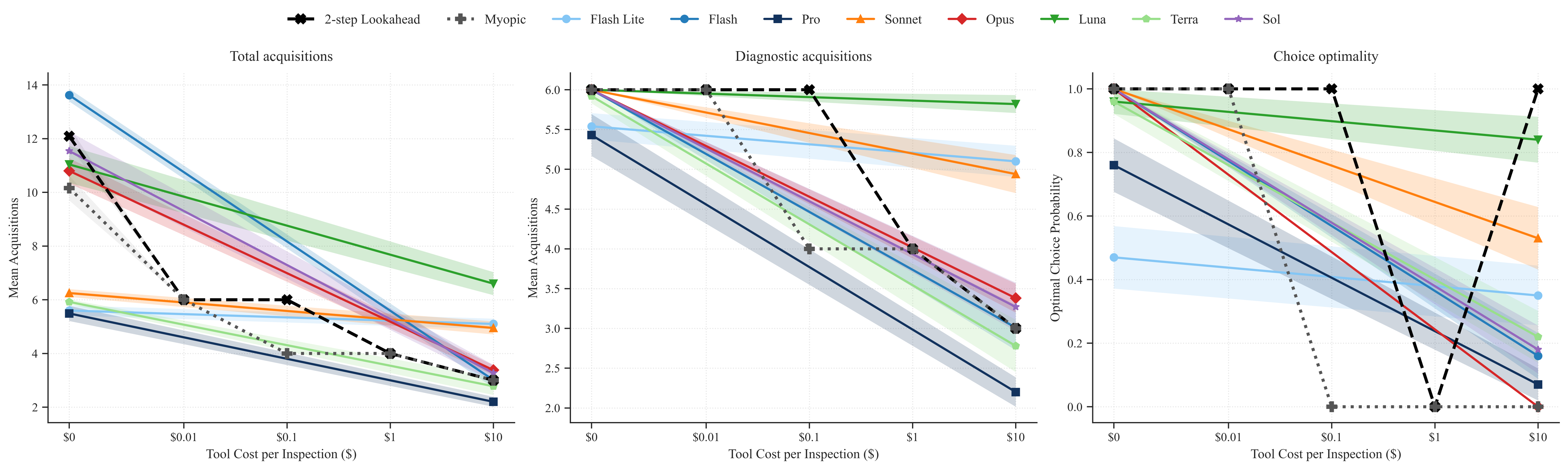}

}

\caption{\label{appfig-expectimax-study1}Information acquisition and
choice optimality across acquisition costs compared against normative
Expectimax benchmarks (Study 1 Tool-Lab).}

\floatnote{\emph{Note.} Left panel: Mean total information acquisitions. Middle
panel: Mean diagnostic acquisitions (dollars, cents, weight; maximum 6).
Right panel: Probability of selecting the optimal alternative
(\emph{Coffee B}: \$5.00, 11 oz, 0.455 \$/oz). Across all panels,
acquisition tool cost
(\(c \in \{\$0.00, \$0.01, \$0.10, \$1.00, \$10.00\}\)) is plotted on
the horizontal axis using a symmetric logarithmic scale. The normative
benchmark policies are represented by dashed and dotted lines: 2-step
Lookahead (black dashed line, Depth 2) and Myopic (gray dotted line,
Depth 1). Empirical LLM trajectories across eight LLM under the vague
prompt are plotted in colored lines. Error bands represent 95\%
confidence intervals across \(N = 100\) independent sessions per LLM and
benchmark cell.}

\end{appfig}%

Given free acquisition (\(c = \$0.00\)) and small cost (\(c = \$0.01\)),
both planning horizons acquire all 6 diagnostic cells (2 dollars, 2
cents, 2 weights) and choose optimally (\(100\%\)). Non-diagnostic cells
are dropped immediately once cost rises to \(\$0.01\) because their EVI
is zero.

Under \$10 cost, (\(c = \$10.00\)) search depth truncates to 3 tool
calls (2 weights, 1 dollar). The 2-step planner halts search and
satisfices on Coffee B (\(\text{Optimality} = 1.00\)).

\subsection{Normative benchmark - Study
2}\label{normative-benchmark---study-2}

In Study 2, the choice environment tests three alternatives where
promotional cues and value point in opposing directions: \emph{Coffee C}
(lowest list price: \$8.00, 0\% discount, 8 oz; 1.00 \$/oz),
\emph{Coffee D} (promotional cue: \$15.00, 20\% discount, 12 oz; 1.00
\$/oz), and \emph{Coffee E} (optimal unit price: \$10.00, 0\% discount,
10.9 oz; 0.917 \$/oz).

\quartoapptblref{apptbl-expectimax-study2} presents the search
trajectories across costs.

{\def\LTcaptype{apptbl}
\begin{longtable}[]{@{}
  >{\raggedright\arraybackslash}p{(\linewidth - 16\tabcolsep) * \real{0.1342}}
  >{\raggedright\arraybackslash}p{(\linewidth - 16\tabcolsep) * \real{0.0872}}
  >{\raggedleft\arraybackslash}p{(\linewidth - 16\tabcolsep) * \real{0.0940}}
  >{\raggedleft\arraybackslash}p{(\linewidth - 16\tabcolsep) * \real{0.0805}}
  >{\raggedleft\arraybackslash}p{(\linewidth - 16\tabcolsep) * \real{0.0671}}
  >{\raggedleft\arraybackslash}p{(\linewidth - 16\tabcolsep) * \real{0.1342}}
  >{\raggedleft\arraybackslash}p{(\linewidth - 16\tabcolsep) * \real{0.1611}}
  >{\raggedleft\arraybackslash}p{(\linewidth - 16\tabcolsep) * \real{0.1007}}
  >{\raggedleft\arraybackslash}p{(\linewidth - 16\tabcolsep) * \real{0.1409}}@{}}
\caption{{\label{apptbl-expectimax-study2}}Expectimax optimal search
depth and choice optimality across acquisition costs (Study 2)}\tabularnewline
\toprule\noalign{}
\begin{minipage}[b]{\linewidth}\raggedright
Lookahead policy
\end{minipage} & \begin{minipage}[b]{\linewidth}\raggedright
Tool cost
\end{minipage} & \begin{minipage}[b]{\linewidth}\raggedleft
List price
\end{minipage} & \begin{minipage}[b]{\linewidth}\raggedleft
Discount
\end{minipage} & \begin{minipage}[b]{\linewidth}\raggedleft
Weight
\end{minipage} & \begin{minipage}[b]{\linewidth}\raggedleft
Diagnostic calls
\end{minipage} & \begin{minipage}[b]{\linewidth}\raggedleft
Non-diagnostic calls
\end{minipage} & \begin{minipage}[b]{\linewidth}\raggedleft
Total calls
\end{minipage} & \begin{minipage}[b]{\linewidth}\raggedleft
Choice optimality
\end{minipage} \\
\midrule\noalign{}
\endfirsthead
\toprule\noalign{}
\begin{minipage}[b]{\linewidth}\raggedright
Lookahead policy
\end{minipage} & \begin{minipage}[b]{\linewidth}\raggedright
Tool cost
\end{minipage} & \begin{minipage}[b]{\linewidth}\raggedleft
List price
\end{minipage} & \begin{minipage}[b]{\linewidth}\raggedleft
Discount
\end{minipage} & \begin{minipage}[b]{\linewidth}\raggedleft
Weight
\end{minipage} & \begin{minipage}[b]{\linewidth}\raggedleft
Diagnostic calls
\end{minipage} & \begin{minipage}[b]{\linewidth}\raggedleft
Non-diagnostic calls
\end{minipage} & \begin{minipage}[b]{\linewidth}\raggedleft
Total calls
\end{minipage} & \begin{minipage}[b]{\linewidth}\raggedleft
Choice optimality
\end{minipage} \\
\midrule\noalign{}
\endhead
\bottomrule\noalign{}
\endlastfoot
Myopic & \$0.00 & 3.00 & 2.22 & 3.00 & 8.22 & 7.38 & 15.60 & 1.00 \\
Myopic & \$0.01 & 3.00 & 1.00 & 3.00 & 7.00 & 0.00 & 7.00 & 1.00 \\
Myopic & \$0.10 & 3.00 & 1.00 & 3.00 & 7.00 & 0.00 & 7.00 & 1.00 \\
Myopic & \$1.00 & 3.00 & 0.00 & 3.00 & 6.00 & 0.00 & 6.00 & 1.00 \\
Myopic & \$50.00 & 0.00 & 0.00 & 0.00 & 0.00 & 0.00 & 0.00 & 0.32 \\
2-step Lookahead & \$0.00 & 3.00 & 2.90 & 3.00 & 8.90 & 10.79 & 19.69 &
1.00 \\
2-step Lookahead & \$0.01 & 3.00 & 2.48 & 3.00 & 8.48 & 0.00 & 8.48 &
1.00 \\
2-step Lookahead & \$0.10 & 3.00 & 2.00 & 3.00 & 8.00 & 0.00 & 8.00 &
1.00 \\
2-step Lookahead & \$1.00 & 3.00 & 1.00 & 3.00 & 7.00 & 0.00 & 7.00 &
1.00 \\
2-step Lookahead & \$50.00 & 0.00 & 0.00 & 1.00 & 1.00 & 0.00 & 1.00 &
0.27 \\
\end{longtable}
}

\emph{Note.} \(N = 100\) independent simulations per cell. Ground truth
realization: Coffee C (\$8.00, 0\%, 8 oz), Coffee D (\$15.00, 20\%, 12
oz), Coffee E (\$10.00, 0\%, 10.9 oz). Choice optimality is coded 1 if
Coffee E is selected.

\begin{appfig}

\centering{

\includegraphics[width=0.85\linewidth,height=\textheight,keepaspectratio]{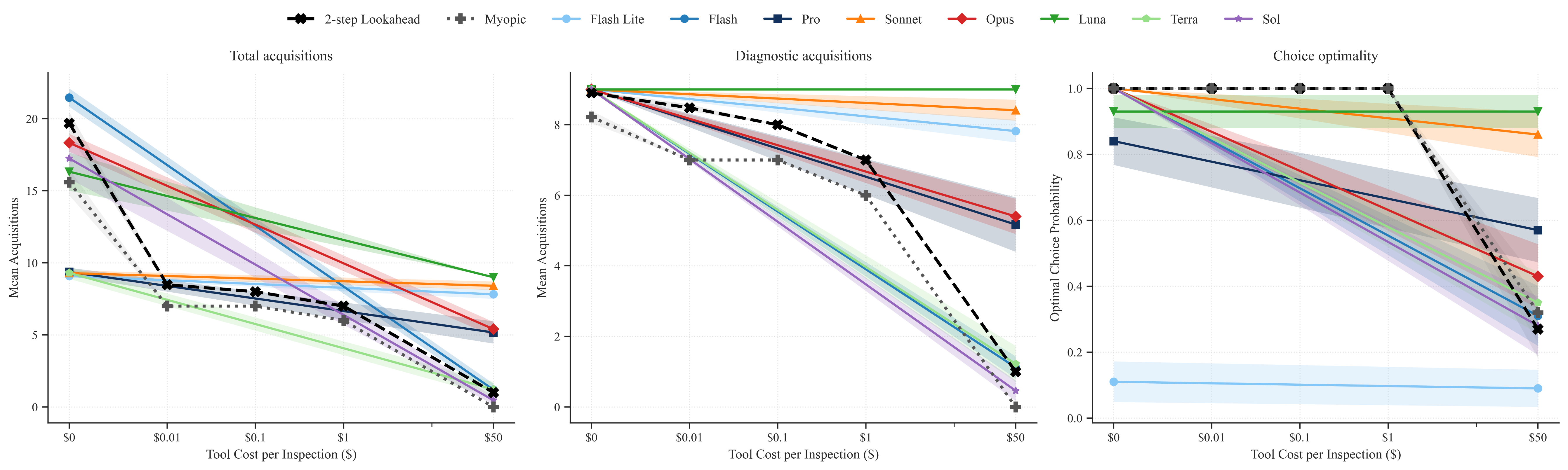}

}

\caption{\label{appfig-expectimax-study2}Promotional information
acquisition and choice optimality across acquisition costs compared
against normative Expectimax benchmarks (Study 2 Tool-Lab).}

\floatnote{\emph{Note.} Left panel: Mean total information acquisitions. Middle
panel: Mean diagnostic acquisitions (list price, discount percentage,
weight; maximum 9). Right panel: Probability of selecting the optimal
alternative (\emph{Coffee E}: \$10.00, 0\%, 10.9 oz, 0.917 \$/oz).
Across all panels, acquisition tool cost
(\(c \in \{\$0.00, \$0.01, \$0.10, \$1.00, \$50.00\}\)) is plotted on
the horizontal axis using a symmetric logarithmic scale. Benchmark
policies are plotted in black (2-step Lookahead, Depth 2) and gray
(Myopic, Depth 1). Empirical LLM data for eight LLMs under the vague
prompt are plotted in colored solid lines. Error bands represent 95\%
confidence intervals across \(N = 100\) independent trials per LLM and
benchmark cell.}

\end{appfig}%

As cost increases from zero to \$50, the optimal policies first truncate
acquisition of discount percentage and then, under higher costs, list
price and weight.

\subsection{Normative benchmark - assigned information
states}\label{normative-benchmark---assigned-information-states}

The exogenously assigned states of Web Appendix F present the surrogate
consumer with a fixed subset of cells and no means of acquiring the
remainder. The legal action space at such a state contains only choose
actions, since \(\mathcal{A}_{\text{reveal}}(s) = \varnothing\) by
construction. The indicator on the reveal branch of the Bellman
recursion is therefore never satisfied, and

\[V(s, d) = \max_{j \in \mathcal{J}} \mathcal{U}_{\text{choose}}(j, s)\]

for every \(d\). The value function is invariant to the planning
horizon, and the myopic and 2-step lookahead policies coincide at every
assigned state. The chance enumeration limit \(L\) is likewise
inoperative, since it governs the quantization of outcomes at reveal
branches only. We accordingly report a single normative choice per state
rather than one per planning horizon.

Because no inspections occur, \(C_{\text{tool}}(s) = 0\) at every
assigned state. The terminal utility reduces to

\[\mathcal{U}_{\text{choose}}(j, s) = - \mathbb{E}[P_j \mid s] \cdot \mathbb{E}\left[ \frac{1}{W_j} \;\Bigg|\; s \right]\]

evaluated over the empirical priors of the preceding section. Where
weight is revealed, \(\mathbb{E}[1/W_j \mid s] = 1/W_j\); where it is
withheld, the expected inverse weight over the empirical domain is
\(\mathbb{E}[1/W] = 0.400\), which is a common factor across
alternatives and does not affect the ranking. Where the cents are
withheld, \(\mathbb{E}[\text{Cents}] = 52.93¢\) enters all alternatives
equally; where the discount is withheld,
\(\mathbb{E}[\text{Discount}] = 20.0\%\) does the same. The whole-dollar
price and the list price are revealed at every assigned state, so their
priors never bind and the benchmark is insensitive to those two domains.

{\def\LTcaptype{apptbl}
\begin{longtable}[]{@{}
  >{\raggedright\arraybackslash}p{(\linewidth - 12\tabcolsep) * \real{0.1429}}
  >{\raggedright\arraybackslash}p{(\linewidth - 12\tabcolsep) * \real{0.1429}}
  >{\raggedright\arraybackslash}p{(\linewidth - 12\tabcolsep) * \real{0.1429}}
  >{\raggedleft\arraybackslash}p{(\linewidth - 12\tabcolsep) * \real{0.1429}}
  >{\raggedleft\arraybackslash}p{(\linewidth - 12\tabcolsep) * \real{0.1429}}
  >{\raggedleft\arraybackslash}p{(\linewidth - 12\tabcolsep) * \real{0.1429}}
  >{\raggedright\arraybackslash}p{(\linewidth - 12\tabcolsep) * \real{0.1429}}@{}}
\caption{{\label{apptbl-fixed-reveal-normative}}Expected cost per ounce
and normative choice at each assigned information state}\tabularnewline
\toprule\noalign{}
\begin{minipage}[b]{\linewidth}\raggedright
Environment
\end{minipage} & \begin{minipage}[b]{\linewidth}\raggedright
State
\end{minipage} & \begin{minipage}[b]{\linewidth}\raggedright
Revealed attributes
\end{minipage} & \begin{minipage}[b]{\linewidth}\raggedleft
\(\mathbb{E}[\$/\text{oz}]\)
\end{minipage} & \begin{minipage}[b]{\linewidth}\raggedleft
\end{minipage} & \begin{minipage}[b]{\linewidth}\raggedleft
\end{minipage} & \begin{minipage}[b]{\linewidth}\raggedright
Normative choice
\end{minipage} \\
\midrule\noalign{}
\endfirsthead
\toprule\noalign{}
\begin{minipage}[b]{\linewidth}\raggedright
Environment
\end{minipage} & \begin{minipage}[b]{\linewidth}\raggedright
State
\end{minipage} & \begin{minipage}[b]{\linewidth}\raggedright
Revealed attributes
\end{minipage} & \begin{minipage}[b]{\linewidth}\raggedleft
\(\mathbb{E}[\$/\text{oz}]\)
\end{minipage} & \begin{minipage}[b]{\linewidth}\raggedleft
\end{minipage} & \begin{minipage}[b]{\linewidth}\raggedleft
\end{minipage} & \begin{minipage}[b]{\linewidth}\raggedright
Normative choice
\end{minipage} \\
\midrule\noalign{}
\endhead
\bottomrule\noalign{}
\endlastfoot
& & & A & B & & \\
Study 1 & \$ & \emph{price\_dollars} & 1.8120 & 2.2121 & & Coffee A \\
Study 1 & \$+roast & \emph{price\_dollars}, \emph{roast\_date} & 1.8120
& 2.2121 & & Coffee A \\
Study 1 & \$+oz & \emph{price\_dollars}, \emph{weight} & 0.4529 & 0.5027
& & Coffee A \\
Study 1 & \$+¢+oz & \emph{price\_dollars}, \emph{price\_cents},
\emph{weight} & 0.4990 & 0.4545 & & Coffee B \\
& & & C & D & E & \\
Study 2 & LP & \emph{list\_price} & 2.5604 & 4.8008 & 3.2005 & Coffee
C \\
Study 2 & LP+roast & \emph{list\_price}, \emph{roast\_date} & 2.5604 &
4.8008 & 3.2005 & Coffee C \\
Study 2 & LP+\% & \emph{list\_price}, \emph{discount\_percentage} &
3.2005 & 4.8008 & 4.0007 & Coffee C \\
Study 2 & LP+oz & \emph{list\_price}, \emph{weight} & 0.8000 & 1.0000 &
0.7339 & Coffee E \\
Study 2 & LP+\%+oz & \emph{list\_price}, \emph{discount\_percentage},
\emph{weight} & 1.0000 & 1.0000 & 0.9174 & Coffee E \\
\end{longtable}
}

\emph{Note.} Entries are \(-\mathcal{U}_{\text{choose}}(j, s)\), the
expected effective unit cost of each alternative given the revealed
cells, in dollars per ounce. Lower values are preferred, and the
normative choice is the alternative minimizing the entry in its row.
Values are comparable within a row but not across rows: where weight is
withheld, the entry is multiplied by \(\mathbb{E}[1/W] = 0.400\), and
where weight is revealed it is divided by the observed weight, so the
two sets of rows are on different scales. Coffee C and Coffee D are
exactly indifferent at the complete Study 2 state by construction, both
at 1.0000, and both are inferior to Coffee E. The normative choice is
invariant to the planning horizon at every state for the reason given
above. State labels follow \quartoapptblref{apptbl-fixed-reveal-design}.

At three of the four Study 1 states and three of the five Study 2
states, the normative choice given the revealed cells is not the
alternative with the superior price-per-ounce at the true attribute
values. Selecting the lower whole-dollar alternative when the cents are
withheld, or the lowest-list-price alternative when weight is withheld,
is optimal conditional on the state while producing an economically
inferior outcome. The two criteria are distinguished throughout
Web Appendix F as state consistency and choice optimality respectively.

\section*{Web appendix I: LLM
configurations}\label{sec-appendix-configurations}
\addcontentsline{toc}{section}{Web appendix I: LLM configurations}

\subsection{Experimental parameters and default decoding
configuration}\label{experimental-parameters-and-default-decoding-configuration}

We retain the provider's default decoding configuration because our
estimand is the behavior of LLMs as deployed in realistic agentic
settings. This choice is increasingly not ours to make, as in July 2026
Google deprecated sampling parameters (e.g., temperature, top-p, and
top-k), with the API ignoring these parameters and rejecting them in
future releases \citep{google2026sampling}. Anthropic has similarly
deprecated custom sampling parameters \citep{anthropic2026sampling}.
Therefore, throughout all studies reported in the paper, we have used
the provider default sampling parameters.

\subsection{Application programming interfaces and client
libraries}\label{application-programming-interfaces-and-client-libraries}

All experimental queries were dispatched programmatically using official
provider client libraries. Google Gemini models were accessed via the
Google GenAI Python SDK (\emph{google-genai}) utilizing the unified
\emph{models.generate\_content} interface. OpenAI models were queried
via the OpenAI Python SDK (\emph{openai}) using the Responses interface
(\emph{client.responses.create} and \emph{client.responses.parse}),
which integrates structured output validation with native reasoning
controls. Anthropic models were executed via the Anthropic Python SDK
(\emph{anthropic}) through the Messages API
(\emph{client.messages.create}).

\subsection{Model identifiers, sampling parameters, and reasoning
efforts}\label{model-identifiers-sampling-parameters-and-reasoning-efforts}

The experimental suite encompasses eight frontier models across three
major families.

\textbf{Google Gemini family.} Gemini 3.5 Flash Lite (API model
identifier \emph{gemini-3.5-flash-lite}), Gemini 3.8 Flash (API model
identifier \emph{gemini-3.8-flash}), and Gemini 3.1 Pro (API model
identifier \emph{gemini-3.1-pro-preview}) were queried using standard
provider defaults, including a default sampling temperature of
\(T = 1.0\) and nucleus sampling threshold of top-\(p = 0.95\). Gemini
3.5 Flash Lite operates under default \emph{minimal} thinking level,
Gemini 3.8 Flash operates under default \emph{medium} thinking level,
and Gemini 3.1 Pro operates under default \emph{high} thinking level.

\textbf{OpenAI GPT-5.6 family.} GPT-5.6 Luna (API model identifier
\emph{gpt-5.6-luna}), GPT-5.6 Terra (API model identifier
\emph{gpt-5.6-terra}), and GPT-5.6 Sol (API model identifier
\emph{gpt-5.6-sol}) were queried using default sampling parameters with
temperature \(T = 1.0\), top-\(p = 1.0\), and null presence and
frequency penalties. All three GPT-5.6 variants were executed under
default \emph{medium} reasoning effort (\emph{reasoning=\{``effort'':
``medium'', ``summary'': ``auto''\}}), allowing internal
chain-of-thought processing before emitting final choices.

\textbf{Anthropic Claude family.} Claude Sonnet 5 (API model identifier
\emph{claude-sonnet-5}) and Claude Opus 5 (API model identifier
\emph{claude-opus-5}) were evaluated using default generation
parameters, comprising temperature \(T = 1.0\), top-\(p = 1.0\). Both
Claude models operated under adaptive thinking mode
(\emph{thinking=\{``type'': ``adaptive'', ``display'':
``summarized''\}}) with default \emph{high} reasoning effort allocation
(\emph{output\_config=\{``effort'': ``high''\}}).

\subsection{Output structure and replication
controls}\label{output-structure-and-replication-controls}

Across all discrete choice and interactive Tool-Lab experiments, outputs
were constrained to valid JSON structures matching Pydantic schemas
without restricting the internal latent sampling distribution. Because
autoregressive sampling produces stochastic variation across independent
queries at default temperature, unless otherwise noted, each
experimental cell was evaluated over 100 independent Monte Carlo
sessions with zero state retention between trials.

{\def\LTcaptype{apptbl}
\begin{longtable}[]{@{}
  >{\raggedright\arraybackslash}p{(\linewidth - 16\tabcolsep) * \real{0.1111}}
  >{\raggedright\arraybackslash}p{(\linewidth - 16\tabcolsep) * \real{0.1111}}
  >{\raggedright\arraybackslash}p{(\linewidth - 16\tabcolsep) * \real{0.1111}}
  >{\raggedright\arraybackslash}p{(\linewidth - 16\tabcolsep) * \real{0.1111}}
  >{\raggedright\arraybackslash}p{(\linewidth - 16\tabcolsep) * \real{0.1111}}
  >{\raggedright\arraybackslash}p{(\linewidth - 16\tabcolsep) * \real{0.1111}}
  >{\raggedright\arraybackslash}p{(\linewidth - 16\tabcolsep) * \real{0.1111}}
  >{\raggedright\arraybackslash}p{(\linewidth - 16\tabcolsep) * \real{0.1111}}
  >{\raggedright\arraybackslash}p{(\linewidth - 16\tabcolsep) * \real{0.1111}}@{}}
\caption{{\label{apptbl-model-specifications}}Large language model
specifications, default sampling parameters, and reasoning efforts}\tabularnewline
\toprule\noalign{}
\begin{minipage}[b]{\linewidth}\raggedright
Provider
\end{minipage} & \begin{minipage}[b]{\linewidth}\raggedright
Model name
\end{minipage} & \begin{minipage}[b]{\linewidth}\raggedright
API identifier
\end{minipage} & \begin{minipage}[b]{\linewidth}\raggedright
Client library
\end{minipage} & \begin{minipage}[b]{\linewidth}\raggedright
API endpoint
\end{minipage} & \begin{minipage}[b]{\linewidth}\raggedright
Default reasoning effort
\end{minipage} & \begin{minipage}[b]{\linewidth}\raggedright
Temperature
\end{minipage} & \begin{minipage}[b]{\linewidth}\raggedright
Top-\(p\)
\end{minipage} & \begin{minipage}[b]{\linewidth}\raggedright
Max output tokens
\end{minipage} \\
\midrule\noalign{}
\endfirsthead
\toprule\noalign{}
\begin{minipage}[b]{\linewidth}\raggedright
Provider
\end{minipage} & \begin{minipage}[b]{\linewidth}\raggedright
Model name
\end{minipage} & \begin{minipage}[b]{\linewidth}\raggedright
API identifier
\end{minipage} & \begin{minipage}[b]{\linewidth}\raggedright
Client library
\end{minipage} & \begin{minipage}[b]{\linewidth}\raggedright
API endpoint
\end{minipage} & \begin{minipage}[b]{\linewidth}\raggedright
Default reasoning effort
\end{minipage} & \begin{minipage}[b]{\linewidth}\raggedright
Temperature
\end{minipage} & \begin{minipage}[b]{\linewidth}\raggedright
Top-\(p\)
\end{minipage} & \begin{minipage}[b]{\linewidth}\raggedright
Max output tokens
\end{minipage} \\
\midrule\noalign{}
\endhead
\bottomrule\noalign{}
\endlastfoot
Google & Gemini 3.5 Flash Lite & \emph{gemini-3.5-flash-lite} &
\emph{google-genai} & \emph{models.generate\_content} & Minimal & 1.0 &
0.95 & - \\
Google & Gemini 3.8 Flash & \emph{gemini-3.8-flash} &
\emph{google-genai} & \emph{models.generate\_content} & Medium & 1.0 &
0.95 & - \\
Google & Gemini 3.1 Pro & \emph{gemini-3.1-pro-preview} &
\emph{google-genai} & \emph{models.generate\_content} & High & 1.0 &
0.95 & - \\
OpenAI & GPT-5.6 Luna & \emph{gpt-5.6-luna} & \emph{openai} &
\emph{client.responses.create} & Medium & 1.0 & 1.0 & - \\
OpenAI & GPT-5.6 Terra & \emph{gpt-5.6-terra} & \emph{openai} &
\emph{client.responses.create} & Medium & 1.0 & 1.0 & - \\
OpenAI & GPT-5.6 Sol & \emph{gpt-5.6-sol} & \emph{openai} &
\emph{client.responses.create} & Medium & 1.0 & 1.0 & - \\
Anthropic & Claude Sonnet 5 & \emph{claude-sonnet-5} & \emph{anthropic}
& \emph{client.messages.create} & High (adaptive) & 1.0 & 1.0 &
10,000 \\
Anthropic & Claude Opus 5 & \emph{claude-opus-5} & \emph{anthropic} &
\emph{client.messages.create} & High (adaptive) & 1.0 & 1.0 & 10,000 \\
\end{longtable}
}

\section*{Web appendix J: Software details}\label{sec-appendix-software}
\addcontentsline{toc}{section}{Web appendix J: Software details}

Parallel mediation models were estimated in R (version 4.6.1) using brms
\citep[version 2.23.0;][]{burkner2017brms} with the \emph{cmdstanr}
backend and Stan (version 2.39.0).

\end{document}